\documentclass[saps,prb,twocolumn,showpacs,10pt,floatfix]{revtex4-1}
\usepackage{color}
\usepackage{bbm}
\usepackage{units}
\usepackage{graphicx}
\usepackage{float}
\usepackage{natbib}
\usepackage{setspace}

\begin{document}

\title{The metal-insulator transition in disordered solids: 
\protect\linebreak How theoretical prejudices influence its characterization
\protect\linebreak {\it A critical review of analyses of experimental data}}

\author{Arnulf M\"obius}
\email{a.moebius@ifw-dresden.de}
\affiliation{Institute for Theoretical Solid State Physics,
IFW Dresden, D-01171 Dresden, Germany}

\begin{abstract}
In a recent experimental study, Siegrist et al.\ [{\it Nature
Materials} {\bf 10}, 202--208 (2011)] investigated the metal-insulator
transition (MIT) induced by annealing in GeSb$_2$Te$_4$ and related
phase-change materials. The authors concluded that these materials
exhibit a discontinuous MIT with a finite minimum metallic
conductivity, and that they violate the Mott criterion for the
critical charge carrier concentration. The striking contrast between
their work and reports on many other disordered substances from the
last 35 years motivates the present in-depth study of the influence of
the MIT criterion used on the character of the MIT derived.

First, we discuss in detail the inherent biases of various approaches
to locating the MIT, that is to discriminating between just metallic
samples and weakly insulating ones which exhibit hopping conduction
with very low characteristic temperatures. Second, reanalyzing the
GeSb$_2$Te$_4$ data, we show that this material resembles other
disordered solids to a large extent: According to a widely-used
approach, the temperature dependences of the conductivity,
$\sigma(T)$, of GeSb$_2$Te$_4$ may likewise be interpreted in terms of
a continuous MIT. Careful checking the justification of the
corresponding fits, however, uncovers discrepancies which currently
render an unambiguous interpretation impossible. Moreover, we give
several arguments against the violation of the Mott criterion stated
by Siegrist et al.; it is traced back to the authors' inappropriate
considering shallow instead of deep impurity states. Third, examining
previous experimental studies of crystalline Si:As, Si:P, Si:B, Ge:Ga,
CdSe:In, Cd$_{0.95}$Mn$_{0.05}$Se,
Cd$_{0.95}$Mn$_{0.05}$Te$_{0.97}$Se$_{0.03}$:In, disordered Gd, and
nanogranular Pt\hbox{-}C, we show that substantial problems in the
interpretation of $\sigma(T)$ can also be detected in numerous studies
which claim the MIT to be continuous: Evaluating the logarithmic
derivative $\mbox{d} \ln \sigma / \mbox{d} \ln T$ highlights serious
inconsistencies. In part, they are common to all such studies and seem
to be generic, in part, they vary from experiment to experiment.
Fourth, for four qualitatively different phenomenological models of
the temperature and control parameter dependence of the conductivity,
we present the respective flow diagrams of this logarithmic
derivative. In consequence, the likely generic inconsistencies seem to
originate from the MITs being discontinuous and occurring when
$\mbox{d} \sigma / \mbox{d} T = 0$ at infinitely small $T$, in
contradiction to most of the original interpretations. The
experiment-dependent inconsistencies, however, cannot be understood
in this way; they may arise from measurement problems.

Because of the large number and diversity of the experiments
considered, the inconsistencies uncovered in this review
provide overwhelming evidence against the common, localization theory
motivated interpretations, which are based on the assumption that
$\sigma(T) \propto T^p$ with $p = 1/2$ or $1/3$ at the MIT. Thus, the
question about the character of the MIT in disordered solids has to
be considered as still open. The primary challenges now lie in
improving the measurement precision and accuracy, rather than in 
extending the temperature range, and in developing a microscopic
theory which explains the seemingly generic features of
$\mbox{d} \ln \sigma / \mbox{d} \ln T$.
\end{abstract}
\date{May 10, 2017}
\maketitle


\tableofcontents

\section{Introduction}

For more than fifty years, localization in disordered systems, in
particular the corresponding metal-insulator transition (MIT), has
attracted a lot of interest from both theoreticians and 
experimentalists.\cite{Lag.etal.09,Abr.10,Mott.90,Mott.Davi,Lee.Rama,
Beli.Kirk, Edw.95,Edw.etal.95,L98,Moe.Adk,Ever.Mirl} Milestones on 
this way have been Anderson noting the absence of diffusion in certain
lattices with disorder,\cite{Ande.58} Mott's concept of the minimum
metallic conductivity,\cite{Mott.72} the scaling theory of
localization,\cite{Abra.etal} and the renormalization group approach
incorporating electron-electron interaction into localization
theory.\cite{Fink.83a} Experimentally, localization in
three-dimensional systems has been studied in a large number of
disordered solids, such as heavily doped crystalline semiconductors
(in which the disorder arises from the randomly positioned impurities),
amorphous transition-metal semiconductor alloys, granular metals, and
nanocrystalline substances.\cite{Edw.95,Edw.etal.95,L98,Moe.Adk} Many
of these solids are or may become application relevant; therefore they
are often considered to be among the materials.\cite{Mott.Davi} In
various experiments, the MIT has been triggered by diverse control
parameters: composition / doping, stress, magnetic field, light, as
well as structure, see, for example, Refs.\
\onlinecite{Yama.etal.67,Zabr.Zino,Shaf.etal.89,Her.etal.83,
Moe.etal.83,Moe.etal.99,Tho.etal.83,Waf.etal.99,Bis.etal.83,
vMo.etal.83,Wojt.etal.86,Dietl.etal.86,Katsu.87,Glod.etal.93,
Lei.etal.98,Sie.etal.11,Mis.etal.11,Giv.Ova.12}; in part, these
publications substantially contradict each other.

The MIT in disordered solids is primarily a zero-temperature
phenomenon. -- We do not consider here the case of an MIT which is 
interrelated with a structural or magnetic phase transition. Such
transitions usually occur at some finite (i.e., nonzero) temperature,
for example in VO$_2$ and V$_2$O$_3$.\cite{Mott.Zina} By means of
doping or applying pressure, the critical temperature may be reduced
to zero.\cite{McW.etal.73} -- Therefore, in the field of localization,
a sample is said to be metallic if its dc conductivity, $\sigma$, is
expected to tend to some finite value as the temperature, $T$, goes to
zero, and it is called insulating if $\sigma$ is expected to tend to
zero as $T \rightarrow 0$. Of course, for the insulating samples,
$\sigma$ is finite at any finite $T$ due to thermally activated
non-metallic transport, in particular variable-range
hopping.\cite{cla}

Hence, in studying the MIT, each evaluation of experimental data 
includes some $T \rightarrow 0$ extrapolation: Early work on two- and
three-dimensional systems judged the $\sigma(T)$ curves from a rather
global perspective. Only samples for which $\sigma$ drops
exponentially with decreasing $T$ were classified as insulating,
while all other samples were regarded as
metallic.\cite{Mott.72a,Adk.78} Later, for three-dimensional systems,
when more dense sets of control parameter values were considered, the
focus concentrated on $\sigma(T \rightarrow 0)$ extrapolations of
low-temperature data from the control parameter region thought of as
metallic.\cite{Lee.Rama} These analyses are based on microscopic
theories yielding augmented power laws, $\sigma = a + b\,T^p$, as
derived in Refs.\ \onlinecite{Alt.Aro.79} and \onlinecite{News.Pepp}.
In this approach, samples with positive extrapolated $\sigma(T = 0)$
value are regarded as metallic, while all other samples are classified
as insulating.

Simultaneously with this change in the data analysis approach, the
majority opinion on the character of the MIT in three-dimensional 
samples turned: It shifted from initially supporting Mott's idea of a
discontinuous transition with a finite minimum metallic conductivity
toward favoring a continuous transition in accord with the scaling
theory of localization.\cite{Lee.Rama} Nowadays, most of the experts
in this field seem to be certain about the continuity of the MIT, see,
for example, Ref.\ \onlinecite{HvL.11}. Nevertheless, two repeatedly
observed phenomena are in conflict with the continuity hypothesis:
the scaling of the $T$ dependences of $\sigma$ in the hopping
region\cite{Moe.85,Sara.Dai.02} and the existence of specific
low-temperature minima in the $T$ dependences of the logarithmic $T$
derivative of $\sigma$, \linebreak
$\mbox{d} \ln \sigma / \mbox{d} \ln T$.\cite{Moe.Adk,Moe.etal.99}
-- Because of the focus on properties of individual samples, we here
mostly use total derivative symbols although $\sigma$ depends not only
on $T$ but also on a control parameter. --

The concurrence of the changes in the data analysis approach and in
the majority opinion on the character of the MIT provokes a naive
question: {\it May it be that this opinion change was caused merely by
the difference between the biases inherent to the analysis approaches
rather than by improvements in the experiments?}

To support our question, we recall the following: In the literature,
two different MIT criteria have been used in confirming the scaling
theory of localization for the three-dimensional case and in
disproving it for the two-dimensional case, respectively. In the
former situation, the augmented power law extrapolation criterion has
been applied to determine the critical value of the control
parameter.\cite{Lee.Rama,HvL.11} In contrast, in the latter case, the
sign change of the derivative, $\mbox{d} \sigma / \mbox{d} T$, at the
lowest measuring temperature has been considered as indicator of the
change in the nature of conduction.\cite{Krav.etal.94,Krav.Sara} One
exception within the studies of two-dimensional systems acts as
additional motivation for our question: Reference
\onlinecite{Feng.etal.01} used the $T \rightarrow 0$ extrapolation
ansatz $\sigma = a + b\,T^2$ and obtained an empirical model of a
\linebreak continuous MIT.

At this point, the reader is invited to scroll down for a moment to
take a brief look at Figures \ref{w_fig_cond}, \ref{castner},
\ref{sip_1}, \ref{sip_2}, \ref{sarachik}, \ref{dietl_c}, \ref{dietl3},
and \ref{ptc}, which regard various disordered solids: All these
diagrams contrast $T$ dependences of
$\mbox{d} \ln \sigma / \mbox{d} \ln T$ obtained by direct numerical
differentiation of experimental data with dashed-dotted curves derived
from augmented power laws, $\sigma = a + b\,T^p$. The slopes of 
the former and of the latter relations differ qualitatively from each
other, again and again! This observation, together with the historical
remarks above, will surely awaken or strengthen the reader's interest
in a detailed analysis of the $T \rightarrow 0$ extrapolation problem,
which is central to this review.

Further motivation for our above question about the role of the
interpretation bias comes from a recent investigation of the MIT in
specific three-dimensional disordered systems. The report by Siegrist
et al.\ on GeSb$_2$Te$_4$ and similar phase-change
materials\cite{Sie.etal.11} claims to obtain surprising results:
Annealing amorphous films of such substances induces a crystallization
process with increasing temperature whereby a nanocrystalline
structure is formed.\cite{Sie.etal.11} During this transformation,
$\sigma$ increases by orders of magnitude while $\sigma(T)$ changes
qualitatively, which indicates an MIT.\cite{Sie.etal.11} Classifying
the samples according to the sign of $\mbox{d} \rho / \mbox{d} T$ at
the measuring temperature, Siegrist et al.\ conclude that the studied
phase-change materials exhibit a finite minimum metallic conductivity,
in contrast to various other disordered substances. Moreover, the
authors state that the phase-change materials violate the Mott
criterion for the critical charge carrier concentration. They
interpret these features as originating from an ``unparalleled quantum
state of matter'' resulting from ``pronounced disorder but weak
electron correlation'',\cite{Sie.etal.11} see also Ref.\
\onlinecite{Sch.11}.

Remarkably, the work by Siegrist et al.\cite{Sie.etal.11} differs from
previous publications which claim continuity of the MIT not only
concerning the substance investigated but also with respect to the
data evaluation approach used. This again raises the question about
the influence of the choice of the data analysis method on the
character of the MIT obtained. 

Therefore, we here scrutinize the justification of the conclusions of
Ref.\ \onlinecite{Sie.etal.11}, reanalyzing data from this work. Due
to the interpretation uncertainties mentioned above, we take a
neutral, phenomenological perspective and avoid, as far as possible,
any bias caused by focusing on a particular microscopic theory. Our
present study shows that the $\sigma(T)$ for GeSb$_2$Te$_4$ resemble
results from previous studies on disordered solids to a large extent.
A part of the data can be well approximated by the ansatz
$\sigma = a + b\,T^{1/2}$, so that, in the same way as in other
investigations, the MIT could be characterized as continuous. However,
when checking the justification of this approach by studying the
behavior of $\mbox{d} \ln \sigma / \mbox{d} \ln T$, new insight is
gained: Both the sample classifications according to the sign of
$\mbox{d} \rho / \mbox{d} T$, on the one hand, and according to the
fit of the augmented power law $\sigma = a + b\,T^{1/2}$ to the
measured data, on the other hand, are called into question.

The situation gets even more complicated when data from a subsequent
GeSb$_2$Te$_4$ study, Ref.\ \onlinecite{Vol.etal.15}, by three of the
authors of Ref.\ \onlinecite{Sie.etal.11} are additionally taken into
account: Therein, the $T$ range was extended by one order of
magnitude, and continuity of the MIT was concluded from augmented
power law approximations of the measured $\sigma(T)$. However, as will
be shown in our analysis of the $T$ dependences of
$\mbox{d} \ln \sigma / \mbox{d} \ln T$, also these augmented power law
approximations substantially fail close to the MIT.

The observations described in the previous two paragraphs suggest to
take a broader view. Thus, in the following, we here examine various
publications on the MIT in other solids, Refs.\
\onlinecite{Shaf.etal.89, Stu.etal.93, Ros.etal.80, Ros.etal.83,
Tho.etal.83, Waf.etal.99, Sara.Dai.02, Dai.etal.91, Wata.etal.98,
Wojt.etal.86, Dietl.etal.86, Glod.etal.93, Mis.etal.11, Sac.etal.11},
from the same perspective. In doing so, one is confronted with similar
problems as for GeSb$_2$Te$_4$: In numerous cases, the behavior of
$\mbox{d} \ln \sigma / \mbox{d} \ln T$ obtained numerically from the
measurements without making any assumptions is not consistent with
the common interpretation in terms of a continuous MIT with, on the
metallic side, $\sigma(T)$ following $\sigma = a + b\,T^p$ where
$p = 1/2$ or $1/3$. More precisely, for many samples which are
classified as metallic in the publications cited above, 
$\mbox{d} \ln \sigma / \mbox{d} \ln T$ in fact increases with
decreasing $T$ over a wide $T$ range, whereas a decrease is expected
according to the ansatz $\sigma(T) = a + b\,T^p$.

One result of this examination is particularly striking: For various
solids, there are even allegedly metallic samples for which the $T$
dependence of $\mbox{d} \ln \sigma / \mbox{d} \ln T$ has negative
slope in spite of $\sigma(T)$ decreasing so slowly with $T$ that 
$\mbox{d} \ln \sigma / \mbox{d} \ln T \ll 1/3$. -- Remarkably, below
about 2~K, also one of the CdSe:In samples from Refs.\
\onlinecite{Zha.etal.90,Aha.etal.92,Zha.Sara.95} shows exactly this
correlation, although it is claimed to exhibit hopping conduction in
those publications. --

These findings provide valuable information about the character of the
MIT: First, the disproofs of the finite minimum metallic conductivity
hypothesis in the analyzed publications, which are based on augmented
power law fits with $p = 1/2$ or $1/3$, cannot be considered
conclusive. Second, the here observed correlation between value and
sign of the slope of $\mbox{d} \ln \sigma / \mbox{d} \ln T(T)$
indicates that, in the limit $T \rightarrow 0$, the control parameter
dependence of $\sigma$ is very likely discontinuous at the MIT. -- We
substantiate this implication by comparing flow diagrams of
$\mbox{d} \ln \sigma / \mbox{d} \ln T(T)$ obtained from four
qualitatively different phenomenological models in a separate section
of the present review. --

In several cases, however, the $T$ dependence of
$\mbox{d} \ln \sigma / \mbox{d} \ln T$ exhibits a maximum which is
incompatible with the seemingly generic behavior of this quantity, as
it has been described and interpreted in the previous three
paragraphs. Since these maxima are experiment-specific, further, very
careful investigations of the same $T$ range are needed. In this way,
we identify key points for the design of future related experiments. 

The present review is organized as follows: Section 2 discusses
various approaches to the precise determination of the transition
point between metallic and insulating phases, the first and most
important difficulty of experiments on the MIT. -- Readers in a hurry
may focus on Subsections 2.2, 2.3, 2.4, and, in particular, 2.6. --
Section 3 is devoted to GeSb$_2$Te$_4$: In its first part, Subsection
3.1, $\sigma(T)$ data from Refs.\ \onlinecite{Sie.etal.11} and
\onlinecite{Vol.etal.15} are reanalyzed by means of alternative
approaches. In doing so, we demonstrate inconsistencies in the data
sets from Ref.\ \onlinecite{Sie.etal.11} which render it impossible to
reach definite conclusions about the nature of conduction for three of
the samples. Moreover, we explain why a part of the sample
classifications of Ref.\ \onlinecite{Vol.etal.15} seems to be
incorrect, so that the characterization of the MIT therein is called
into question. In the second part of Section 3, Subsection 3.2,
several arguments against the hypothetical violation of the Mott
criterion by GeSb$_2$Te$_4$ are presented; this deviation is found to
be not real but to result from an invalid assumption on the
participating states. -- Subsection 3.2 may be skipped on first
reading. -- Section 4 compares our findings on GeSb$_2$Te$_4$ with
results of a multitude of studies on various other disordered solids.
Here we show that, in numerous publications favoring continuity of the
MIT, severe interpretation problems can be uncovered by taking the
behavior of $\mbox{d} \ln \sigma / \mbox{d} \ln T$ into consideration.
Section 5, evaluating simple phenomenological hypotheses, studies how
the character of the MIT determines qualitative properties of sets of
$T$ dependences of $\mbox{d} \ln \sigma / \mbox{d} \ln T$ for various
control parameter values. It demonstrates that such flow diagrams
obtained directly from experimental data can be an informative
fingerprint of the character of the zero-temperature phenomenon MIT.
Finally, Section 6 summarizes our results and draws conclusions for
future studies.

In Appendix A, relations between the effective mass, permittivity,
critical charge carrier concentration of the MIT, charge carrier
concentration at which $\mbox{d} \sigma / \mbox{d} T$ changes sign in
the low-$T$ limit, and corresponding $\sigma$ value are deduced by
means of dimensional analyses. These results are  robust with respect
to a broad class of theoretical approximations, also regarding the
incorporation of the electron-electron interaction. Appendix B is
devoted to mathematical aspects in the interpretation of observations
of scaling of $T$ dependences of $\sigma$, in particular to the hidden
suppositions in this way of concluding the existence of a finite
minimum metallic conductivity. Finally, to ensure that our data
evaluations are easily reproducible, Appendix C explains the
sophisticated numerical differentiation method for functions given by
noisy values at non-equidistant points which is the basic tool for a
large part of the reanalyses presented here.

\section{Criteria for detecting the MIT}

Although, at first glance, the identification of metallic and
insulating phases of disordered solids seems to be a simple task,
it is far from trivial.\cite{Ros.etal.94,Moe.etal.99} Therefore, the
general aspects of various approaches, in particular their implicit
preconditions and consequences, are discussed in detail in this
section.

\subsection{Sign change of $\mbox{d} \rho / \mbox{d} T$ at the
measuring temperature}

Siegrist et al.\cite{Sie.etal.11} describe the current state of
the literature on three-dimensional systems as follows: In
experimental studies, the sign of the temperature derivative of the
resistivity, $\mbox{d} \rho / \mbox{d} T$, is taken as criterion,
where positive and negative $\mbox{d} \rho / \mbox{d} T$ indicate
metallic and insulating behavior, respectively.\cite{Sie.etal.11}
Note that this classification refers to the current measuring
temperature. -- Alternatively, as done below,
$\mbox{d} \sigma / \mbox{d} T$ can be considered which always has the
opposite sign. -- Furthermore, Siegrist et al.\ state that this
approach by experimentalists contrasts with theoretical
investigations.\cite{Sie.etal.11} Those studies consider transport as
metallic if, as $T \rightarrow 0$, the conductivity $\sigma$ tends to
a finite value, and as insulating if $\sigma$
vanishes.\cite{Sie.etal.11} 

That summary of the literature is incomplete: As already mentioned in
Section 1 and discussed in more detail in Subsection 2.3 as well as
Sections 3 and 4, also quite a number of experimental studies have
focused on $\sigma(T \rightarrow 0)$ extrapolations based on
microscopic theories for the metallic phase. They concern not only
doped crystalline semiconductors but also amorphous transition metal
semiconductor alloys and granular systems, see, for example, Refs.\
\onlinecite{Tho.etal.83}, \onlinecite{Sac.etal.11}, and 
\onlinecite{Dod.etal.81}.

More importantly, the above usage of notions by Siegrist et al.,
which, by the way, seems to be rather popular among nonspecialists in
localization still nowadays, is misleading: The transition from
metallic to insulating behavior at $T = 0$ and the sign change of
$\mbox{d} \rho / \mbox{d} T$ at finite $T$ are two different, only
loosely related phenomena. They should not be confused by using the
same term ``metal-insulator transition'' for both of them. This is
illustrated by qualitative considerations in the following three
paragraphs.

Consider $\sigma$ as function of $T$ and some control parameter $x$.
In the limit $T \rightarrow 0$, the conductivity is identical to zero
in the insulating region. -- From the empirical perspective, the
existence of such a region is a hypothesis, although a very plausible
one. Consider, for example, Figure 2 of Ref.\
\onlinecite{Tho.etal.83}: The convergence of the finite $T$ curves to
a sharp transition is not proven. In principle,
$\lim_{T \rightarrow 0} \sigma(T,x)$ might also continuously vanish in
some very rapid manner. -- The onset of metallic conduction happens at
the $x$ value where $\lim_{T \rightarrow 0} \sigma(T,x)$ suddenly
starts to deviate from zero. There, this function of $x$ is not smooth
but has some peculiarity. That means it is either discontinuous or at
least not infinitely often differentiable. This non-analytic behavior
of $\lim_{T \rightarrow 0} \sigma(T,x)$ indicates a phase transition,
more precisely, a quantum phase transition. 

On the contrary, the room-temperature resistivity, 
$\rho(T = 293\ {\rm K},x)$, seems to be a smooth function of $x$ in
the region of the sign change of 
$\mbox{d} \rho / \mbox{d} T(T = 293\ {\rm K},x)$. To the best of our
knowledge, no indication for any peculiarity (non-analyticity) of
$\rho(T = 293\ {\rm K},x)$ correlated with the sign change of 
$\mbox{d} \rho / \mbox{d} T(T = 293\ {\rm K},x)$ has been reported
up to now. Thus, the sign change of
$\mbox{d} \rho / \mbox{d} T(T = 293\ {\rm K},x)$ very likely does not
arise from a phase transition.

Moreover, if there are two interfering mechanisms yielding additive
$T$-dependent resistivity contributions, $\rho_i(T,x)$, with different
signs of $\mbox{d} \rho_i / \mbox{d} T$ (as in the case of the Kondo
effect), $\rho(T, x = {\rm const.})$ may exhibit a maximum or minimum,
as two curves in the shaded transition region in Figure 2 of Ref.\
\onlinecite{Sie.etal.11} do. Although $\mbox{d} \rho / \mbox{d} T = 0$
at such an extremum, this feature cannot be interpreted as transition
between qualitatively different phases since
$\rho(T, x = {\rm const.})$ does not exhibit any non-analyticity
there.

Nevertheless, one might expect the finite-$T$ criterion 
$\mbox{d} \rho / \mbox{d} T = 0$ to yield approximate results for the
critical value of the control parameter $x$ and possibly for the
hypothetical minimum metallic conductivity and the hypothetical
maximum metallic resistivity. To which extent does such an estimate
depend on the measuring temperature? To answer this question, we now
compare room-temperature data with measurements at temperatures of the
order of 1~K for four groups of materials. Thereby, as working
definition, we use the terms room-temperature minimum metallic
conductivity, $\sigma_{\rm rtmm}$, and low-temperature minimum
metallic conductivity, $\sigma_{\rm ltmm}$, to denote the $\sigma$
values which are related to $\mbox{d} \rho / \mbox{d} T = 0$ in the
respective temperature regions.

First, for GeSb$_2$Te$_4$, Ref.\ \onlinecite{Sie.etal.11} obtains a
maximum metallic resistivity of 2 to $3\ {\rm m \Omega cm}$
considering the temperature range from 5 to $290\ {\rm K}$. Compared
to the logarithmic scale in Figure 3 of Ref.\
\onlinecite{Sie.etal.11}, the uncertainty factor of 1.5 is rather
small. The only sample with $\rho$ falling in this interval was
prepared by annealing at $275\ {\rm ^oC}$. Its $\rho(T)$ is almost
constant over the whole temperature range considered; it varies far
less than the respective $\rho(T)$ of the ``neighboring'' samples,
which were annealed at 250 and $300\ {\rm ^oC}$. Thus, the values of
the critical annealing temperature and of the minimum metallic
conductivity depend only slightly on the measuring temperature, so
that in this case $\sigma_{\rm rtmm} \approx \sigma_{\rm ltmm}$.

Second, the Mooij rule,\cite{Moo.73} an empirical relation between the
room-temperature values of $\mbox{d} \rho / \mbox{d} T$ and $\rho$,
states: For a large number of disordered alloys, the sign of
$\mbox{d} \rho / \mbox{d} T$ changes from plus to minus with
increasing $\rho$ at about $150 \pm 50\ {\rm \mu \Omega cm}$.
Disordered solids with
$\rho(300\ \mbox{K}) \lesssim 1000\ {\rm \mu \Omega cm}$, however, do 
usually not exhibit any increase in $\rho(T)$ by several orders of 
magnitude with decreasing $T$, provided they do not undergo some phase
transition at finite $T$. Accordingly, 
$\sigma_{\rm rtmm} > 5\ \sigma_{\rm ltmm}$.

Third, consider the amorphous Si$_{1-x}$Cr$_x$ films from Refs.\ 
\onlinecite{Moe.etal.83} and \onlinecite{Moe.etal.85}, prepared by
electron-beam evaporation. Their Cr contents cover the range from
$x = 0.09$ to 0.26. At low $T$, $\mbox{d} \sigma / \mbox{d} T$ changes
sign when $x \approx 0.15$ and
$\sigma \approx 270\ {\rm \Omega^{-1} cm^{-1}}$, see Figures 3 and 5
of Ref.\ \onlinecite{Moe.etal.83} and Figure 7 of Ref.\
\onlinecite{Moe.etal.85}. At room-temperature, however, up to
$x = 0.26$, where $\sigma = 1360\ {\rm \Omega^{-1} cm^{-1}}$, all
samples exhibit positive $\mbox{d} \sigma / \mbox{d} T$, see Figures 1
and 7 of Ref.\ \onlinecite{Moe.etal.85}. Thus, at room-temperature,
they all are to be regarded as insulating according to the
$\mbox{d} \rho / \mbox{d} T < 0$ criterion. Therefore, when using only
room-temperature data, one would overestimate the critical Cr content 
at least by a factor of 1.7, and 
$\sigma_{\rm rtmm} > 5\ \sigma_{\rm ltmm}$.

Fourth, concerning localization, the best investigated material is
heavily doped crystalline Si:P.\cite{L98,HvL.11} In the mK region,
$\mbox{d} \sigma / \mbox{d} T$ changes sign at a P concentration of
about $4 \times 10^{18}\ {\rm cm^{-3}}$, when 
$\sigma \approx 60\ {\rm \Omega^{-1} cm^{-1}}$, see Figure 3 of Ref.\ 
\onlinecite{Stu.etal.93}, compare Ref.\ \onlinecite{Ros.etal.81}. At
room-temperature, however, due to the electron-phonon scattering in
the exhaustion region, $\mbox{d} \rho / \mbox{d} T$ stays positive
within the very wide $\sigma$ range from $10^{-3}$ to
$10^3\ {\rm \Omega^{-1} cm^{-1}}$, except for a small interval around
$10^2\ {\rm \Omega^{-1} cm^{-1}}$, see Figure 1 of the NIST study
Ref.\ \onlinecite{Bul.etal.68}. (Si:P does not comply with Mooij's
rule.) This $\sigma$ range is related to a P concentration range from
roughly $10^{13}$ to about $10^{20}\ {\rm cm^{-3}}$, see Figure 4 of
Ref.\ \onlinecite{Thu.etal.80}. Hence, for room-temperature, inferring
the character of conduction from the sign of
$\mbox{d} \rho / \mbox{d} T$ fails dramatically: Samples are marked as
metallic even if the P concentration is smaller by a factor of $10^5$
than the critical concentration obtained from low-temperature
measurements. Furthermore, in contrast to the above considered
materials, $\sigma_{\rm rtmm} < 2 \times 10^{-5}\ \sigma_{\rm ltmm}$
for the region of low P concentrations.

In summary, the classification of the conduction character according
to the sign of $\mbox{d} \rho / \mbox{d} T$ at an arbitrarily chosen
measuring temperature is usually highly questionable. The reason is
the absence of any non-analyticity in the control parameter dependence
of the conductivity at such hypothetical MIT points. Moreover, this
criterion suffers from considerable uncertainties. In particular, when
it is applied to room-temperature data, both, false-metallic and
false-insulating classifications in comparison to corresponding
low-temperature evaluations are not unlikely.

\subsection{Sign change of $\mbox{d} \rho / \mbox{d} T$ in the limit
as $T \rightarrow 0$}

In the previous subsection, we demonstrated that the identification of
the nature of conduction according to the sign of
$\mbox{d} \rho / \mbox{d} T$ at some arbitrary measuring temperature
leads to severe problems. Nevertheless, the question arises whether or
not such a differential approach may indicate the MIT at least in the
limit as $T \rightarrow 0$. In order to apply this criterion, one
would have to determine the control parameter value for which
$\mbox{d} \rho / \mbox{d} T = 0$ at various temperatures. Its
$T \rightarrow 0$ limit would identify the MIT.

In practice, in particular for two-dimensional systems, the focus is
on the transport at the lowest experimentally accessible temperature,
$T_{\rm lea}$, often in the mK range.\cite{Krav.etal.94,Krav.Sara}
That means, the sign change of
$\mbox{d} \rho / \mbox{d} T (T_{\rm lea},x)$ is regarded as the
indication of the MIT, instead of a corresponding $T \rightarrow 0$
extrapolation. Of course, this approach can be meaningful only if
$| \mbox{d}^2 \rho / \mbox{d} T^2 |$ is so small at this hypothetical
MIT that the influence of $T_{\rm lea}$ on its location is negligible.
Fortunately, this condition seems to be often fulfilled; the width of
the corresponding $x$ range, however, decreases with $T$.

Nevertheless, one should be very cautious since the issue is far more
profound than it appears at first glance: Using this criterion means
to favor a particular character of the MIT, namely discontinuity, and
thus the existence of a finite minimum metallic conductivity. Our
statement can be substantiated in two ways: It results from a short
mathematical consideration, and from purely experimental experience. 

First, we take the mathematical perspective. Suppose the properties
of the investigated material can be continuously modified where the
extent of the modification is measured by a real control parameter $x$
with $x \in [x_{\rm a},x_{\rm b}]$. 

Focusing now on the $T$ region $(0,T_{\rm lea}]$, we consider
$\sigma(T,x)$ and $\mbox{d} \rho / \mbox{d} T (T,x)$ to be continuous
functions of $T$ and $x$ where $\sigma(T,x) > 0$. Without loss of
generality, we suppose that, when
$T = {\rm const.} \in (0,T_{\rm lea}]$, both these functions increase
strictly monotonically with $x$. Suppose, furthermore, that
$\mbox{d} \rho / \mbox{d} T (T_{\rm lea}, x_{\rm a}) < 0$ and that
$\mbox{d} \rho / \mbox{d} T (T_{\rm lea}, x_{\rm b}) > 0$. Then,
$\mbox{d} \rho / \mbox{d} T (T_{\rm lea}, x)$ changes sign at some
critical control parameter value, 
$x_{\rm c} \in (x_{\rm a}, x_{\rm b})$, and 
$\sigma_{\rm c} = \sigma(T_{\rm lea},x_{\rm c}) > 0$.

In practically applying the hypothetical MIT criterion considered in
this subsection, as pointed to above, we suppose that (i)
$\mbox{d} \rho / \mbox{d} T (T,x_{\rm c}) = 0$ for any
$T \in (0,T_{\rm lea}]$, and hypothesize that (ii) $x_{\rm c}$ marks
the MIT.

From (i) and the suppositions above, we deduce that, in case
$x < x_{\rm c}$ and $T \in (0,T_{\rm lea}]$,
$\mbox{d} \rho / \mbox{d} T (T,x) < 0$ holds, so that
$\mbox{d} \sigma / \mbox{d} T (T,x) > 0$. Thus
$\sigma(T,x) \le \sigma(T_{\rm lea},x) < \sigma_{\rm c}$ and,
therefore, $\lim_{T \rightarrow 0}{\sigma(T,x)} < \sigma_{\rm c}$.
Finally, due to (ii), we even expect that, for $x < x_{\rm c}$, 
$\lim_{T \rightarrow 0}{\sigma(T,x) = 0}$. -- Such a limiting behavior
may result when, on the insulating side of the MIT, the rapid decrease
of $\sigma(T,x = {\rm const.})$  sets in at the lower temperature the
closer $x$ to $x_{\rm c}$, see Subsections 2.4 and 2.5. -- For 
$x \ge x_{\rm c}$ and $T \in (0,T_{\rm lea}]$, we obtain
$\sigma(T,x) \ge \sigma(T_{\rm lea},x) \ge \sigma_{\rm c}$ in an
analogous way, so that 
$\lim_{T \rightarrow 0}{\sigma(T,x)} \ge \sigma_{\rm c}$ in this $x$
region. 

Therefore, under the above physically plausible suppositions, the
hypothetical MIT criterion considered here implies that 
$\lim_{T \rightarrow 0}{\sigma(T,x)}$ is discontinuous at $x_{\rm c}$ 
-- the limit jumps there from 0 to $\sigma_{\rm c}$ -- and
that $\sigma_{\rm c}$ has the meaning of a finite minimum metallic
conductivity. Thus, in the present situation, already by the choice of
the MIT criterion, one determines which character of the MIT one will
infer from the experimental data. 

Now, we consider the experimental experience: For many
three-dimensional systems, the $\sigma$ value which corresponds to
$\mbox{d} \rho / \mbox{d} T$ changing sign at low $T$ was found to
roughly agree with Mott's minimum metallic conductivity
estimate,\cite{Mott.Davi,Mott.72,Mott.82}
\begin{equation}
\sigma_{\rm M} = C \frac{e^2}{\hbar d_{\rm c}}\,.
\label{mmc}
\end{equation}
Here $d_{\rm c}$ denotes the interatomic distance of the transport
enabling constituents at the MIT (e.g., P atoms in crystalline Si:P)
and $C$ a numerical constant of order 0.025 to 0.05;\linebreak
$d_{\rm c} = n_{\rm c}^{\ -1/3}$, where $n_{\rm c}$ is the corresponding
critical density. For example, Ganguly et al.\ explicitly used this
sign change criterion to determine the value of $\sigma_{\rm M}$ for
oxide systems, see Ref.\ \onlinecite{Gan.etal.84}. Further support for
this correlation comes from the original references of the data
compiled in Figure 10 of Ref.\ \onlinecite{Mott.Kaveh}; moreover,
concerning crystalline Si:P and amorphous Si$_{1-x}$Cr$_x$, see Figure
3 of Ref.\ \onlinecite{Stu.etal.93} and Figure 7 of Ref.\
\onlinecite{Moe.etal.85}, respectively. Thus, in the literature, using
the criterion $\mbox{d} \rho / \mbox{d} T = 0$ at $T_{\rm lea}$ for
defining the MIT usually results in support for Mott's idea of the
finite minimum metallic conductivity.

It has to be stressed, however, that the frequently observed
correlation between the features $\mbox{d} \rho / \mbox{d} T = 0$ and
$\sigma = \sigma_{\rm M}$ concerns only measurements at $T_{\rm lea}$.
Without a further assumption, it does neither identify the MIT nor
does it imply that Mott's reasoning is correct. In particular, this
correlation alone does not justify the conclusion that, for each
sample with negative $\mbox{d} \rho / \mbox{d} T (T_{\rm lea},x)$, the
conductivity $\sigma(T)$ cannot saturate at some nonzero value (far)
below $\sigma_{\rm M}$ as $T \rightarrow 0$. In other words, this
experimental experience is not sufficient to exclude that there might
be metallic samples with negative
$\mbox{d} \rho / \mbox{d} T (T_{\rm lea},x)$. Hence, interpreting
solely the correlation between the $T_{\rm lea}$-related features
$\mbox{d} \rho / \mbox{d} T = 0$ and $\sigma = \sigma_{\rm M}$  as
evidence for Mott's minimum metallic conductivity theory overvalues
the experimental findings.  

In this sense, {\it relying only on the MIT criterion
$\mbox{d} \rho / \mbox{d} T = 0$ in the limit as  $T \rightarrow 0$
means to bias the data analysis, and to presume the existence of a
finite minimum metallic conductivity.} That is why the corresponding
conclusion by Siegrist et al.\cite{Sie.etal.11} is not surprising. It
is the natural consequence of these authors focusing on the sign
change of $\mbox{d} \rho / \mbox{d} T$.

Note, furthermore, that the MIT criterion
$\mbox{d} \rho / \mbox{d} T = 0$ in the limit as $T \rightarrow 0$
does not seem to be in accord with any of the currently available
microscopic theories: First, because of the following argument, this
criterion is incompatible with the interpretation in terms of an
Anderson transition caused by the mobility edge crossing the Fermi
energy. According to this model, just at the MIT, the number of
electrons (or holes) excited to extended states would be proportional
to $T$. It is hard to believe that these electrons do not cause a
corresponding increase in $\sigma$ with rising $T$.\cite{Moe.etal.83}
-- Thus, the derivation of Eq.\ (\ref{mmc}) as zero-temperature
evaluation of the Kubo-Greenwood formula by Mott in Ref.\
\onlinecite{Mott.72} is not consistent with his interpretation of
experimental data in terms of the activation energy of hopping tending
to zero as $x \rightarrow x_{\rm c}$ in the same work. -- Our
incompatibility argument, however, does not disprove the hypothetical
MIT criterion considered here because the interpretation in terms of
an Anderson transition is based on a strong simplification; it
neglects the electron-electron interaction. Second, to the best of our
knowledge, also the currently available microscopic theories
incorporating electron-electron interaction cannot provide a
justification for this criterion. The reason is that none of them
yields its logical consequence, that is a discontinuous MIT with a
finite minimum metallic conductivity.

Now, if the sign change of $\mbox{d} \rho / \mbox{d} T$ in the limit
as $T \rightarrow 0$ did not arise from the MIT, how could it be
understood? An alternative explanation was provided by an analysis of
the influence of the electron-electron interaction: According to
Altshuler and Aronov, the electron-electron interaction should yield a
$T^{1/2}$ contribution to $\sigma(T)$.\cite{Alt.Aro.79} Rosenbaum et
al.\ pointed out that its sign should be determined by the relative
importance of exchange and Hartree terms, which in turn is controlled
by the ratio of screening length and Fermi wave
length.\cite{Ros.etal.81,Lee.Rama} Therefore, the variation of the
screening length may cause two sign changes of the $T^{1/2}$
contribution.\cite{Ros.etal.81} This interpretation has been very
influential up to now.

Nevertheless, in our opinion, the applicability of this theoretical
idea by Rosenbaum et al.\cite{Ros.etal.81} is highly questionable: The
point is that the value of the exponent, $1/2$, seems to disagree with
the experimental findings. For samples which exhibit negative
$\mbox{d} \rho / \mbox{d} T$ at $T_{\rm lea}$ and which were
classified as metallic in the respective original publications, a
series of such discrepancies is presented in our Section 4. 

Concerning samples with positive $\mbox{d} \rho / \mbox{d} T$ at
$T_{\rm lea}$, being very likely metallic, doubts about the exponent
value $1/2$ arise from three publications: Already Figure 1 of Ref.\
\onlinecite{Ros.etal.81} shows that, for crystalline Si:P, the
exponent $1/3$ enables a clearly better approximation of the
experimental data than $1/2$. The Figures 3a and 3b of Ref.\
\onlinecite{Dai.etal.92} testify that, for crystalline Si:B, $\sigma$
versus $T^{1/2}$ plots exhibit substantial curvature in the low-$T$
range. -- The authors of Ref.\ \onlinecite{Dai.etal.92} model this
feature by additionally taking into account a localization
contribution to $\sigma(T)$, but the deviation visible in Figure 3b
questions the physical meaning of such fits. -- These observations are
consistent with a study of amorphous Si$_{1-x}$Cr$_x$ in which a power
law contribution to $\sigma(T)$ with the exponent $0.19 \pm 0.03$ was
identified by collective augmented power law fits to sets of
conductivity differences of pairs out of 15 samples.\cite{Moe.90a}
Remarkably, the corresponding decomposition of $\sigma(T)$ ceases to
be applicable when, with decreasing $x$, $\mbox{d} \rho / \mbox{d} T$
at $T_{\rm lea}$ changes sign from plus to minus.\cite{Moe.90a}

Furthermore, in a very recent theoretical investigation, Di Sante et
al.\ studied the interplay of thermal lattice deformations and static
disorder in the vicinity of the MIT.\cite{Di_Sante.etal} Claiming
continuity of the transition, they state the existence of a control
parameter interval within which $\mbox{d} \rho / \mbox{d} T$ is
negative although $\sigma$ tends to finite values when
$T \rightarrow 0$, and denote it as region of ``bad insulators''. --
The choice of this name is not in accord with the definition of an
insulator as we use it here. -- The authors conclude that, in this
control parameter range, transport is governed by the strong $T$
dependence of the density of states, rather than by the $T$ dependence
of the scattering of the electrons.\cite{Di_Sante.etal} However,
Di Sante et al.\ focus on the situation of a half-filled band, and it
seems at least questionable whether their conclusions remain valid for
arbitrary band filling. Moreover, the quantitative comparison of this
theory to experiment is hindered by the following:  Reference
\onlinecite{Di_Sante.etal} does not report any specific exponent value
for the power law governing $\sigma(T)$ at the MIT.

According to the above three paragraphs, the theoretical
interpretations of the sign change of $\mbox{d} \rho / \mbox{d} T$
given in Ref.\ \onlinecite{Ros.etal.81} and \onlinecite{Di_Sante.etal}
should be considered with great caution, so that alternative
hypotheses have to checked too. Therefore, at the current stage, it
seems quite possible that the criterion
$\mbox{d} \rho / \mbox{d} T = 0$ in the limit as $T \rightarrow 0$ may
indeed identify the MIT. 

Due to the described unclear situation, the development of a more
appropriate theory is required.  Fortunately, one of its features can
be easily predicted because, for dimensional reasons, Eq.\ (\ref{mmc})
is robust, also with respect to the incorporation of electron-electron
interactions: If a theory yields a relation which links the distance
$d_{\rm c}$ with any characteristic conductivity value, and if
electron charge, Planck constant, effective mass, and permittivity are
the only dimensioned parameters of this theory, then the so obtained
relation must be universal and have the form of Eq.\ (\ref{mmc}); for
the proof see Appendix A. Independently of the character of the MIT,
the condition $\mbox{d} \rho / \mbox{d} T = 0$ in the limit as
$T \rightarrow 0$ is a natural way to define such a characteristic
conductivity value. In case a finite minimum metallic conductivity
exists, it should equal a universal multiple of this characteristic
conductivity value; it might even coincide with this value.

We now continue with our phenomenological analysis: Currently, to the
best of our knowledge, the MIT criterion
$\mbox{d} \rho / \mbox{d} T = 0$ in the limit as $T \rightarrow 0$ is
only a hypothesis. How could it be justified? It seems natural to
regard all samples with $\mbox{d} \rho / \mbox{d} T > 0$ at
$T_{\rm lea}$ as metallic. However, for the inverse approach, the
classification of all samples with $\mbox{d} \rho / \mbox{d} T < 0$ at
$T_{\rm lea}$ as insulating, further arguments and / or additional
measurements are required. In particular, this concerns the $\rho(T)$
curves with small negative slope, for which the interpretation in
terms of activated transport may seem counterintuitive.

Nevertheless, one can easily imagine a situation where such samples
are indeed insulating. Let us assume that some kind of hopping is the
only relevant conduction mechanism and that, as already surmised by
Mott,\cite{Mott.72} its mean activation energy tends continuously to 0
as the MIT is  approached. In this case, the existence of insulating
samples with quite flat, nonexponential $\rho(T)$ is not
counterintuitive but natural: All the samples close to the MIT for
which the characteristic temperature, corresponding to the mean
hopping energy, is smaller than the lowest measuring temperature
should behave this way, see also Subsections 2.4 and 2.5.

Thus, also the identification of insulating samples in the penultimate
paragraph according to the sign of $\mbox{d} \rho / \mbox{d} T$ for
$T \rightarrow 0$ may be correct. To find out whether or not it is
correct indeed, we have to ask: Which value does the mean hopping
energy tend to as the MIT is approached? Empirical support for the
hypothesis that the mean hopping energy continuously approaches zero
comes from scaling of the $\sigma(T, x = {\rm const.})$ curves for
various values of $x$ on the insulating side of the MIT, see
Subsection 2.5. 

In conclusion, the unclear situation described above poses the
challenge to find out which of the samples with nonexponential
$\rho(T)$ and $\mbox{d} \rho / \mbox{d} T < 0$ at $T_{\rm lea}$ are
metallic and which are insulating. Various approaches to this question
will be discussed in the following subsections.

\subsection{Breakdown of the augmented power law approximation}

An alternative to considering $\mbox{d} \rho / \mbox{d} T$ at
$T_{\rm lea}$ is to detect the MIT by quantitatively analyzing
measurements in some low-temperature range in terms of a microscopic
theory: Starting from a theory-based ansatz for $\sigma(T)$ with
adjustable parameters, one determines the value of the control
parameter at which this equation ceases to be valid. Ideally, such an
analysis should be performed for both sides of the MIT. In this
subsection, we focus on the metallic side. For the corresponding
consideration of the non-metallic region see Subsection 2.4.

Considering metallic conduction, augmented power laws,
\begin{equation}
\sigma(T,x) = a(x) + b(x) \, T^p \,,
\label{apl}
\end{equation}
were derived for different situations. Altshuler and Aronov studied
the superposition of electron-electron interaction and static disorder
and obtained $p = 1/2$, see Ref.\ \onlinecite{Alt.Aro.79}. They
derived, however, Eq.\ (\ref{apl}) from a perturbation theory, so that
its applicability very close to the MIT is at least questionable.
Nevertheless, in various low-$T$ experiments, Eq.\ (\ref{apl}) has
been claimed to describe measured data rather well, see, for example,
Refs.\ \onlinecite{Zabr.Zino} and \onlinecite{Tho.etal.83}.

Newson and Pepper additionally incorporated the $T$ dependence of the
diffusion constant into the result by Altshuler and Aronov; for this
they used the Einstein relation linking conductivity and diffusion
constant.\cite{News.Pepp} Their approach again results in Eq.\
(\ref{apl}), but with a smaller exponent: Now $p = 1/3$, see Ref.\
\onlinecite{News.Pepp} and compare Ref.\ \onlinecite{Mali.etal}. The
derivation, however, presumes $a \ll b \, T^{1/3}$. In several cases,
this version of Eq.\ (\ref{apl}) was claimed to be more appropriate
for describing the experimental data than Eq.\ (\ref{apl}) with
$p = 1/2$, see Refs.\ \onlinecite{Waf.etal.99},
\onlinecite{Giv.Ova.12}, and \onlinecite{Mali.etal}.

In practice, the choice of the value of $p$ seems not to influence the
conclusion about the qualitative character of 
$\lim_{T \rightarrow 0} \sigma(T,x)$ in the vicinity of the
MIT.\cite{Stu.etal.93} Nevertheless, a modification of $p$ as well as
a variation of the $T$ range taken into account in the fit usually
cause a small shift of the resulting MIT point; an example will be
given in Subsection 4.2.

In a few studies, extensions of Eq.\ (\ref{apl}) were used to model
the experimental data. For example, investigating crystalline
Si:(P,B), Hirsch et al.\ included a second $T$-dependent term
accounting for a weak-localization correction due to inelastic
electron-electron collisions,
\begin{equation}
\sigma(T,x) = a(x) + b(x) \, T^p + c(x) \, T^q 
\label{apl_ext}
\end{equation}
with $q = 3/4$, see Ref.\ \onlinecite{Hir.etal.88}. Not surprisingly,
including such a second $T$-dependent term improves the fit quality
even if the model is not physically
justified;\cite{Hir.etal.88,Moe.89,Hir.etal.89} see also Subsection
2.6.

In all data analyses based on Eq.\ (\ref{apl}) or (\ref{apl_ext}), the
diagnosis of the breakdown of the theoretical description is the
crucial point. There are the following two approaches to this problem;
both have to be combined to ensure that the sample classification is
as reliable as possible.

First, relying on the validity of the considered ansatz, one
determines the value of the control parameter at which one of the
adjustable parameters ceases to have a physically reasonable value.
In the present situation, one asks at which value of $x$ the parameter
$a$ reaches 0.

Second, one checks for systematic deviations of the adjusted
theoretical $\sigma(T)$ relation from the experimental data. This,
however, is a fuzzy condition: The precision and accuracy of the
experimental data as well as the width of the considered $T$ interval
have great influence on strength and assessment of deviations.
Presumably because of these uncertainties, the search for systematic
deviations has played only a minor role in the literature. Thus,
essential information was often lost; we will demonstrate this in
Sections 3 and 4.

It is important to note that, when using solely the first approach,
one demands only one of two necessary conditions for the validity of
Eq.\ (\ref{apl}) to be satisfied. Thus, in such an analysis, there is
a considerable risk that some insulating samples very close to the MIT
are misinterpreted as metallic. Therefore, since the function
$\sigma(T = {\rm const.}, x)$ seems to be continuous at the MIT for
any $T > 0$, it is not unlikely that one only gets out what is put
into the model used in the data analysis, that is the continuity of
$\lim_{T \rightarrow 0} \sigma(T,x)$.  For this reason, {\it analyses
based solely on augmented power law fits without careful applicability
checks exhibit a substantial bias in favor of a continuous MIT.}

The way out of this dilemma is to consider another observable
additionally to $\sigma(T)$. To that end, however, it is not necessary
to measure a further transport coefficient or some thermodynamic
observable. Already considering a specific derivative of $\sigma(T)$
can be very helpful as will be explained in Subsections 2.6 and 2.8.
The situation concerning the incorporation of alternative observables
will be discussed in Subsection 2.7.

\subsection{Breakdown of the stretched Arrhenius law approximation}

We now turn to the insulating side of the MIT. There, at low $T$, the
transport proceeds by some kind of hopping conduction. Such mechanisms
cause $\sigma(T,x = {\rm const.})$ to follow a stretched Arrhenius
law,
\begin{equation}
\sigma(T,x) = \sigma_0(x) \, \exp(-(T_0(x)/T)^\nu)\, .
\label{mer}
\end{equation}
The characteristic temperature $T_0$ depends on the distance to the
MIT; it seems to vanish as the MIT is approached,\cite{Mott.72} see
also Subsection 2.2. The exponent $\nu$ is mechanism-dependent:
Thermal excitation over some finite gap implies $\nu = 1$. In case of
variable-range hopping without Coulomb interaction, $\nu$ equals 1/3
and $1/4$ for two- and three-dimensional systems,
respectively,\cite{Mott.69} compare, for example, Figure 8s in the
Supplementary Information of Ref.\ \onlinecite{Sie.etal.11}. If,
however, Coulomb interaction has to be taken into account in
variable-range hopping, $\nu = 1/2$ is expected for two-dimensional as
well as for three-dimensional samples,\cite{Efro.Shkl} see also Refs.\
\onlinecite{Zabr.Zino} and \onlinecite{Moe.etal.85}.

The prefactor $\sigma_0$ may be weakly $T$-dependent, which is
neglected in our Eq.\ (\ref{mer}). A reliable determination of $\nu$
is therefore only possible if the $\sigma$ values cover a very wide
range, ideally several orders of magnitude.

It seems natural to classify all those samples as insulating for which
$\sigma(T)$ approximately obeys Eq.\ (\ref{mer}). -- In doing so, one 
implicitly assumes this relation to be valid down to $T = 0$. -- 
However, simply regarding all other samples as metallic is not
justified for the following two reasons.

First, there is no sharp distinction between exponential and
nonexponential $\sigma(T)$: For example, assume the ratio
$\sigma(6\, {\rm K}) / \sigma(4.2\, {\rm K})$ amounts to 256, 16, 4,
2, 1.4, 1.2, 1.1, 1.05, 1.02, or 1.01. At which value could metallic
behavior set in?

Second, in its pure form, an exponential law such as Eq.\ (\ref{mer})
is usually only an approximation valid for $T \ll T_0$. However, as
already pointed to in Subsection 2.2, $T_0(x)$ seems to vanish 
continuously as the MIT is approached. This limiting behavior, which
is very likely but not completely certain, has an important
consequence: Consider $\sigma(T, x = {\rm const.})$ measurements by
means of some cryostat down to its lowest experimentally accessible
temperature, $T_{\rm lea}$. Then, for any value of $T_{\rm lea}$,
there is a finite interval of $x$ adjacent to the MIT within which
$T_0(x)$ is smaller than $T_{\rm lea}$. Thus, for all samples
belonging to this interval of $x$, it is impossible to observe an
exponential decrease in $\sigma(T)$ by orders of magnitude using the
specific cryostat although these samples are insulating. Compare
Figure 1 of Ref.\ \onlinecite{Moe.etal.99} and the explanation in its
caption. 

Note that the second argument holds regardless of the experimental
technique used, for studies down to 4.2~K as well as for measurements
in a dilution refrigerator. With respect to the continuous variation
of the control parameter $x$, one can even state: {\it Since the $T$
scale is set by $T_0(x)$, it is very likely impossible to stay at low
temperatures while passing the MIT.}

Because of this problem, and due to the possible weak $T$ dependence
of $\sigma_0$, fits based on Eq.\ (\ref{mer}) do not allow the
reliable identification of insulating samples very close to the MIT.
Therefore, these samples may easily be misinterpreted as metallic in a
corresponding analysis. In evaluating experimental data, one is quite
often confronted with such interpretation ambiguities since the
critical exponent of the characteristic temperature, $T_0(x)$, seems
to be rather large: In studies of crystalline n-Ge:(As,Ga) and
amorphous Si$_{1-x}$Cr$_x$, in which, at low $T$, the
$\sigma(T, x = {\rm const.})$ could be well approximated by Eq.\
(\ref{mer}) with $\nu = 1/2$, critical exponent values of
$2.1 \pm 0.1$ and $3.0 \pm 0.4$, respectively, were
obtained.\cite{Zabr.Zino,Moe.90b}

Note, moreover, that an unambiguous classification of metallic and
insulating samples based solely on describing $\sigma(T)$
alternatively by Eqs.\ (\ref{apl}) or (\ref{mer}) is in principle
impossible if $\sigma(T)$ is proportional to $T^p$ at the MIT: The two
analytic functions modeling $\sigma(T)$ on both sides of the MIT do
not fit together at the critical value of $x$, in contradiction to the
experimental experience that $\sigma(T = {\rm const.}, x)$ seems to be
continuous for any $T > 0$.

\subsection{Breakdown of scaling of $T$ dependences of $\sigma$}

In several studies of two- and three-dimensional
systems,\cite{Zabr.Zino,Moe.etal.83,Moe.85,Sara.Dai.02,Moe.etal.85,
Liu.etal.92,Krav.etal.95} but by far not in all such investigations,
the identification of insulating samples has been facilitated by
empirical scaling of $\sigma(T, x = {\rm const.})$ curves, 
\begin{equation}
\sigma(T,x) = \sigma_{\rm scal}(T/T_0(x))\,.
\label{scaling}
\end{equation}
Here, $T_0(x)$ denotes an $x$-dependent characteristic temperature;
$T_0(x) > 0$. This equation links measurements on different samples.
In particular, it relates samples with quite flat $\sigma(T)$ to
samples with exponential $\sigma(T)$. We remark that, for
three-dimensional systems, the scaling relation Eq.\ (\ref{scaling})
is incompatible with basic ideas of the scaling theory of
localization, despite the similarity of the terms used.\cite{Moe.86}

The important point is that, provided Eq.\ (\ref{scaling}) remains
valid when $T \rightarrow 0$, $\lim_{T \rightarrow 0} \sigma(T,x)$ is
independent of $x$. Thus, in all the samples for which the
$\sigma(T, x = {\rm const.})$ data sets are linked to each other by
this relation, the nature of electrical conduction must be the same.
We stress that this statement does not rely on specific assumptions of
any particular microscopic theory, in contrast to the analysis methods
discussed in Subsections 2.3 and 2.4. 

Deep in the non-metallic region where $\sigma$ follows a stretched 
exponential $T$ dependence according to Eq.\ (\ref{mer}), scaling is
reflected by the prefactor $\sigma_0(x)$ being a constant independent
of $x$. For $\nu = 1/2$, such a behavior has been observed quite 
often.\cite{Zabr.Zino,Moe.etal.83,Moe.85,Sara.Dai.02,Moe.etal.85} This
exponential limit of Eq.\ (\ref{scaling}) has a twofold relevance: It
indicates that {\it all} the samples with $\sigma(T,x)$ satisfying
Eq.\ (\ref{scaling}) belong to the insulating phase. Furthermore, it
enables to define an absolute scale of $T_0$.

Therefore, when the MIT is approached from the insulating side, the
breakdown of the validity of Eq.\ (\ref{scaling}) very likely
indicates the transition; for a detailed reasoning see Appendix B.
This criterion has a great advantage over the approaches discussed in
the previous subsections: In a scaling analysis, the reduced
temperature, $T/T_0$, may vary over several orders of magnitude,
compare Fig.\ 4 in Ref.\ \onlinecite{Moe.etal.83} and Fig.\ 2 in Ref.\
\onlinecite{Liu.etal.92}. Thus, the reliability of
$\sigma(T \rightarrow 0)$ extrapolations and the trustworthiness of
sample classifications close to the MIT can be considerably improved
by utilizing the scaling relation Eq.\ (\ref{scaling}).

Note, however, that there are essential preconditions for Eq.\ 
(\ref{scaling}) being possibly valid. First, it can only hold if all
but one conduction mechanisms are frozen out. Therefore, its
applicability is always limited by some material-dependent upper
temperature threshold, $T_{\rm utt}$, the value of which is determined
by the maximum tolerable contribution of the second most relevant
mechanism to $\sigma(T)$.\cite{Moe.etal.85} -- In the case of 
amorphous Si$_{1-x}$Cr$_x$, it amounts to roughly
50~K;\cite{Moe.etal.85} for doped crystalline semiconductors, it is
far smaller and may even fall below 1~K, compare Subsection 4.5. --
Second, Eq.\ (\ref{scaling}) can only be valid over a wide range of
the reduced temperature $T/T_0$ if there is merely a single
transport-relevant length scale, that means, if the disordered solid
considered is sufficiently homogeneous. To ensure this, the influences
of macroscopic and mesoscopic inhomogeneities have to be effectively
suppressed. Here, we remind that such a scaling of
$\sigma(T, x = {\rm const.})$ curves seems to be destroyed by
clustering.\cite{Moe.85,Moe.etal.85}

In its general form, the scaling relation Eq.\ (\ref{scaling}) to hold
can be concluded from experimental data for a set of $k$ samples
either by a master curve
construction\cite{Moe.etal.83,Liu.etal.92,Krav.etal.95} or by checking
whether or not $\mbox{d} \ln \sigma / \mbox{d} \ln T$ is a function of
$\sigma$ alone.\cite{Moe.etal.85} The former approach requires
adjusting the quotients of the $T_0$ values for $(k-1)$ pairs of
samples by means of rescaling $T$, while the latter procedure is
parameter-free. The degree to which Eq.\ (\ref{scaling}) can be
experimentally verified depends not only on the set of control
parameter values considered as well as on precision and accuracy of
resistance, temperature, and geometry measurements, but also on how
well the preconditions specified in the previous paragraph are
fulfilled.

The scaling of $\sigma(T, x = {\rm const.})$ curves according to Eq.\
(\ref{scaling}) has been interpreted as indication of the existence of
a finite minimum metallic conductivity, $\sigma_{\rm mm}$, that means
of the MIT being
discontinuous.\cite{Moe.etal.83,Zha.Sara.95,Moe.85,Moe.etal.85,
Krav.etal.95} This conclusion, however, is based on several more or
less hidden assumptions. In order to identify them, we illuminate the
details of this scaling analysis by means of a short mathematical
consideration in Appendix B. There, we show that, under quite natural
suppositions, the scaling of the $T$ dependences of $\sigma$ for
several values of $x$ implies $T_0(x) \rightarrow 0$ as $x$ tends to
the MIT value $x_{\rm c}$ and, moreover, that the limit
$\lim_{t \rightarrow \infty} \sigma_{\rm scal}(t)$ coincides with the
minimum metallic conductivity $\sigma_{\rm mm}$.

Furthermore, Appendix B explains why, because of the finiteness of
$\sigma_{\rm mm}$, the \linebreak $\sigma(T,x  = {\rm const.})$ must
become more and more flat when the MIT is approached, in accord with
the MIT criterion ``Sign change of $\mbox{d} \rho / \mbox{d} T$ in the
limit as $T \rightarrow 0$'' considered in Subsection 2.2. This
feature of $\sigma(T,x)$, however, has an unpleasant consequence:
There is always some close vicinity of the MIT where it is impossible
to prove scaling by constructing a mastercurve
$\sigma_{\rm scal}(T/T_0(x))$ out of overlapping pieces.

As a side remark, we mention here that the observation of scaling
according to Eq.\ (\ref{scaling}) together with the plausible
suppositions made in Appendix B imply a surprising conclusion: At any
fixed $T \in (0,T_{\rm utt}]$, when varying $x$, one should cross a
sharp boundary, located at $x_{\rm c}$, between non-metallic and
metallic conduction. In other words, the zero-temperature phenomenon
MIT -- a discontinuous quantum phase transition -- should be the
endpoint of a line of continuous phase transitions in the finite-$T$
part of the $(x,T)$-plane. Therefore, $\sigma(T = {\rm const.}, x)$
should exhibit a non-analyticity at $x_{\rm c}$; see Section 4 of
Ref.\ \onlinecite{Moe.85}, in particular Figure 9 therein, and the
introduction of Ref.\ \onlinecite{Moe.08}. In consequence,
$\partial \sigma / \partial T (T = {\rm const.}, x)$, too, 
should have a non-analyticity at the MIT, just where it changes sign.

In the case of amorphous Si$_{1-x}$Cr$_x$, this hypothesis of a line
of continuous phase transitions is supported by two observations:
(i) The optimal exponent of a power law contribution to $\sigma(T,x)$
changes qualitatively just when $\mbox{d} \sigma / \mbox{d} T = 0$,
see Ref.\ \onlinecite{Moe.90a}. (ii) The relation between two
coefficients of a phenomenological model of $\sigma(T,x)$ for the
metallic side seems to exhibit a non-analyticity coinciding with
$\mbox{d} \sigma / \mbox{d} T = 0$, see Refs.\
\onlinecite{Moe.etal.85} and \onlinecite{Moe.90b}.

The direct identification of such a composition tuned phase transition
in $\sigma(T = {\rm const.}, x)$ would likely be hindered by the
requirement of an extremely high degree of sample-to-sample
reproducibility; presumably, $\Delta \sigma / \sigma$ would have to
amount to one to a very few percent. However, if shallow impurity
states play the dominant role, this problem can be circumvented by
tuning the critical doping concentration by means of stress or
magnetic field instead of tuning the doping concentration itself. In
the two-dimensional case, this problem can be avoided by tuning the
charge carrier concentration via a gate voltage; by reanalyzing such
an experiment, first indications of the hypothesized peculiarity in
$\sigma(T = {\rm const.}, x)$ separating regions of metallic and
non-metallic conduction at finite $T$ were detected in Ref.\
\onlinecite{Moe.08}. Moreover, focusing on correlations between
$\sigma(T, x = {\rm const.})$ and appropriate derivatives of this
function or between $\sigma(T, x = {\rm const.})$ and the exponent of
an augmented power law approximation may help; compare Ref.\
\onlinecite{Moe.90a}.

Nevertheless, in all these approaches, inhomogeneities can easily
smooth out the hypothetical finite-$T$ phase transition.
Paradoxically, their influence should be less disturbing when the
measurements are performed at intermediate temperatures rather than at
extremely low ones.\cite{Moe.08} The reason is that, in the insulating
$x$ range, when $T < T_{\rm utt}$, the following holds for the
immediate neighborhood of the MIT: The smaller $T$, the steeper is
$\sigma(T = {\rm const.},x)$, and the wider is the $\sigma$ range
affected by a given, fixed uncertainty of $x$.

Finally, motivated by a dimensional consideration, we point to a
possible universality of homogeneous disordered solids: Consider
samples of two kinds of three-dimensional disordered solids. Suppose
that the measured $\sigma(T,x)$ satisfy Eq.\ (\ref{scaling}) with two
separate $\sigma_{\rm scal}(t)$. Let us define both the respective
scales of $T_0$ by means of Eq.\ (\ref{mer}). Then, although the
minimum metallic conductivity value
$\sigma_{\rm mm} = \lim_{t \rightarrow \infty} \sigma_{\rm scal}(t)$
is a material-dependent quantity, the ratio
$\sigma_{\rm scal}(t) / \sigma_{\rm mm}$ should be a universal
function of the reduced temperature $t = T/T_0$. This should hold not
only when $\sigma_{\rm scal}$ depends on $t$ in an exponential manner,
but also in the range of nonexponential behavior of
$\sigma_{\rm scal}(t)$.  Thus, the quotient of the characteristic
conductivity values $\sigma_0$, defined by Eq.\ (\ref{mer}), and
$\sigma_{\rm mm}$, corresponding to
$\mbox{d} \sigma / \mbox{d} T = 0$, should be material-independent,
too; for amorphous Si$_{1-x}$Cr$_x$, it amounts to $0.46 \pm 0.14$,
compare Refs.\ \onlinecite{Moe.etal.83} and \onlinecite{Moe.etal.85}.

\subsection{Bounds obtained from the logarithmic derivative of
$\sigma(T)$}

Both the analyses of $\sigma(T)$ data based on Eqs.\ (\ref{apl}) and
(\ref{mer}), which were discussed in Subsections 2.3 and 2.4,
respectively, have an inherent disadvantage: They may misinterpret
insulating samples very close to the MIT as metallic. The risk of
such a misclassification can be largely reduced by additionally
evaluating the logarithmic derivative,
\begin{equation}
w(T) = \frac{\mbox{d} \ln \sigma}{\mbox{d} \ln T}\,,
\label{logderiv}
\end{equation}
see Refs.\ \onlinecite{Moe.etal.99}, \onlinecite{Moe.89}, and 
\onlinecite{Hir.etal.89}. Here, $\ln \sigma$ instead of $\sigma$ is
considered in order to simplify the examination of an exponentially
wide conductivity range. This quantity is differentiated with respect
to $\ln T$ instead of with respect to $T$ to enable the clear
discrimination between metallic and non-metallic $\sigma(T)$ which is
explained in the following.

The logarithmic derivative $w(T)$ exhibits qualitatively different
behavior for the augmented power law, Eq.\ (\ref{apl}) with $p > 0$,
and for the stretched Arrhenius relation, Eq.\ (\ref{mer}) with
$\nu > 0$. In the former case, that is for metallic samples, 
\begin{equation}
w(T) = \frac{p \, b \, T^p}{a + b \, T^p} 
\label{derivapl}
\end{equation}
holds, where $a + b \, T^p = \sigma(T) > 0$ and $a = \sigma(0) > 0$.
In case the MIT is continuous, the logarithmic derivative stays
constant at the MIT itself, $w(T) = p$ due to $a = 0$.

Equation (\ref{derivapl}) looks simple but provides several
experimentally checkable conclusions. Thus, for metallic samples, as
$T$ vanishes, $w(T)$ becomes proportional to $T^p$ and tends to 0. If,
moreover, $b > 0$, then, as $T$ increases far beyond $(a/b)^{1/p}$,
$w(T)$ approaches $p$. Therefore, plotting $w$ versus $T^p$ can test
the validity of Eq.\ (\ref{apl}) over a wide $T$ range.

Furthermore, if measurements yield $w > 0$ for a hypothetically
metallic sample, then the parameter $b$ has to be positive as well.
Hence, $0 < w(T) < p$ must be fulfilled.

Even more important, for $w > 0$ and thus $b > 0$, the logarithmic
derivative $w(T)$ always decreases with diminishing $T$. This is
obvious from rewriting Eq.\ (\ref{derivapl}) as 
\begin{equation}
w(T) = p \, \left( 1 - \frac{a}{a + b \, T^p} \right) \,.
\label{derivapl_r}
\end{equation}
In other words, for a sample with $w > 0$ to be metallic,
${\mbox d} w / {\mbox d} T$ must be positive. 

Consequently, if measurements yield simultaneously a positive value of
$w$ and a negative value of ${\mbox d} w / {\mbox d} T$, then
$\sigma(T)$ cannot be described by some augmented power law with a
positive exponent. In this case, the sample considered is very likely
not metallic but insulating. -- Focusing on the slope of $w(T)$, we
make use of a second derivative (in a generalized sense) of
$\sigma(T)$ instead of a first derivative as it is considered in the
${\mbox d} \rho / {\mbox d} T = 0$ criteria of the MIT discussed in
Subsections 2.1 and 2.2. --

Assume now, $\sigma(T)$ follows the stretched Arrhenius relation Eq.\
(\ref{mer}) with positive $T_0$ and positive $\nu$, indicating that
the considered sample is insulating. In this case, we obtain
\begin{equation}
w(T) = \nu \, (T_0/T)^\nu \,.
\label{derivmer}
\end{equation}
Thus, $w(T)$ is always positive. It increases with diminishing $T$, so
that ${\mbox d} w / {\mbox d} T < 0$, and it diverges as
$T \rightarrow 0$.

It has to be stressed, however, that Eq.\ (\ref{derivmer}) can be
expected to be a good quantitative description of experimental data
only for the exponential limit. For this, $T$ has to be smaller than
$T_0$, so that $w$ must exceed $\nu$. 

Nevertheless, for any positive $w$, even if $w \ll \nu$, negative
${\mbox d} w / {\mbox d} T$ indicates non-metallic transport as,
starting from Eq.\ (\ref{apl}), we concluded in the above
considerations of $w(T)$ for the metallic side. Therefore, in this
way, also a large part of the samples with quite flat $\sigma(T)$ can
be unambiguously classified. -- If $\sigma(T,x)$ obeys scaling
according to Eq.\ (\ref{scaling}), then ${\mbox d} w / {\mbox d} T$
should be negative even for any positive $w$;\cite{Moe.etal.85}
compare Appendix B. -- 

In addition to its simplicity and unbiasedness for broad classes of
conceivable $\sigma(T)$, the discrimination between metallic and
insulating samples based on the sign of $\mbox{d} w / \mbox{d} T$ has
still another big advantage: Graphs of $w(T)$ for sample sets close to
the MIT contain direct, valuable information on the character of the
MIT. This can be understood as follows.

Suppose the MIT is continuous and, on its metallic side,
$\sigma(T, x)$ obeys Eq.\ (\ref{apl}) with some positive exponent $p$.
Thus, $w(T) = p$ would hold at the MIT itself.  Now, according to
experimental experience, $w(T = {\rm const.},x)$ always seems to
decrease strictly monotonically when the MIT is approached from the
insulating side and then crossed. Hence, as soon as the MIT has been
crossed, that means as soon as $w < p$, $\mbox{d} w / \mbox{d} T$
should be positive. Therefore, for any supposed value of $p$, the
existence of samples with simultaneously $0 < w < p$ and
$\mbox{d} w / \mbox{d} T < 0$ is a strong argument against the
hypothetical continuity of the MIT, that means against the continuity
of $\lim_{T \rightarrow 0} \sigma(T,x)$.\cite{Moe.etal.99,Moe.Adk}

The latter argument will play a central role in our reanalysis of
numerous experimental studies of the MIT in Section 4. It will be
supported by an  exploration of possible structures of sets of $w(T)$
curves in Section 5. Therein, we compare four qualitatively different
phenomenological models and exemplify how the character of the 
zero-temperature phenomenon MIT determines qualitative features of the
separatrix between metallic and insulating $w(T)$ at finite $T$.

At this point, however, we have to emphasize a physical restriction on
utilizing the simply structured Eqs.\ (\ref{derivapl}) and
(\ref{derivmer}): They can only be valid if the temperature is so low
that all but one conduction mechanisms are frozen out. There are many
cases where this condition is not met. For example, in amorphous
semiconductor transition-metal alloys, some high-temperature mechanism
seems to considerably enhance the conductivity on both sides of the
MIT above a threshold of the order of 
50~K.\cite{Moe.etal.85,Moe.85,Moe.Adk} It yields a positive
contribution to $w(T)$ which increases with $T$. For insulating
samples, its superposition with the hopping contribution, Eq.\
(\ref{derivmer}), causes $w(T)$ to exhibit a minimum; the
corresponding $T$ value tends to 0 as the MIT is
approached.\cite{Moe.etal.99,Moe.Adk} 
Such a behavior of $w(T)$ is not restricted to amorphous solids, it
can also be observed for nanocrystalline GeSb$_2$Te$_4$ above about
50~K, see Subsection 3.1, as well as for crystalline CdSe:In above
about 2~K, see Subsection 4.5.

Over the last two and a half decades, evaluating the logarithmic
derivative $w(T)$ for individual samples has proven to be very
effective in checking the validity of the augmented power law, Eq.\ 
(\ref{apl}), as well as of its extended version, Eq.\ (\ref{apl_ext});
see, for example, Refs.\ \onlinecite{Moe.Adk},
\onlinecite{Ros.etal.94}, \onlinecite{Moe.89},
\onlinecite{Hir.etal.89}, and \onlinecite{YY.etal.11}. For this aim,
the $w(T)$ data which are obtained by numerical differentiation
directly from the measured $\sigma(T)$ are contrasted with the
approximations of $w(T)$ resulting from analytic differentiation of
Eqs.\ (\ref{apl}) or (\ref{apl_ext}) with adjusted parameter values.
This comparison highlights small but possibly qualitative deviations
between experimental points and $\sigma(T)$
fits.\cite{Moe.89,Hir.etal.89} In particular, as pointed to above, for
slowly varying $\sigma(T)$, the repeatedly observed increase in $w(T)$
with decreasing $T$ is not compatible with the metallic nature of
transport; such samples should be insulating.\cite{Moe.89,Hir.etal.89}
This data analysis approach will be basic to the reconsideration of a
series of experimental key investigations of the MIT in various
disordered solids in our Section 4.

In several cases, a second type of deviation between the $w(T)$
obtained by numerical differentiation and the corresponding analytical
differentiation of the augmented power law, Eq.\ (\ref{apl}) fitted to
$\sigma(T)$, has been observed: According to these $w$ versus $T^p$
plots, the numerically obtained $w(T)$ reaches 0 or rapidly tends to 0
already at some finite $T$ in disagreement with Eq.\ (\ref{derivapl}).
Such a behavior may arise from a superposition of two mechanisms, see,
for example, Figures 8 and 9 in Ref.\ \onlinecite{Moe.etal.99}.
Alternatively, the rapid decrease at finite $T$ may occur due to the
measured $\sigma(T)$ data following an augmented power law with an
exponent which is considerably larger than $p$; this seems to be the
case in our Figure \ref{sip_1} below roughly 50~mK. Thermal decoupling
of sample and thermometer is a possible reason for such behavior of
$w(T)$. This explanation is particularly likely if, for different
samples, the rapid decrease of $w(T)$ always sets in at roughly the
same temperature. For example, Figure 1b of Ref.\
\onlinecite{Sac.etal.11} seems to indicate a related problem; more
such cases will be discussed in our Section 4.

Finally, we take up the universality hypothesis on the scaling of the
$T$ dependences of $\sigma$ in different homogeneous disordered solids
which was formulated in the last paragraph of the previous subsection.
It implies that $w(T/T_0)$ may be a material-independent function. In
this case, for a series of insulating samples of different such
solids, in the region of sufficiently low $T$, all the $w(T)$ curves
together would form a common flow diagram without intersections. More
specifically, for arbitrary two of these $w(T)$ curves, provided the
respective codomains overlap each other, it would even be possible to
collapse them by rescaling $T$ for one of the two samples by an
appropriate factor.

\subsection{Behavior of other observables near the MIT}

In the preceding subsections, we discussed six MIT criteria which all
focus on specific features of the temperature dependence of the
conductivity (or resistivity). As we will see in Section 3 and 4, the
respective classifications of samples into metallic and insulating
ones often contradict each other. This begs the question of what
additional information on the location of the MIT can be gained from
studying other observables. At first glance, the idea of using such
additional information to reach a less ambiguous classification may
look promising. However, one has to be aware of two fundamental
difficulties.

First, the MIT definition is based on the limit of $\sigma$ as
$T \rightarrow 0$. -- Furthermore, it presupposes that electric field
strength and frequency are infinitely small and that the sample size
is infinitely large; these three conditions are sufficiently well
fulfilled in most experiments. -- Likewise, when, instead of $\sigma$,
another observable is considered, it is essential to determine its
limiting value as $T \rightarrow 0$ for each sample; often,
additionally, the limit concerning a further measurement parameter
tending to zero has to be taken. These tasks may be very demanding, in
particular in case the alternative observable diverges or tends from
finite values to zero at one side of the MIT: One has to be aware
that, in the non-metallic region, nonexponential and exponential
$\sigma(T)$ may be correlated with qualitatively different $T$
dependences of the alternative observable.

Second, we will see in  Sections 3 and 4 how reanalyses of
$\sigma(T,x)$ data from various publications uncover systematic,
apparently generic inconsistencies in the respective interpretations
based on current localization theory. Thus, a sound and successful
microscopic theory on the $T$ dependence of the observable $\sigma$
seems not to be available at present. Since, however, this quantity is
fundamental in discriminating between metals and insulators, it is
therefore unlikely that present theories on other observables can
yield more reliable results than the available theories on
$\sigma(T)$. Consequently, we primarily take an empirical perspective
also in this subsection. 

In the following, we discuss in detail studies of two alternative 
observables and illuminate the biases inherent in the data analyses of
the respective publications. First, we turn to the behavior of the
Hall coefficient, $R_{\rm H}$. At crystalline Ge:Sb, the dependence of
this quantity on the donor concentration, $n$, was studied by Field
and Rosenbaum.\cite{Fie.Ros} The authors stated that conductivity,
$\sigma$, and inverse Hall coefficient, $1/R_{\rm H}$, simultaneously
tend to zero as the MIT is approached from the metallic side. This
seems to be substantiated by Figures 1 and 3 in Ref.\
\onlinecite{Fie.Ros}. Discussing these measurements, Field and
Rosenbaum pointed out that their conclusion conflicts with theoretical
work by Shapiro and Abrahams who expected $R_{\rm H}$ to be almost
constant close to the MIT.\cite{Fie.Ros,Sha.Abr} 

At first glance, the mentioned graphs look very convincing, but great
caution is advised in interpreting them for four reasons: (i) On
$\sigma$ as well as on $R_{\rm H}$, only data taken at one particular
temperature, 8~mK, were published in Ref.\ \onlinecite{Fie.Ros}.
Because they do not allow any $T \rightarrow 0$ extrapolation, these
data alone are of limited value, see the second paragraph of this
subsection. (ii) The smallest shown value of $\sigma(8\ {\rm mK},n)$
is not clearly smaller than Mott's estimate of the minimum metallic
conductivity $\sigma_{\rm M}$, but it roughly agrees with
$\sigma_{\rm M}$. Hence, also for this reason, the authors' conclusion
that $\lim_{T \rightarrow 0}\sigma(T,n)$ is continuous at the MIT has
to be considered as localization theory biased interpretation, and
their identification of the MIT point needs to be questioned.
(iii) The reliability of the authors' evaluation of the critical
behavior of $1/R_{\rm H}$ is strongly restricted by the uncertainty in
locating the MIT which has been uncovered in the previous points.
(iv) In case the MIT was indeed continuous, the samples might not be
close enough to the MIT to observe the theoretically expected
asymptotic behavior of $R_{\rm H}$, as was already discussed in Ref.\
\onlinecite{Fie.Ros}.

Therefore, the conclusion by Field and Rosenbaum that, for Ge:Sb,
$1/R_{\rm H}$ tends to zero as the MIT is approached cannot be
regarded as well founded. Later, however, their interpretation was
supported by Dai et al.\ in Ref.\ \onlinecite{Dai.etal.93}. The latter
authors investigated the $T$ dependence of the Hall coefficient of
crystalline Si:B for a series of samples, all of which they considered
to be metallic. Unfortunately, this publication itself does not
include any corresponding $\sigma(T,n)$ data. However, one can relate
these results on $R_{\rm H}$ to the measurements of $\sigma(T,n)$ in
Ref.\ \onlinecite{Dai.etal.91} performed by the same authors; samples
of same origin as well as the same concentration scale were used in
both cases. 

Dai et al.\ extrapolated the $T$ dependence of $1/R_{\rm H}$ to
$T = 0$ by adjusting the parameters of the ansatz
$1 / R_{\rm H}(T) = 1 / R_{\rm H}(0) + m_{\rm H}\,T^{1/2}$, see Figure
2 of Ref.\ \onlinecite{Dai.etal.93}. In checking the authors'
conclusions from this plot, doubts arise from two problems: (i) For
the two samples with the lowest dopant concentrations, $n = 4.16$ and
$4.11 \times 10^{18}\ {\rm cm}^{-3}$, the adjusted augmented power
laws well approximate the experimental data only within rather small
$T$ intervals. For the former sample, the width of the corresponding
$T^{1/2}$ range roughly equals the extrapolation gap, while, for the
latter sample, it is even considerably smaller than the gap. (ii) As
we will see in Subsection 4.3, at least the sample with
$n = 4.11 \times 10^{18}\ {\rm cm}^{-3}$ is presumably in fact
insulating, so that, in this case, $1/R_{\rm H}$ would likely tend to
zero if $T$ were diminished further, in contradiction to the
extrapolation by Dai et al..

Because of these two problems, one cannot be sure of the
interpretation of the $R_{\rm H}(T,n)$ data for Si:B in Ref.\
\onlinecite{Dai.etal.93}, too. In our opinion, these data are not
sufficient to exclude that $\lim_{T \rightarrow 0} 1/R_{\rm H}(T,n)$
might tend to a nonzero value when the MIT is approached from the
metallic side.

Finally, let us consider the study of the Hall effect in Si:As by Koon
and Castner.\cite{Koo.Cas.88,Koo.Cas.90} This investigation addresses
both sides of the MIT rather than only the metallic side as Refs.\
\onlinecite{Fie.Ros} and \onlinecite{Dai.etal.93} do: Figure 4b of
Ref.\ \onlinecite{Koo.Cas.90} presents $R_{\rm H}(T)$ for three
metallic and one insulating samples. According to this plot, it seems
not unlikely that $|R_{\rm H}|$ depends on $T$ and $n$ qualitatively
in the same way as the resistivity $\rho$. 

This similarity motivates the following hypothesis: In case the MIT is
discontinuous (in contrast to the perspective taken by Koon and
Castner), $|\lim_{T \rightarrow 0} 1/R_{\rm H}(T,n)|$ may tend to a
finite minimum metallic value when the MIT is approached from the
metallic side and jump to zero at the MIT itself. For dimensional
reasons, compare Appendix A, this minimum metallic value should be
proportional to the critical dopant concentration, $n_{\rm c}$. The
above hypothesis may be supported by the compilation of data presented
in Figure 3 of Ref.\ \onlinecite{Koo.Cas.88}, although with the
exception of the Ge:Sb points from Ref.\ \onlinecite{Fie.Ros} included
therein. Because, however, Ref.\ \onlinecite{Fie.Ros} does not report
$T$ dependences of $R_{\rm H}$, see above, it remains open whether or
not these Ge:Sb data are a valid counterexample to our hypothesis.

Hence, as shown in the above paragraphs, at the current stage,
studying the Hall coefficient does not substantially simplify the MIT
identification task. Instead, similar severe extrapolation problems
for $T \rightarrow 0$ as they impede the identification of the
transition point and the study of its vicinity on the basis of
$\sigma(T)$ data hamper also the analysis of $R_{\rm H}(T)$. Moreover,
trying to identify the MIT on the basis of the Hall coefficient is
hindered by far less information on this quantity being available in
the literature than on the conductivity.

As a second alternative observable, we now consider the
single-particle density of states obtained from tunneling experiments,
more precisely, the control parameter dependence of the zero-bias
density of states.  Investigating amorphous Nb$_x$Si$_{1-x}$ in Ref.\
\onlinecite{Her.etal.83}, Hertel et al.\ reported the formation of a
square-root zero-bias anomaly as the MIT is approached from the
metallic side and the opening of a gap on the insulating side. The
authors claimed a coincidence of, on the one hand,
$\lim_{T \rightarrow 0} \sigma(T,x)$ reaching zero and, on the other
hand, the zero-bias density of states vanishing.\cite{Her.etal.83} 

Unfortunately, Ref.\ \onlinecite{Her.etal.83} contains only the
extrapolated $\sigma(T = 0,x)$ data and not the original $T$ 
dependences of $\sigma$. However, a part of the latter data is
displayed in Figures 1 and 8 of the follow-up publication by Bishop et
al., Ref.\ \onlinecite{Bish.etal.85}. According to Figure 8 of Ref.\
\onlinecite{Bish.etal.85}, at low $T$, the experimental $\sigma(T)$
substantially deviate from the $\sigma = a + b\,T^{1/2}$ behavior
stated in Ref.\ \onlinecite{Her.etal.83}. Hence, the corresponding
$T \rightarrow 0$ extrapolations as well as the location of the MIT
derived therefrom are called into question. (The published information
is not sufficient for a detailed reanalysis such as presented for
various other substances in our Sections 3 and 4.)  

For the following four reasons, also the MIT identification by means
of tunneling experiments in Ref.\ \onlinecite{Her.etal.83} is not
conclusive: (i) These measurements of the density of states were
performed only at a single temperature, 2~K, see Figure 2 of Ref.\
\onlinecite{Her.etal.83}, which again renders a serious
$T \rightarrow 0$ extrapolation impossible, compare the second
paragraph of this subsection. (ii) To estimate the density of states
for zero tunneling voltage based on values taken at finite tunneling
voltages, the authors extrapolated an augmented power law
approximation, $N(E) = A + B\,E^{1/2}$. However, the $N(E)$ versus
$E^{1/2}$ plots in Figure 2 of Ref.\ \onlinecite{Her.etal.83} exhibit
a significant curvature in the energy range taken into account in
adjusting $A$ and $B$. \linebreak (iii) Without presenting any
evidence, the authors interpreted the deviations between these fits
and the measuring results for tunneling voltages below about 1~mV as
an effect of thermal fluctuations. However, without comparison with
data for other $T$ values, such a specification of the corresponding
voltage range remains speculation, and so do the zero-voltage
extrapolations in Figure 2 of Ref.\ \onlinecite{Her.etal.83}. -- For
a Coulomb gap occurring on the insulating side of the MIT, the
finite-$T$ effect can be quite large and influence the density of
states at energies far above $k_{\rm B} T$, see Figures 1 in Refs.\
\onlinecite{Wolf.etal.75}, \onlinecite{Sand.etal.01}, and
\onlinecite{Sarv.etal.95}. -- (iv) The permittivity diverges as the
MIT is approached from the insulating side.\cite{Hess.etal} Thus, for
$T = 0$, the width of the Coulomb gap is expected to tend to zero
there. Hence, the density increase caused by thermal fluctuations
should be particularly large at the MIT. These expectations are not
met by that tunneling-voltage dependence of the density of states
which is ascribed to the allegedly critical Nb content in Figure 2 of
Ref.\ \onlinecite{Her.etal.83}. 

The problems discussed in the two paragraphs above invalidate the main
points of the interpretation in Ref.\ \onlinecite{Her.etal.83}: In our
opinion, neither the extrapolated conductivity values nor the
tunneling data which were published therein can reliably locate the
MIT. Therefore, this experiment cannot be regarded as conclusive
support for the MIT being continuous.

In addition to Hall coefficient and single-particle density of states,
various other alternative observables have been studied close to the
MIT over the last decades, in particular, dielectric
susceptibility,\cite{Hess.etal,Paa.etal} magnetic 
susceptibility,\cite{Oot.Mat} thermopower,\cite{Lak.Loe} and
noise.\cite{Coh.etal.92,Coh.etal.94} Reconsidering all these
experiments is beyond the scope of our work. Nevertheless, we
emphasize that the general remarks on the interpretation difficulties
of such investigations made in the second and third paragraphs of this
subsection are valid in all these cases.

\subsection{Combination of criteria}

Let us turn back to the temperature and control parameter dependences
of the conductivity or resistivity on which the MIT definition is
based.  The discussion in Subsections 2.2 to 2.6 provided us with the
following two apparently generally applicable, sufficient but not
necessary classification conditions: On the one hand, all samples with
${\mbox d} \rho / {\mbox d} T > 0$, and thus $w < 0$, at the lowest
experimentally accessible temperature, $T_{\rm lea}$, are very likely
metallic. On the other hand, all samples with simultaneously $w > 0$
and ${\mbox d} w / {\mbox d} T < 0$ at $T_{\rm lea}$ are with high
likelihood insulating. Only for the very few samples that do not fall
into one of these two categories, additional measurements at
temperatures below $T_{\rm lea}$ are necessary for a reliable
decision.

This cautious classification is denoted as 
${\mbox d} \ln \sigma / {\mbox d} \ln T$ approach in the following. 
It has the invaluable advantage to be almost unbiased with respect to
the character of the MIT. The price one pays is that it can only
bracket the location of the MIT. However, the corresponding control
parameter gap is far narrower than the uncertainty interval in an
analysis by means of stretched Arrhenius law fits. The reason is that,
simultaneously, the conditions $w > 0$ and
${\mbox d} w / {\mbox d} T < 0$ are fulfilled also by a great fraction
of those insulating samples for which only nonexponential $\sigma(T)$
can be observed above $T_{\rm lea}$, see Subsection 2.6.

In summary, mainly three approaches are available for determining the
MIT point: the criterion ${\mbox d} \rho / {\mbox d} T = 0$ in the
limit as $T \rightarrow 0$ (or at the lowest accessible $T$), the
criterion $a = 0$ for augmented power law approximations
$\sigma = a + b\,T^p$, and the
${\mbox d} \ln \sigma / {\mbox d} \ln T$ approach described in the
previous paragraphs. For low-temperature measurements at a few K or
below, the resulting critical values of the control parameter may be
close to each other. Simultaneously, however, the conclusions about
the character of the MIT can differ qualitatively due to the biases
inherent in the first and second criteria, which favor discontinuity
and continuity of the MIT, respectively. Therefore, only if the sample
classification according to the
${\mbox d} \ln \sigma / {\mbox d} \ln T$ approach turns out to agree
either with the corresponding results from the first criterion or with
those from the second, one will be able to state that consistency of
the data evaluation has been reached.

In the next two sections, we combine the three approaches to
critically examine first the statement of Ref.\
\onlinecite{Sie.etal.11} on the specific character of the MIT in
phase-change materials and then the common wisdom on the continuity of
the MIT in broad classes of disordered solids.

\section{Character of the MIT in GeSb$_2$Te$_4$}

\subsection{Temperature dependences of the conductivity}

We now examine the recent Ref.\ \onlinecite{Sie.etal.11} in which
Siegrist et al.\ investigated the MIT in GeSb$_2$Te$_4$ and related
phase-change materials on increasing annealing temperature. In this
work, the authors claim to have detected the existence of a finite
minimum metallic conductivity. They conclude this result from
$\log \rho$ versus $T$ plots, Figures 2 and 3 in Ref.\
\onlinecite{Sie.etal.11}, using the sign change of
${\mbox d} \rho / {\mbox d} T$ at the measuring temperature as MIT
criterion.

The authors emphasize their observation to be surprising. They do so
despite various related prior work,\cite{Edw.etal.95,Sch.11} see also
Subsection 2.2. In the literature, one finds several studies of the
MIT in disordered systems which contain graphs resembling Figure 3 of
Ref.\ \onlinecite{Sie.etal.11}. In particular, already more than four
decades ago, Yamanouchi et al.\ obtained similar results for
crystalline Si:P, see the log-log plot Figure 1 of Ref.\
\onlinecite{Yama.etal.67}. This investigation was one of the
experiments to which Mott referred as support for his hypothesis of
the minimum metallic conductivity.\cite{Mott.72a,Mott.Kaveh} 

Later, however, the $\sigma(T)$ studies of crystalline Si:P were
extended to the mK range.\cite{Tho.etal.83,Stu.etal.93,Waf.etal.99}
These measurements were evaluated by $T \rightarrow 0$ extrapolations
based on $\sigma = a + b \, T^p$ fits, see Subsection 2.3. In these
analyses, $p = 1/2$ was supposed by Thomas et al.\cite{Tho.etal.83}
and by Stupp et al.\cite{Stu.etal.93}, whereas Waffenschmidt et
al.\cite{Waf.etal.99} started from $p = 1/3$. All these three groups
inferred that the MIT is continuous. This raises the question whether
the conclusion drawn by Siegrist et al.\ could merely be a consequence
of the perspective chosen, i.e.\ of using the sign change of
${\mbox d} \rho / {\mbox d} T$ as MIT criterion.

\begin{figure}
\includegraphics[width=0.85\linewidth]{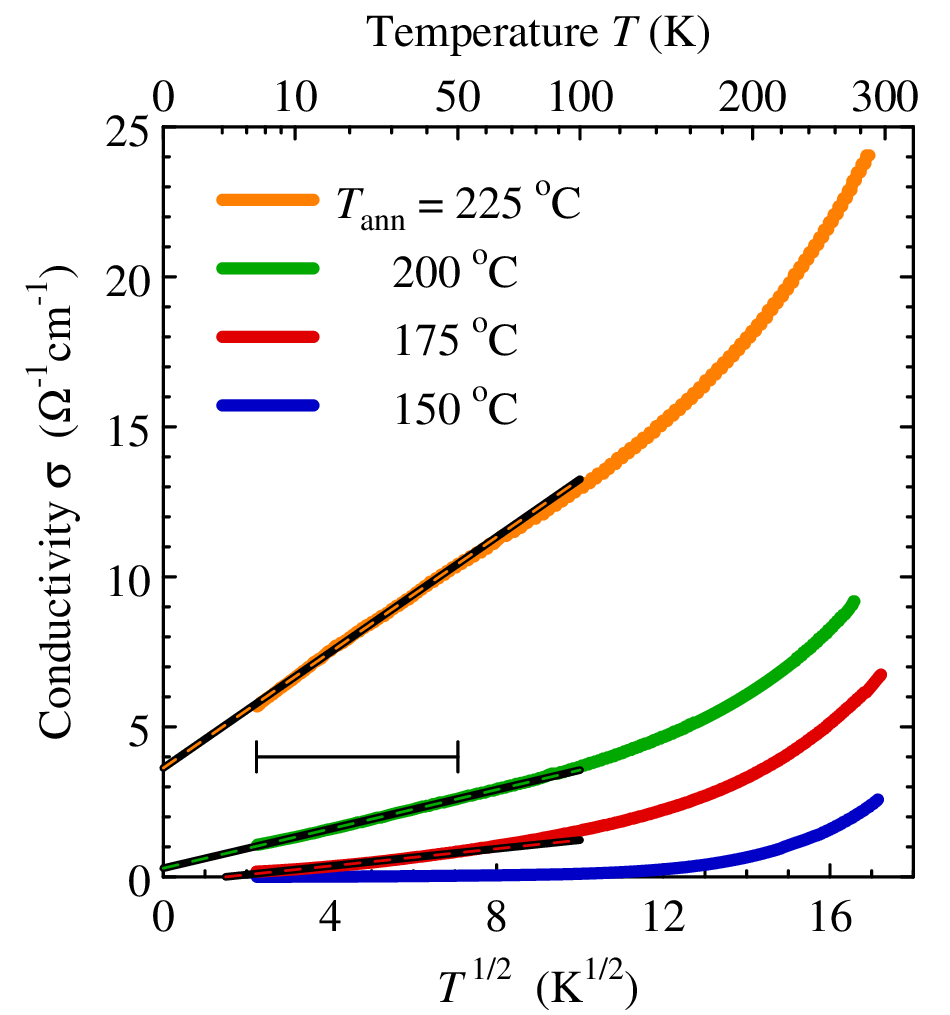}
\caption{(Color online) Temperature dependences of the conductivity of
four GeSb$_2$Te$_4$ films redrawn from Figure 3 of Ref.\ 
\onlinecite{Sie.etal.11}. According to that plot, all these samples
are considered as insulating in Ref.\ \onlinecite{Sie.etal.11}. The
black straight lines which are overlaid with colored dashes show
corresponding $\sigma = a + b \, T^{1/2}$ approximations. The $T$
interval 5 to 50~K taken into account in adjusting the respective
parameters $a$ and $b$ is marked by a horizontal bar.} 
\label{AA_fig}
\end{figure}

To elucidate this problem, we now evaluate in detail $\rho(T)$ data
for GeSb$_2$Te$_4$ from Figure 3 of Ref.\ \onlinecite{Sie.etal.11}.
For this aim, we digitized the curves for the four lowest annealing
temperatures, $T_{\rm ann} = 150$, 175, 200, and $225\ {\rm ^oC}$.
They are redrawn in a $\sigma$ versus $T^{1/2}$ plot in our Figure
\ref{AA_fig}. This graph tests for the relation
$\sigma(T) = a + b \, T^{1/2}$, the Altshuler-Aronov case of the
augmented power law, Eq.\ (\ref{apl}).\cite{Alt.Aro.79} Corresponding
regression lines which incorporate all data points from one
temperature decade, 5--50~K, are included in Figure \ref{AA_fig}. 

For $T_{\rm ann} = 200$ and $225\ {\rm ^oC}$, the regression lines
seem to nicely approximate the experimental data, not only in the
temperature range that was used in adjusting the parameters, but even
beyond it up to roughly 100~K. (One should not overestimate the fit
quality for $T_{\rm ann} = 175\ {\rm ^oC}$, since these $\sigma$
values are so small that the substantial relative deviations are not
noticeable in Figure \ref{AA_fig}.) Note that, for
$T_{\rm ann} = 225\ {\rm ^oC}$, the $\sigma = a + b \, T^{1/2}$
extrapolation yields $\sigma(T = 0) = 3.6\ {\rm \Omega^{-1} cm^{-1}}$.
This value is smaller by a factor of 100 than the estimate of the
minimum metallic conductivity for GeSb$_2$Te$_4$ by means of the
$\mbox{d} \rho / \mbox{d} T = 0$ criterion in Ref.\
\onlinecite{Sie.etal.11}, 300 to $500\ {\rm \Omega^{-1} cm^{-1}}$. 

In the augmented power law approach, the MIT is indicated by $a = 0$,
see Subsection 2.3. Thus, according to our Figure \ref{AA_fig}, the
critical value of $T_{\rm ann}$ should amount to roughly
$200\ {\rm ^oC}$. This value deviates considerably from the result
$275\ {\rm ^oC}$ which was obtained in Ref.\ \onlinecite{Sie.etal.11}
by means of the $\mbox{d} \rho / \mbox{d} T = 0$ criterion. Therefore,
the question arises how to classify the samples annealed at
temperatures in between these two values, in particular the sample
obtained by annealing at $225\ {\rm ^oC}$: According to our Figure
\ref{AA_fig}, it should be metallic, but Figure 3 of Ref.\
\onlinecite{Sie.etal.11} suggests that it is presumably insulating. 

One and a half years after a first preliminary version of this review
had been made available at arXiv.org, see Ref.\ \onlinecite{Moe.13},
three of the authors of Ref.\ \onlinecite{Sie.etal.11} submitted a
subsequent study of the MIT in GeSb$_2$Te$_4$, Ref.\
\onlinecite{Vol.etal.15}. -- Apparently, in Refs.\
\onlinecite{Sie.etal.11} and \onlinecite{Vol.etal.15}, (almost)
identical sample preparation procedures were used, compare the
preparation details given in the Supplementary Information of Ref.\
\onlinecite{Sie.etal.11} and in Section 2 of Ref.\
\onlinecite{Vol.etal.15} with each other. -- The authors of Ref.\
\onlinecite{Vol.etal.15} not only extended the explored $T$ range by
one order of magnitude down to 0.35~K but also changed their
interpretation approach: In locating the MIT, they attached particular
importance to the $\sigma = a + b \, T^{1/2}$ approximation, now
trusting in it as ``the most widely accepted extrapolation
method''.\cite{Vol.etal.15} Indeed, Figure 2 of Ref.\
\onlinecite{Vol.etal.15} shows that, again, this ansatz works rather
well. Thus it supports our above conclusion on the good quality of
$\sigma = a + b \, T^{1/2}$ approximations of part of the data
published in Ref.\ \onlinecite{Sie.etal.11}. Simultaneously, it yields
that the critical annealing temperature should fall between 200 and
$225\ {\rm ^oC}$, but closer to the former value. This result only
slightly exceeds our above estimate of roughly $200\ {\rm ^oC}$
obtained from the $\sigma(T)$ in Ref.\ \onlinecite{Sie.etal.11}.

Hence, together, Refs.\ \onlinecite{Sie.etal.11} and
\onlinecite{Vol.etal.15} form a prime example of the decisive
influence of the MIT criterion used on the character of the MIT
derived. However, not mentioning the reinterpretation of their
previous data in Ref.\ \onlinecite{Moe.13}, although citing that
source, the authors of Ref.\ \onlinecite{Vol.etal.15} implicitly
ascribe their altering the MIT characterization which they had
concluded in Ref.\ \onlinecite{Sie.etal.11} only to the focus on the
lowest $T$ decade. This way, they hide the interpretation ambiguity
from the reader. 

Nevertheless, the $\sigma = a + b \, T^{1/2}$ approximations are not
perfect. The authors of Ref.\ \onlinecite{Vol.etal.15} point to the
weak curvature of their $\sigma$ versus $T^{1/2}$ plots limiting the
applicability range of the augmented power law; they mention similar
findings for the alternative exponent $1/3$. This feature is not
specific to the lowest $T$ decade of their study, which was not taken
into consideration in Ref.\ \onlinecite{Sie.etal.11}, and thus also
not in the above examination of that work. One can already note it
carefully inspecting the curve for $T_{\rm ann} = 225\ {\rm ^oC}$ in
our Figure \ref{AA_fig}. A detailed investigation of such weak
deviations will be presented for the example of crystalline Si:P in
Subsection 4.2.

\begin{figure}
\includegraphics[width=0.85\linewidth]{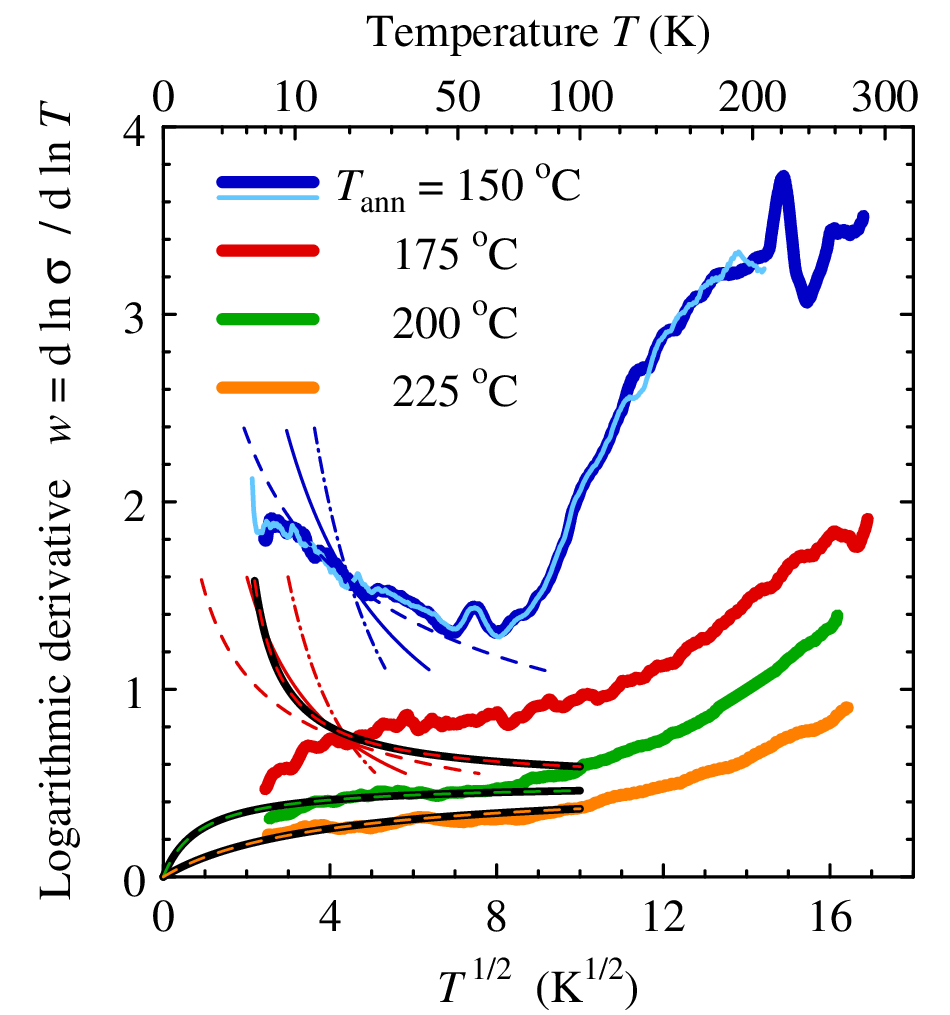}
\caption{(Color online) Temperature dependences of the logarithmic
derivative of the conductivity, $w$, for the four GeSb$_2$Te$_4$ data
sets shown in Figure \ref{AA_fig}. For $T_{\rm ann} = 150\ {\rm ^oC}$,
these values (dark-colored) are compared to data resulting from Figure
8s of the Supplementary Information of Ref.\ \onlinecite{Sie.etal.11}
(light-colored), see text. Black curves overlaid with colored dashes
correspond to the regression lines in Figure \ref{AA_fig}. They
indicate metallic behavior if $w \rightarrow 0$ as $T \rightarrow 0$.
To check for activated transport, dashed, full, and dashed-dotted thin
lines give approximations of $w(T)$ by Eq.\ (\ref{derivmer}) for
$\nu = 1/4$, $1/2$ and 1, respectively, see text.}
\label{w_fig}
\end{figure}

To gain additional information for the controversial classification of
the $\sigma(T)$ data from Ref.\ \onlinecite{Sie.etal.11}, we now study
the $T$ dependence of the logarithmic derivative $w$, defined by Eq.\
(\ref{logderiv}) above. Figure \ref{w_fig} shows $w(T)$ for the four
samples already considered in Figure \ref{AA_fig}. It compares results
obtained by direct numerical differentiation of the experimental data
with functions derived from analytic differentiations of various fits
to the experimental $\sigma(T)$.

The numerically calculated $w(T)$ points were obtained from the
$\ln T (\ln \sigma)$ relations as follows: Sliding a $\ln \sigma$
window along the curve, we calculated the slope by means of linear
regression. In this procedure, we related the respective value of the
slope to that $\ln \sigma$ value where the truncation error is
particularly small, that means, where it is only of second order in
the width of the $\ln \sigma$ window; for a detailed explanation see
Appendix C. To determine the corresponding value of $\ln T$, we used
linear regression. In the present case, numerically calculating 
$1 / w =  \mbox{d} \ln T / \mbox{d} \ln \sigma$ instead of $w$ was
advantageous with respect to finding a window width which is almost
optimal simultaneously at low and high $T$. This width was chosen as
small as possible, but so large that digitization effects and random
errors are of negligible influence. The values 0.30, 0.15, 0.15, and
0.10 were used for $T_{\rm ann} = 150$, 175, 200, and
$225\ {\rm ^oC}$, respectively. To ensure that the width of the
related $T$ interval does not fall below 1.75~K, the $\ln \sigma$
window was correspondingly expanded if necessary.

The reliability of this procedure was checked in two different ways:
First, for $T_{\rm ann} = 150\ {\rm ^oC}$, additionally, the
$\sigma(T)$ data were also reconstructed from the $\log \rho$ versus
$T^{-1/4}$ plot in Figure 8s of the Supplementary Information of Ref.\
\onlinecite{Sie.etal.11}. At low $T$, this graph has a far better
resolution than Figure 3 of Ref.\ \onlinecite{Sie.etal.11}. Thus,
$w(T)$ could be determined using a $\ln \sigma$ window of width 0.3
over the whole $T$ range, without the occasional interval expansion
mentioned above. Figure \ref{w_fig} shows that both the independently
obtained $w(T)$ curves for $T_{\rm ann} = 150\ {\rm ^oC}$ nicely agree
with each other. Second, several unusual patterns of the curves were
analyzed. For example, in the case of $T_{\rm ann} = 150\ {\rm ^oC}$,
the minimum at 63~K turned out not to be an artifact of the
digitization but to originate from an offset of the parts of the curve
below 61~K and above 65~K by roughly 1~K. This offset is also visible
in a large magnification of the original graph. Hence, although the
published $\sigma(T)$ data are sufficiently precise for an overview,
they exhibit artifacts which substantially influence $w(T)$. Therefore
one should be cautious in judging the abnormal features of $w(T)$
discussed in the following.

Figure \ref{w_fig} compares the curves obtained by numerical
differentiation from the reconstructed experimental data with several
theoretical approximations: The dashed, full, and dashed-dotted thin
lines relate to activated transport in insulating samples according to
Eq.\ (\ref{mer}) with $\nu = 1/4$, $1/2$, and 1, respectively. The
corresponding expressions for $w(T)$ include only one adjustable
parameter, $T_0$, see Eq.\ (\ref{derivmer}). Thus, to compare with the
experimental data, the values of $T_0$ were chosen such that
$w(20\ {\rm  K})$ agrees with the result of numerical differentiation.
The black curves overlaid with colored dashes are based on the
Altshuler-Aronov approximation for metallic samples, Eq.\ (\ref{apl})
with $p = 1/2$.  They result from analytic differentiation of the
regression curves in Figure \ref{AA_fig}. 

We now discuss Figure \ref{w_fig} in detail. For $T_{\rm ann} = 200$
and $225\ {\rm ^oC}$, within a wide $T$ range, the curves obtained by
numerical differentiation of the $\sigma(T)$ data agree rather well
with the analytically differentiated $\sigma =\  a + b \, T^{1/2}$
regressions. Thus, the impression from Figure \ref{AA_fig} is
supported that already annealing at 225 ${\rm ^oC}$ should be
sufficient to reach clearly metallic behavior. Hence, from this
perspective, the MIT seems to be continuous, at variance with Ref.\
\onlinecite{Sie.etal.11}. 

For $T_{\rm ann} = 150\ {\rm ^oC}$, the low-$T$ part of $w(T)$ clearly
indicates activated transport due to its negative slope, as expected.
In this case, below roughly 50~K, the interpretation in terms of
variable-range hopping of Mott type (Eq.\ (\ref{mer}) with
$\nu = 1/4$) by Siegrist et al.\ seems to be confirmed. However,
significant deviations are visible below 9~K. The shape of $w(T)$
suggests that they are presumably not real but originate from
experimental inaccuracies. Note, moreover, the qualitative change of
$w(T)$ at roughly 50~K. The positive slope of $w(T)$ above 50~K is not
compatible with any description of a hopping mechanism by Eq.\
(\ref{mer}). However, such a behavior is not unusual: It points to a
second mechanism contributing,\cite{Moe.etal.85,Moe.85,Moe.etal.99}
compare also Subsection 4.5, devoted to crystalline CdSe:In.

So far so good, but what about the measurements on the sample annealed
at $175\ {\rm ^oC}$? For this sample, the findings are confusing: It
should exhibit activated transport according to the unphysical
$\sigma(T \rightarrow 0)$ extrapolation in Figure \ref{AA_fig}. (The
corresponding $\sigma = a + b \, T^{1/2}$ fit implies a divergence of
$w(T)$ at finite $T$, indicated by the black line overlaid with red
dashes in Figure \ref{w_fig}.) Hence, in the low $T$ region, the slope
of $w(T)$ should be negative as for the three shown approximations for
activated transport according to Eq.\ (\ref{derivmer}). In fact,
however, the slope of the $w(T)$ curve obtained by numerical
differentiation of the $\sigma(T)$ data is positive down to the
smallest $T$ considered. This discrepancy raises serious doubts about
the consistency of all the sample classifications obtained from the
$\sigma = a + b \, T^{1/2}$ fits in Figure \ref{AA_fig}.

How could one understand the contradiction? On the one hand, the
temperature might not be low enough. This would be the case if the
above mentioned high-$T$ mechanism had a strong influence even at
temperatures of a few K. Consequently, for the samples annealed at 200
and $225\ {\rm ^oC}$, the usual $\sigma = a + b \, T^{1/2}$ fits in
Figure \ref{AA_fig} would be applicable only by chance. Moreover, also
the Mott variable-range hopping behavior for
$T_{\rm ann} = 150\ {\rm ^oC}$ would not be generic but result from
the superposition of two mechanisms.  

Another possibility is that the exponent of the augmented power law
might increase when the MIT is approached. Thus, the simple
$\sigma = a + b \, T^{1/2}$ extrapolation procedure might erroneously
classify a metallic sample as insulating. However, such a considerable
increase of $p$ in the vicinity of the MIT seems to contradict the
common experience with other disordered solids, compare Subsection 2.3
and Section 4.

On the other hand, the positive slope of $w(T)$ at low temperatures
for $T_{\rm ann} = 175\ {\rm ^oC}$ might also be caused by various
experimental problems such as incomplete thermalization, Joule
heating, or sample inhomogeneities. The low-$T$ deviations for
$T_{\rm ann} = 150\ {\rm ^oC}$ mentioned above point in this
direction. They may originate from a continuous resistance measurement
with too fast sliding temperature. A more clear picture will probably
be gained when, for a comparably small number of $T$ values, the
resistance is always measured only after careful
thermalization.\cite{Moe.etal.99} Of course, the effect of such an
improvement of the measurement has to be checked for all samples. In
particular, it will be intriguing to see how, for
$T_{\rm ann} = 200\ {\rm ^oC}$ and $225\ {\rm ^oC}$, the seemingly
good agreement between measured data and regression lines in Figure
\ref{AA_fig} and between numerically and analytically obtained $w(T)$
curves in Figure \ref{w_fig} will be influenced. Will it persist or
disappear?

\begin{figure}
\includegraphics[width=0.85\linewidth]{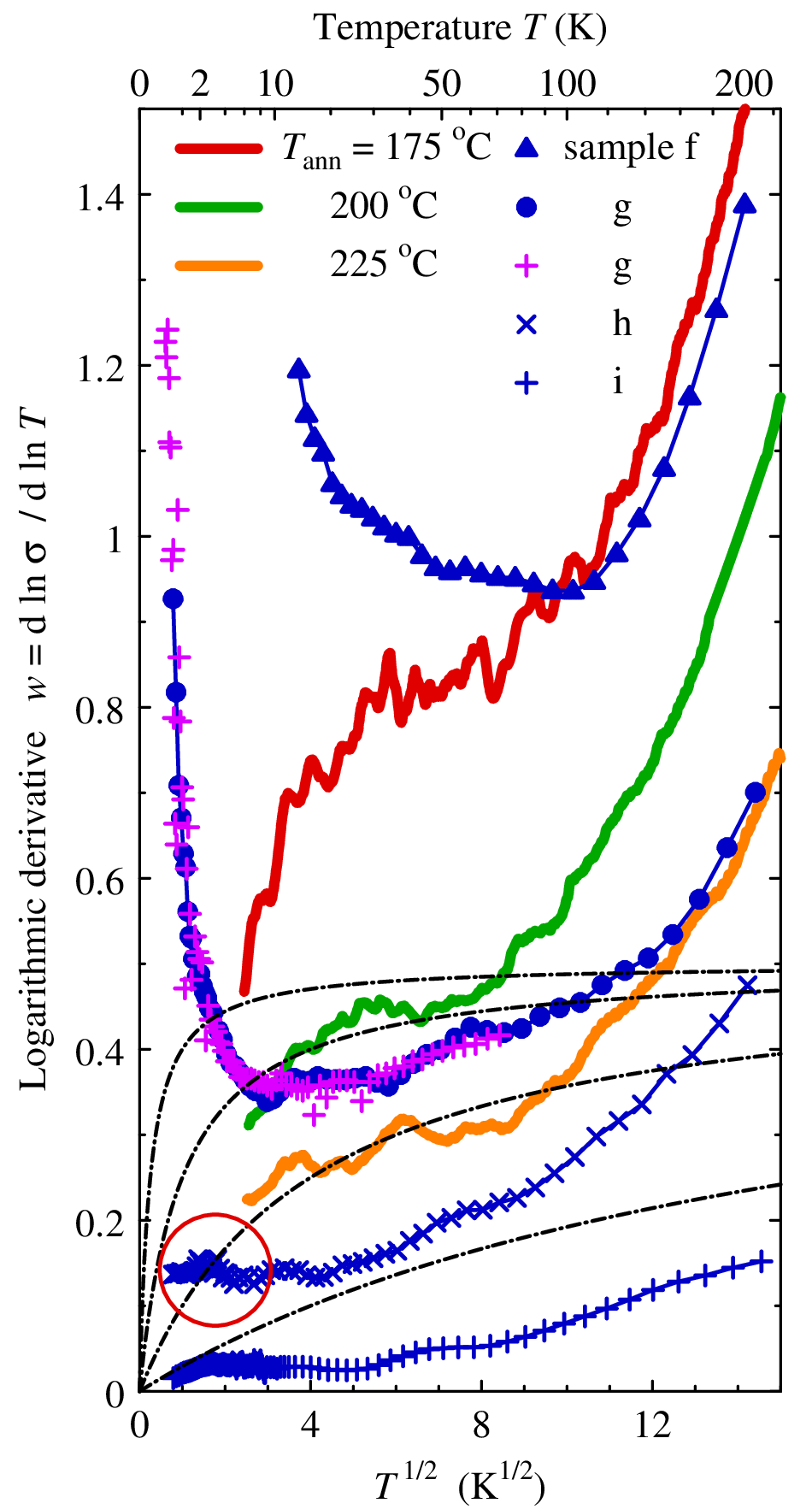}
\caption{(Color online) Comparison of temperature dependences of the
logarithmic derivative of the conductivity, $w$, obtained from two
consecutive studies of GeSb$_2$Te$_4$. The three curves marked by the
respective annealing temperatures are a zoom-in of Figure \ref{w_fig}.
They were obtained from Ref.\ \onlinecite{Sie.etal.11}; for
explanations see the caption of Figure \ref{w_fig} and the related
text. The data sets for the samples f, g, h, and i, marked blue, were
obtained from the subsequent publication Ref.\ \onlinecite{Vol.etal.15}
by means of digitizing Figures 1 and 2 therein and calculating $w(T)$
numerically, see text; as precision check, data for sample g, marked by
magenta $+$ here, were redrawn from Figure 3 of Ref.\ \onlinecite{Vol.etal.15}.
These four samples had been annealed at 175, 200, 225, and
$250\ {\rm ^oC}$, respectively. For comparison with theory, the
dashed-dotted lines represent $w(T)$ resulting from hypothetical
$\sigma = a + b \, T^{1/2}$ with $a / b = 1/4$, 1, 4, and 16 (from
top to bottom).}
\label{w_fig_cond}
\end{figure}

A first answer to this question is given by Figure 3 of the subsequent
publication Ref.\ \onlinecite{Vol.etal.15}, already pointed to above.
It presents $w(T)$ curves derived from measurements on GeSb$_2$Te$_4$
films down to 0.35~K. To simplify the comparison of our Figure
\ref{w_fig} with that diagram, we here condense $w(T)$ data sets
obtained from both the Refs.\ \onlinecite{Sie.etal.11} and
\onlinecite{Vol.etal.15} in a common plot, Figure \ref{w_fig_cond}.
For this aim, we digitized Figures 1 and 2 of Ref.\
\onlinecite{Vol.etal.15} and calculated the corresponding $w(T)$
sliding windows containing eight neighboring data points along the
$\ln \sigma (\ln T)$ curves of the samples f, g, h, and i; these films
had been annealed at 175, 200, 225, and $250\ {\rm ^oC}$,
respectively. Additionally, as precision check of our procedure,
Figure \ref{w_fig_cond} includes data points of sample g obtained by
digitizing the $w(T)$ presented in Figure 3 of Ref.\
\onlinecite{Vol.etal.15}.

Our Figure \ref{w_fig_cond} contains a lot of valuable information as
we will see in this and the next four paragraphs. Note, above about
140~K, both the $w(T)$ for $T_{\rm ann} = 175\ {\rm ^oC}$ from Refs.\
\onlinecite{Sie.etal.11} (old) and \onlinecite{Vol.etal.15} (new)
agree rather well with each other. The same can be said about the two
$w(T)$ curves for $T_{\rm ann} = 225\ {\rm ^oC}$ from Ref.\
\onlinecite{Sie.etal.11} and for $T_{\rm ann} = 200\ {\rm ^oC}$ from
Ref.\ \onlinecite{Vol.etal.15}. On the contrary, below about 100~K,
the respective old and new curves qualitatively differ from each
other: While the old $w(T)$ seem to decrease with $T$ down to the
lowest $T$ taken into account in such a way that both the
corresponding samples should be metallic, see above, the new $w(T)$
clearly increase with decreasing $T$ at the lower end of the $T$ range
considered which indicates non-metallic conduction. Based on the
common assumption that the new data are more precise than the old
ones, we conclude that experimental artifacts are the most likely
origin of the unusual behavior of the old $w(T)$ for
$T_{\rm ann} = 175\ {\rm ^oC}$, discussed above. Simultaneously, we
mention that, according to the $w(T)$ of sample g, doubts also arise
about the low-$T$ decrease of the $w(T)$ of the old sample annealed at
$200\ {\rm ^oC}$.

In consequence of their low-$T$ upturns, both the $w(T)$ of samples f
and g exhibit minima. They occur at about 100~K for sample f, annealed
at $175\ {\rm ^oC}$, and at roughly 12~K for sample g, annealed at
$200\ {\rm ^oC}$. Thus, this $T$ value seems to decrease as the MIT is
approached. Such a behavior is already known from a-Si$_{1-x}$Cr$_x$
and a-Si$_{1-x}$Ni$_x$;\cite{Moe.Adk,Moe.etal.99} moreover, it is
present also in the case of crystalline CdSe:In as will be shown in
Section 4.5.

The crucial point now is the following: According to $w(T)$, sample g
is clearly non-metallic. Its minimum $w$ value, however, falls
considerably below 1/2; it amounts to about 0.35. This finding is
incompatible with the hypothesis of a continuous MIT at which
$\sigma \propto T^{1/2}$ since $w(T = {\rm const.}, T_{\rm ann})$
decreases monotonically with increasing $T_{\rm ann}$; see Subsection
2.6.

This conclusion is supported by the behavior of $w(T)$ for sample h.
In this case, $w(T)$ seems to be roughly constant below about 20~K;
$w$ amounts to 0.14 in this $T$ range. Focusing on the diagram region
highlighted by a red circle, one sees that this feature is clearly
inconsistent with the properties of the $w(T)$ curves resulting from
hypothetical $\sigma = a + b \, T^{1/2}$ for several values of the
quotient $a/b$, which is the only adjustable parameter here.
Therefore, the $T \rightarrow 0$ extrapolation by means of this ansatz
in Ref.\ \onlinecite{Vol.etal.15}, yielding
$a = 11.2\ {\rm \Omega^{-1} cm^{-1}}$ so that sample h should be
metallic, cannot be trusted. -- It is strange that this problem was
not mentioned in that work. -- Even worse, according to our Figure
\ref{w_fig_cond}, it is not justified to conclude that the $w(T)$ of
sample h vanishes as $T \rightarrow 0$. For both these reasons, the
classification of this sample as metallic in Ref.\
\onlinecite{Vol.etal.15} is presumably not correct. Because, however,
just this classification is basic to the central statement of Ref.\
\onlinecite{Vol.etal.15} that the MIT is most likely continuous, this
main conclusion of that publication has to be called into question as
well.

Finally, concerning sample i, annealed at $250\ {\rm ^oC}$ and
considered as clearly metallic in Ref.\ \onlinecite{Vol.etal.15}, we
point to the apparent knee in $w(T)$ at about 4 K. It could be
interesting to check this unusual feature by repeating the
measurements with lower current and increased precision.

Concluding, taken as a whole, the situation is paradoxical: On the one
hand, our analysis of the $\sigma(T)$ data published in the original
study, Ref.\ \onlinecite{Sie.etal.11}, in which Siegrist et al.\ had
claimed the MIT to be discontinuous, showed that the ansatz
$\sigma = a + b \, T^{1/2}$ seems to work nicely: It yields reasonable
descriptions of experimental $w(T)$ for two samples which were
classified as insulating in Ref.\ \onlinecite{Sie.etal.11}. This
finding is not consistent with the sample classification in that work
and might be an indication of the MIT being continuous. On the other
hand, in the subsequent publication, Ref.\ \onlinecite{Vol.etal.15},
the explored $T$ range was extended by one order of magnitude down to
0.35~K and the measurement precision was apparently improved. Therein,
the authors state the MIT to be most likely continuous and make use of
the ansatz $\sigma = a + b \, T^{1/2}$.  However, the $w(T)$ curves
presented in Figure 3 of that work and the ones in our Figure
\ref{w_fig_cond} clearly imply that this ansatz is not applicable in
the immediate vicinity of the hypothetical MIT. Thus the continuity of
the MIT is called into question.

These contradictions highlight how the ``common wisdom''
interpretation bias can play a subtle but decisive role in MIT
studies. In Section 4, we will show that also numerous publications on
the MIT in various other disordered solids suffer from such bias
problems. In doing so, considering different choices of the control
parameter, $x$, we will identify the likely generic features of the
$w(T,x = {\rm const.})$ flow diagrams for the vicinity of an MIT.

\subsection{Critical charge carrier concentration}

The second argument based on which Siegrist et al.\cite{Sie.etal.11}
ascribe an ``unparalleled quantum state of matter'' to GeSb$_2$Te$_4$
is the allegedly strong deviation of their data from the Mott
criterion,\cite{Edw.Sie.78} 
\begin{equation}
n_{\rm c}^{\ 1/3} a_{\rm H}^{\ *} = 0.26 \pm 0.05\,.
\label{mott_crit}
\end{equation}
This universal relation links the critical charge carrier
concentration, $n_{\rm c}$, at the MIT to the effective Bohr radius,
$a_{\rm H}^{\ *}$, of the localized states. Its structure -- without
specific value at the right-hand side -- follows already from a
dimensional analysis, see Appendix A. Originally, such a criterion was
derived by Mott.\cite{Mott.56} Later, Edwards and Sienko extended its
applicability by incorporating more realistic wave functions of the
localized states, see Ref.\ \onlinecite{Edw.Sie.78}, especially Table
I as well as Figure 1 therein. In that work, the simple estimate of
$a_{\rm H}^{\ *}$ originally used, which incorporates only the effective
electron mass and the permittivity, was refined.

Figure 5 of Ref.\ \onlinecite{Sie.etal.11} depicts the claimed
discrepancy between Eq.\  (\ref{mott_crit}) and the experimental
finding. Siegrist et al.\ comment on this deviation in a firm manner:
``This estimate fails completely for GeSb$_2$Te$_4$, where the 
observed critical carrier concentration at the MIT determined by Hall
measurements is $2.0 \times 10^{20}\ {\rm cm}^{-3}$. This carrier 
concentration reproduces the observed minimum metallic conductivity. 
This consistently indicates that the critical charge carrier 
concentration is more than a factor of 25,000 larger than the Mott 
prediction! We are not aware of any other solid where a similar
deviation has been observed.''\cite{Sie.etal.11}

This statement, however, is based on a highly questionable assumption:
Siegrist et al.\ presume that the transport proceeds via shallow
impurity states. Only the wave functions of such states may have an
extension of more than 100 \AA. In contrast, the wave functions of
defect states deep in the gap are far more localized. Unlike Siegrist
et al., we think it is very likely that these deep states govern the
electrical conduction process in GeSb$_2$Te$_4$. Our interpretation is
supported by the following series of independent arguments.

According to Yu and Cardona, see chapter 4.2.2 in Ref.\ 
\onlinecite{Yu.Ca}, impurity levels tend to be shallow if the cores
(atom minus its outer valence electrons) of host and impurity atoms
resemble each other. If, however, the defect induces a substantial
strongly localized potential, the arising central cell corrections
usually cause the corresponding impurity level to be a deep one.
Vacancies belong to the second of these two classes of crystal
imperfections. How they give rise to deep levels is discussed in
chapter 4.3 of Ref.\ \onlinecite{Yu.Ca}, see also Figure 4.5 therein.

In this context, it is noteworthy that Siegrist et al.\ conclude from
their measurements that the concentration of vacancies in
GeSb$_2$Te$_4$ is high, see Figure 6a in Ref.\
\onlinecite{Sie.etal.11} and the last but one paragraph on page 4 of
the Supplementary Information of Ref.\ \onlinecite{Sie.etal.11}. In
the latter, the authors state ``... the empty lattice sites play a
crucial role in reducing the electrical conductivity in these
phase-change materials.'' Hence, the concentration of deep levels in
the gap should be high and substantially affect the electronic
transport.
   
The influence of vacancies on the electronic structure of various
configurations of GeSb$_2$Te$_4$ was recently investigated by Zhang
et al.\ by means of density functional theory
calculations.\cite{Zhang.etal.12} These authors showed that a high
concentration of vacancies implies a high density of states at the
Fermi energy. Studying the inverse participation ratio of the cubic
GeSb$_2$Te$_4$ phase with random occupation, they observed that these
states are localized to regions of only 25--60 atoms. Their volume is
smaller by orders of magnitude than the spatial extension of a shallow
impurity state.

The annealing process in Ge$_2$Sb$_2$Te$_5$ films, a related material,
was studied by Kato and Tanaka.\cite{Kato.Tana} They concluded that,
in the face-centered cubic phase, the chemical potential is situated
0.15~eV above the valence band edge.\cite{Kato.Tana} This value is
considerably larger than the excitation energy of a shallow doping
level, estimated as 8.6~meV in Ref.\ \onlinecite{Sie.etal.11}. Thus,
this position of the chemical potential supports the hypothesis that
deep levels with strongly localized wave functions play an essential
role.

Furthermore, for the following two reasons, it is unlikely that
conventional shallow impurity states played an important role in the
experiment by Siegrist et al.: (i) Textbook derivations of the
properties of a shallow impurity state consider a single impurity in
an otherwise perfect infinite crystal. It is therefore questionable to
make use of these results in a situation with a very large number of
crystal imperfections within the range of the hypothetical shallow
donor or acceptor wave function. (ii) Due to the nanocrystalline
structure, strain as well as boundary effects may considerably
influence the electronic properties of the crystallites. Thus,
additional doubts on the applicability of these textbook derivations
arise from the grain size of the nanocrystalline GeSb$_2$Te$_4$ films
being only of the order of 10 to 20~nm, see p.\ 4 of the Supplementary
Information of Ref.\ \onlinecite{Sie.etal.11}. In the case of an 
effective Bohr radius of 100~\AA, the probability density would still
have substantial finite values at a grain surface for almost all
shallow impurity states.

Our counter-arguments to the interpretation in Ref.\
\onlinecite{Sie.etal.11} which have been presented in the previous
paragraphs raise the following question: How does a ``standard''
material behave in an analogous annealing experiment? Such a study was
performed by Song and co-workers for amorphous Si very heavily doped
with P or B.\cite{Song.etal.10} (For crystalline Si, these elements
are typical shallow donors and acceptors, respectively.) These authors
deposited a-Si:H films with P and B concentrations of 0.41 and
0.42~atomic~percent, respectively. Thus, in both cases, the spatial
concentration of the dopants amounts to about
$2 \times 10^{20}\ {\rm cm}^{-3}$. This value is roughly a factor of
50 larger than the critical concentrations for the MIT in crystalline
Si:P and Si:B. Nevertheless, the amorphous films showed clearly
activated conduction at room-temperature. Their conductivities
amounted to $3.4 \times 10^{-4}$ and
$1.28 \times 10^{-3}\ \Omega^{-1} {\rm cm}^{-1}$, respectively.

In the annealing, the films were dehydrogenated and heavily doped 
nanocrystalline Si was formed. In this process, the conductivity 
increased by orders of magnitude up to 5.3 and 
$130\ \Omega^{-1}{\rm cm}^{-1}$ for the P- and B-doped films,
respectively. Simultaneously, the ``conductivity activation energy'',
$T \mbox{d} \ln \sigma / \mbox{d} \ln T$, decreased by a factor of 10
for P doping and the conductivity almost completely lost its $T$
dependence in the case of B doping. In terms of classifying the
character of conduction according to the sign of
$\mbox{d} \rho / \mbox{d} T$ at the measuring temperature as in Ref.\
\onlinecite{Sie.etal.11}, this means the MIT had still not been
reached in the case of P doping, although the P concentration was so
much higher than the critical concentration for crystalline Si:P.
However, for B doping, the MIT had probably already almost been
reached or even crossed. The findings described here resemble to a
large extent the observations in GeSb$_2$Te$_4$ by Siegrist et al..
This similarity is a further clear argument against their claim that
this phase-change material exhibits an ``unparalleled quantum state of
matter''.

Song et al.\ interpret their results in terms of a shift of the Fermi
energy due to annealing.\cite{Song.etal.10} Simultaneously with the
change of the activation energy, the character of the participating
states should alter, from deep, strongly localized states toward
shallow, far more extended states. In this process, according to Eq.\
(\ref{mott_crit}), the critical charge carrier concentration,
$n_{\rm c}$, decreases. Hence, such an MIT happens primarily due to a
variation of $n_{\rm c}$, similar to the MIT caused by applying stress
to heavily doped crystalline Si,\cite{Tho.etal.83,Waf.etal.99} see
also Subsection 4.2.  The same mechanism should govern the MIT in
GeSb$_2$Te$_4$, where the transition occurs sometime in the
recrystallization process, not at its end.

Summarizing the above arguments, it seems to be very likely that 
strongly localized states deep in the gap play the crucial role in the
transport close to the annealing-induced MIT in GeSb$_2$Te$_4$. These
states have a far smaller spatial extension than the shallow states.
Hence, it is not meaningful to relate the critical charge carrier
concentration to the effective Bohr radius of shallow states. In this
way, the discrepancy between critical charge carrier density and
effective Bohr radius stressed by Siegrist et al.\ in Ref.\
\onlinecite{Sie.etal.11} is traced back to the authors using an
inappropriate value for the latter quantity. Therefore, also
concerning the Mott criterion, GeSb$_2$Te$_4$ is not special.

\section{Comparison with other solids}

One might shrug off the ambivalence of the GeSb$_2$Te$_4$ measurements
discussed above considering them to be ambiguous results for a
special, complex material. However, as we will see, contradictory
information is also contained in various publications on the MIT in
disordered solids which conclude that, at the transition,
$\lim_{T \rightarrow 0} \sigma(T,x)$ is a continuous function of the
control parameter $x$. To illustrate these contradictions, we now
examine several older investigations of crystalline elemental and
compound semiconductors, heavily doped with different impurities, as
well as recent studies of disordered Gd and nanogranular Pt-C. In this
process, to uncover the bias inherent to the usual data evaluations,
we analyze in particular the respective $w(T,x = {\rm const.})$ flow
diagrams; moreover, we demonstrate how, in the augmented power law
approach, exponent and temperature range influence the sample
classification.

\subsection{Crystalline Si:As}

First, we turn to crystalline Si:As, an n-type semiconductor, which
was studied in great detail by Shafarman et al..\cite{Shaf.etal.89}
For a series of uncompensated samples with charge carrier
concentration, $n$, between $6.85 \times 10^{18}$ and
$32.8 \times 10^{18}\ {\rm cm}^{-3}$, these authors investigated
$\sigma(T,n)$ down to 0.5~K. They claimed that, at sufficiently low
$T$, the conductivity of their allegedly metallic samples can be well
described by $\sigma(T,n) = a(n) + b(n) \, T^{1/2}$. Supposing
continuity of the MIT and
$\lim_{T \rightarrow 0} \sigma(T,n) \propto (n - n_{\rm c})^\mu$ on
its metallic side, where $n_{\rm c}$ denotes the critical As
concentration and $\mu$ the corresponding critical exponent, Shafarman
et al.\ obtained $\mu = 0.60 \pm 0.05$; this value is consistent with
previous results for Si:P.\cite{Shaf.etal.89,Tho.etal.83} Close to the
MIT, on the insulating side, below 8~K, the authors observed
$\sigma(T)$ to be well approximated by Eq.\ (\ref{mer}) with
$\nu = 1/4$, which seems to indicate Mott type variable-range hopping.

From the perspective of Section 2, the following points of Ref.\
\onlinecite{Shaf.etal.89} are particularly interesting: According to
its Figure 8, at low $T$, while $n$ is varied,
$\mbox{d} \sigma / \mbox{d} T$ changes sign when
$\sigma \approx 50\ {\rm \Omega^{-1} cm^{-1}}$, in agreement with
corresponding Si:P data.\cite{Stu.etal.93,Ros.etal.81} However, in
contrast to previous findings on Si:P,\cite{Ros.etal.80,Tho.etal.83}
none of the possibly metallic Si:As samples has an extrapolated
$\sigma(T = 0)$ which is smaller than half this value, despite the
dense distribution of As concentration values considered: For
$n = 8.67 \times 10^{18}\ {\rm cm}^{-3}$, Figure 8 of Ref.\
\onlinecite{Shaf.etal.89} depicts 
$\sigma(T = 0) \approx 28\ {\rm \Omega^{-1} cm^{-1}}$. For the samples
with $n = 8.63$ and $8.59 \times 10^{18}\ {\rm cm}^{-3}$, whose
character Shafarman et al.\ regarded as not decidable, it is obvious
from Figure 11a of Ref.\ \onlinecite{Shaf.etal.89} that
$\sigma = a + b \, T^{1/2}$ extrapolations to $T = 0$ are problematic
due to the considerable curvature of the $\sigma$ versus $T^{1/2}$
plots. Anyway, focusing on the data below 1~K for these two samples,
we obtained the estimates $\sigma(T = 0) \approx 30$ and
$28\ {\rm \Omega^{-1} cm^{-1}}$, respectively. Finally, the sample
with the next lower value of the charge carrier concentration,
$n = 8.48 \times 10^{18}\ {\rm cm}^{-3}$, was classified as insulating
in Ref.\ \onlinecite{Shaf.etal.89}.

Note, moreover, that for the two samples with $n = 8.63$ and
$8.59 \times 10^{18}\ {\rm cm}^{-3}$, $\sigma$ varies only by roughly
15~\% between 0.5 and 4~K. As already remarked in Ref.\
\onlinecite{Shaf.etal.89}, such a $\sigma$ range is much too small to
draw reliable conclusions from the alternative stretched exponential
fits presented in Figure 11b therein.

\begin{figure}
\includegraphics[width=0.85\linewidth]{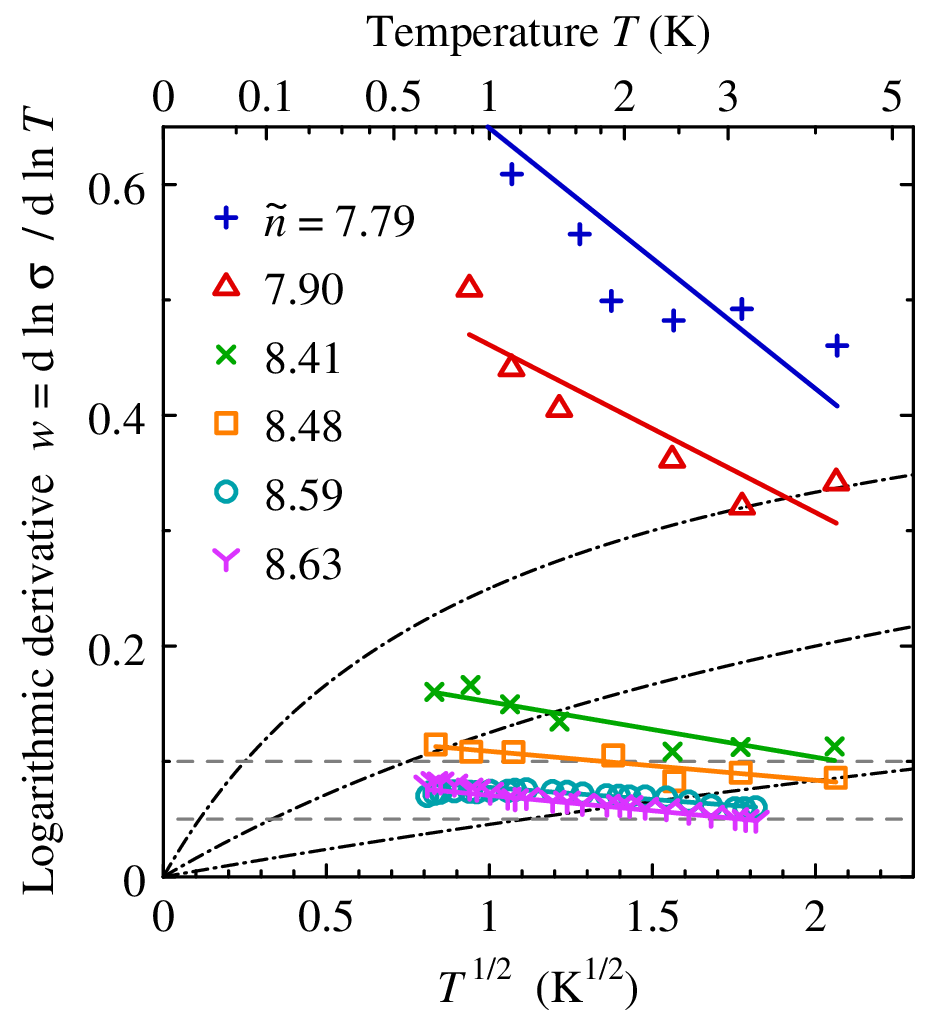}
\caption{(Color online) Temperature dependences of the logarithmic
derivative of the conductivity, $w$, for six crystalline Si:As samples
investigated in Ref.\ \onlinecite{Shaf.etal.89}. Here, $\tilde{n}$
indicates the donor concentration as multiple of
$10^{18}\ {\rm cm}^{-3}$. The points for $\tilde{n} = 7.79$ to 8.48
are redrawn from Figure 4 of Ref.\ \onlinecite{Shaf.etal.89}; we use
the same symbols as in the original plot but another $T$ scale. In
addition, points for $\tilde{n} = 8.59$ and 8.63 are included which
we obtained from the data published in Figure 11b of Ref.\
\onlinecite{Shaf.etal.89}; see text for details. The straight lines
only serve as guides to the eye. The dashed-dotted lines represent
$w(T)$ resulting from hypothetical $\sigma = a + b \, T^{1/2}$ with
$a / b = 1$, 3, and 10 (from top to bottom). To facilitate judging the
slope for the samples with the smallest values of $w$, dashed gray
lines mark constant $w = 0.05$ and 0.1.
}
\label{castner}
\end{figure}

In this context, it is very instructive to explore the logarithmic
derivative $w(T,n)$ for the Si:As data from Ref.\
\onlinecite{Shaf.etal.89}. Our Figure \ref{castner} presents
$w(T,n = {\rm const.})$ for six samples with $n = 7.79$ to
$8.63 \times 10^{18}\ {\rm cm}^{-3}$. This interval includes the four
samples which Ref.\ \onlinecite{Shaf.etal.89} considered as the
clearly insulating ones closest to the MIT as well as the two samples
which remained unclassified in that study, see above. Partly, the data
were redrawn from Figure 4 of Ref.\ \onlinecite{Shaf.etal.89}, partly,
they were obtained by digitizing Figure 11b of Ref.\
\onlinecite{Shaf.etal.89} and subsequent numerical differentiation of
$\ln \sigma (\ln T)$ based on Appendix C. In these calculations,
always, eight neighboring data points were taken into account.

For all samples considered in Figure \ref{castner}, the inequalities
$w > 0$ and $\mbox{d} w / \mbox{d} T < 0$, on sliding average, hold
simultaneously, in clear contradiction to the examples of $w(T)$
curves for hypothetical $\sigma(T)$ following pure augmented power
laws. -- In fitting the theoretical $w(T)$ to experimental data, there
would be only a single adjustable parameter, the ratio $a / b$.
Therefore, such a comparison is very meaningful. -- Thus, according to
Section 2.6, all these samples, including the two samples which could
not be classified in Ref.\ \onlinecite{Shaf.etal.89}, are very likely
insulating.

Moreover, since $w(T = {\rm const.},n)$ decreases with increasing $n$
in the proximity of the MIT, our Figure \ref{castner} allows for a
conclusion on the separatrix between metallic and non-metallic
regions: It is very unlikely that $w(T) \equiv 1/2$ or 1/3 holds at
the MIT itself. Hence, for Si:As, it is also very unlikely that
$\sigma(T)$ follows a pure power law with exponent 1/2 or 1/3 just at
the MIT.

We emphasize that $\mbox{d} w / \mbox{d} T < 0$ for all the samples
considered in Figure \ref{castner} despite of the very weak $T$
dependences of $\sigma$ which a part of them exhibits within the $T$
range studied; for four of them, $w < 0.2$, for two of them, even
$w < 0.1$. Thus, as we will see in Section 5, this diagram is in
accord with the hypothesis that $\lim_{T \rightarrow 0} \sigma(T,n)$
is discontinuous at the MIT, whereas, for any $T > 0$,
$\sigma(T = {\rm const.},n)$ is continuous. This hypothesis conflicts
with the data analysis by Shafarman et al.\cite{Shaf.etal.89}, but it
is supported by the above stated lack of possibly metallic Si:As
samples with finite $\sigma(T = 0)$ extrapolations below half the
$\sigma$ value at which, at low $T$, $\mbox{d} \sigma / \mbox{d} T$
changes sign when $n$ is varied.

\subsection{Crystalline Si:P}

We now consider a similar material, crystalline Si:P, also an n-type
semiconductor. It was intensively investigated by groups from Bell
Laboratories and from Karls\-ruhe University down to temperatures of
the order of 10~mK, far lower than in the Si:As study considered
above. In these Si:P investigations, the MIT was tuned by changing the
P concentration, $n$,\cite{Ros.etal.80,Ros.etal.83,Stu.etal.93} and,
alternatively, by varying its critical P concentration, $n_{\rm c}$,
by applying stress, $S$; thus
$n_{\rm c} = n_{\rm c}(S)$.\cite{Tho.etal.83,Waf.etal.99} In the
literature, this series of Si:P studies has been regarded as key
experiments confirming the continuity of
$\lim_{T \rightarrow 0} \sigma(T,n - n_{\rm c})$ at
$n = n_{\rm c}$.\cite{Lee.Rama,Beli.Kirk,Ever.Mirl}

A discontinuity of $\lim_{T \rightarrow 0} \sigma(T,n - n_{\rm c})$
was still not definitely ruled out in the first of these works, Ref.\
\onlinecite{Ros.etal.80}. In the subsequent publications, however, the
continuity of $\lim_{T \rightarrow 0} \sigma(T,n - n_{\rm c})$
\linebreak was implicitly presumed when the measurements were analyzed
only by means of the augmented power law ansatz, Eq.\ (\ref{apl}). In
doing so, these analyses paid little attention to the considerable
curvature of the $\sigma$ versus $T^{1/2}$ plots being obvious in the
respective Figures 1 of Refs.\ \onlinecite{Tho.etal.83},
\onlinecite{Waf.etal.99}, and \onlinecite{Stu.etal.93}. -- In this
regard, these diagrams resemble a corresponding plot for crystalline
Si:As, Figure 11a of Ref.\ \onlinecite{Shaf.etal.89}, compare previous
subsection. -- Due to the curvature, the $\sigma(T = 0)$
extrapolations significantly depend on which $T$ interval is
considered in the $\sigma = a + b \, T^{1/2}$ fits.

\begin{figure}
\includegraphics[width=0.85\linewidth]{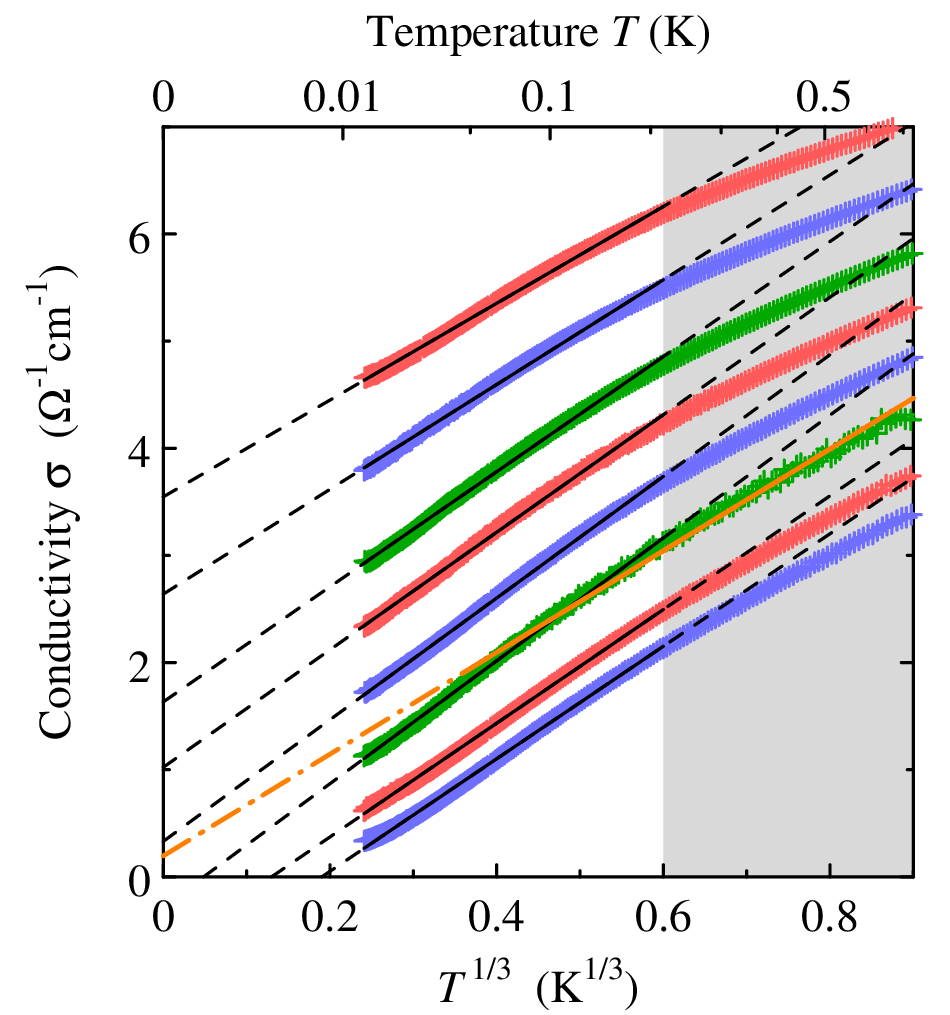}
\caption{(Color online) Temperature dependences of the conductivity of
a Si:P sample with P concentration
$3.21 \times 10^{18}\ {\rm cm}^{-3}$ under uniaxial stress, $S$, in
$\sigma$ versus $T^{1/3}$ representation. The data are redrawn from
Figure 1 of Ref.\ \onlinecite{Waf.etal.99}. Compared to Figure 2a of
that work, a wider $T$ range is displayed; the part additionally taken
into account, 0.216 -- 0.729~K, is marked by shading. The same
interval of $S$ values as in Figure 2a of Ref.\
\onlinecite{Waf.etal.99} is considered here; from top to bottom,
$S = 2.00$, 1.94, 1.87, 1.82, 1.77, 1.72, 1.66, and 1.61~kbar;
different colors are used to facilitate the inspection of the diagram.
(The data of $\sigma(T,1.66\ {\rm kbar})$ are left out in Figure 2a
of Ref.\ \onlinecite{Waf.etal.99}.) The black straight lines show
respective $\sigma = a + b \, T^{1/3}$ approximations: They are given
as full lines within the $T$ interval considered in the fits,
0.014 -- 0.216~K as in Figure 2a of Ref.\ \onlinecite{Waf.etal.99},
and as dashed lines in the extrapolation regions outside of it. For
comparison, the $a + b \, T^{1/3}$ approximation of
$\sigma(T,1.72\ {\rm kbar})$ within the alternative $T$ interval
0.047 -- 0.729~K is presented as orange line, full within this
interval and dashed-dotted in the extrapolation region.
}
\label{sip_a}
\end{figure}

\begin{figure}
\includegraphics[width=0.85\linewidth]{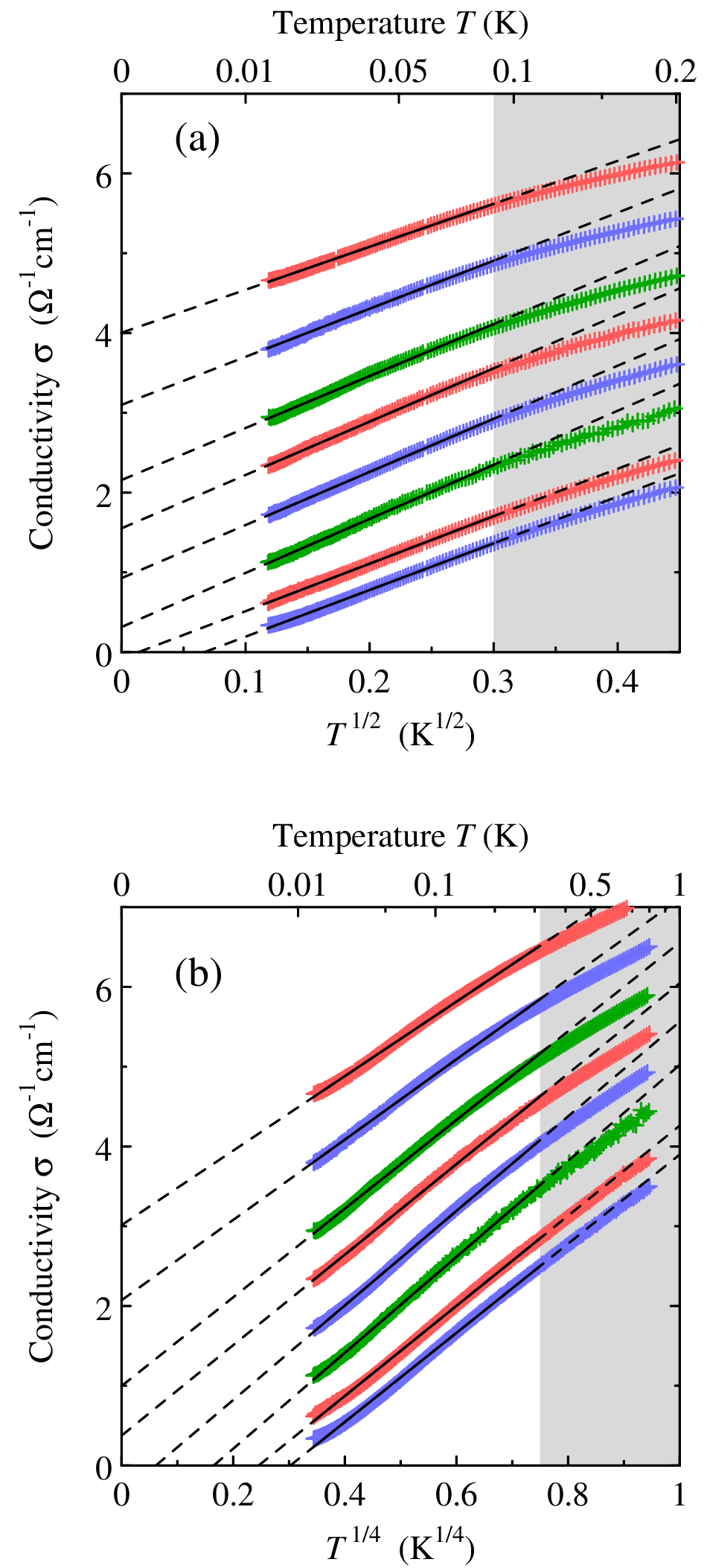}
\caption{(Color online) Temperature dependences of the conductivity of
a Si:P sample under uniaxial stress, redrawn from Figure 1 of Ref.\
\onlinecite{Waf.etal.99}. The same data as in our Figure \ref{sip_a}
are shown, but now in (a) $\sigma$ versus $T^{1/2}$ and (b) $\sigma$
versus  $T^{1/4}$ representations. Here, the straight lines refer to
$\sigma = a + b \, T^p$ approximations within the intervals (a)
0.014 -- 0.090~K and (b) 0.014 -- 0.316~K for $p = 1/2$ and $1/4$,
respectively; the high-temperature regions not taken into account in these fits are
shaded. For further details see caption of Figure \ref{sip_a}.
}
\label{sip_b}
\end{figure}

There is one exception among the Si:P studies mentioned above: In the
report on their stress tuning of the MIT in Ref.\
\onlinecite{Waf.etal.99}, Waffenschmidt et al.\ state that, close to
the MIT, at low $T$, $\sigma = a + b \, T^{1/3}$ would better describe
the experimental data than $\sigma = a + b \, T^{1/2}$. At first
glance, this seems to be confirmed by their Figure 2a which shows
almost linear $\sigma$ versus $T^{1/3}$ plots for
$0.014\ {\rm K} \le T \le 0.216\ {\rm K}$; in this representation, the
interpolation interval is by a factor of about 1.5 wider than the
extrapolation gap. One should note, however, that the impression of a
substantial curvature which one gains from the $\sigma$ versus
$T^{1/2}$ plots in Figure 1 of Ref.\ \onlinecite{Waf.etal.99} arises
under quite other conditions: Therein, the width of the $T^{1/2}$ data
range exceeds that of the corresponding low-temperature gap by a
factor of about 6.5. Thus, on this basis, a fair comparison of both
approximations is impossible.

Therefore, to examine whether the ansatz $\sigma = a + b \, T^{1/3}$
truly better describes the Si:P data published in Ref.\
\onlinecite{Waf.etal.99} than the ansatz $\sigma = a + b \, T^{1/2}$,
we now reproduce and extend the augmented power law analysis of these
measurements. In doing so, we also aim to find out to what extent the
sample classification depends, first, on the $T$ range taken into
account in the fits and, second, on the exponent of the augmented
power law used.

Our Figure \ref{sip_a} replots $\sigma(T,S)$ data from Figure 1 of
Ref.\ \onlinecite{Waf.etal.99} in a $\sigma$ versus $T^{1/3}$ diagram
considering the same interval of stress, $S$, as Figure 2a of Ref.\
\onlinecite{Waf.etal.99}, but a wider $T$ range. For clarity, we
shaded the $T$ region additionally included into consideration.

Figure \ref{sip_a} shows clearly that substantial deviations from the
$\sigma = a + b \, T^{1/3}$ approximations given in Figure 2a of Ref.\
\onlinecite{Waf.etal.99} set in immediately above the $T$ interval
taken into account in that diagram,
$0.014\ {\rm K} \le T \le 0.216\ {\rm K}$.  A more detailed inspection
leads to the conclusion that precursors of these deviations are
already detectable below 0.216~K for $S = 2.00$, 1.94, 1.87, 1.82,
1.77, and 1.72~kbar, that means, in particular for the stress values
to which the presented augmented power law fits ascribe metallic
conduction. Furthermore, our diagram demonstrates for the case
$S = 1.72\ {\rm kbar}$ that, due to the curvature of
$\sigma(T^{1/3})$, the resulting sample classification may depend on
the $T$ interval considered in the augmented power law fit. Finally,
we draw the reader's attention to the weak s-shaped deviations of the
measured data from the augmented power law approximations in our
Figure \ref{sip_a}; we will come back to this point later.

As alternative plots, our Figures \ref{sip_b}a and \ref{sip_b}b
present the same data sets in analogous $\sigma$ versus $T^{1/2}$ and
$\sigma$ versus $T^{1/4}$ diagrams, respectively; $p = 1/4$ is a
purely empirical choice. Comparing them with our Figure \ref{sip_a}
and with each other is very instructive as is detailed in the
following three paragraphs.

First, this comparison makes clear that the sample classification
depends to a substantial extent on the exponent value presumed; two,
three, and four of the considered $S$ values are classified as
belonging to the insulating range for $p = 1/2$, 1/3, and 1/4,
respectively.

Second, it illustrates the problem that any attempt to find out which
value of the exponent, $p$, of the augmented power laws yields the
most reliable estimate of $\lim_{T \rightarrow 0} \sigma(T,S)$ is
hindered by the uncertainty about the choice of the rating method.
There are (at least) three easily accessible figures of merit for such
a comparison: mean square deviation, width of the $T$ region
considered, and ratio of the width of the interpolation interval to
that of the extrapolation gap in the $\sigma$ versus $T^p$ plot. -- A
mathematically strict approach would have to be based on assumptions
about $\sigma(T, S = {\rm const.})$ which cannot be verified at this
stage. -- For example, on the one hand, the mean square deviation of
the approximations shown in Figure \ref{sip_b}a is significantly
smaller than that of the fits presented in Figure \ref{sip_a}. On the
other hand, in Figure \ref{sip_a}, the $T$ range taken into account in
the fits is considerably wider than that in Figure \ref{sip_b}a. The
above mentioned ratio of interpolation interval to extrapolation gap,
however, is the same in both cases.

Third, the comparison of Figures \ref{sip_a}, \ref{sip_b}a, and
\ref{sip_b}b shows  that the applicability range of the augmented
power law approximation is the wider the smaller the value of $p$. We
got support for this finding when we adjusted the parameter $p$ for
the individual $\sigma(T,S = {\rm const.})$ by numerically minimizing
the mean square deviation: For all data sets displayed in Figure
\ref{sip_a}, the optimum value of $p$ considerably decreases with
increasing upper bound of the considered $T$ interval. Moreover, it is
noteworthy that this optimum $p$ value varies substantially with $S$,
too.

Summarizing the above three points, we have demonstrated how biased
interpretations based solely on a single $\sigma$ versus $T^p$ diagram
are. Furthermore, concerning the examined data sets, we have seen that
it is impossible to draw any definite conclusion on the optimum
exponent, in contradiction to the interpretation in Ref.\
\onlinecite{Waf.etal.99}. In consequence, it is also not possible to
locate the transition point from metallic to insulating conduction in
an unambiguous manner by means of the augmented power law approach.

\begin{figure}
\includegraphics[width=0.85\linewidth]{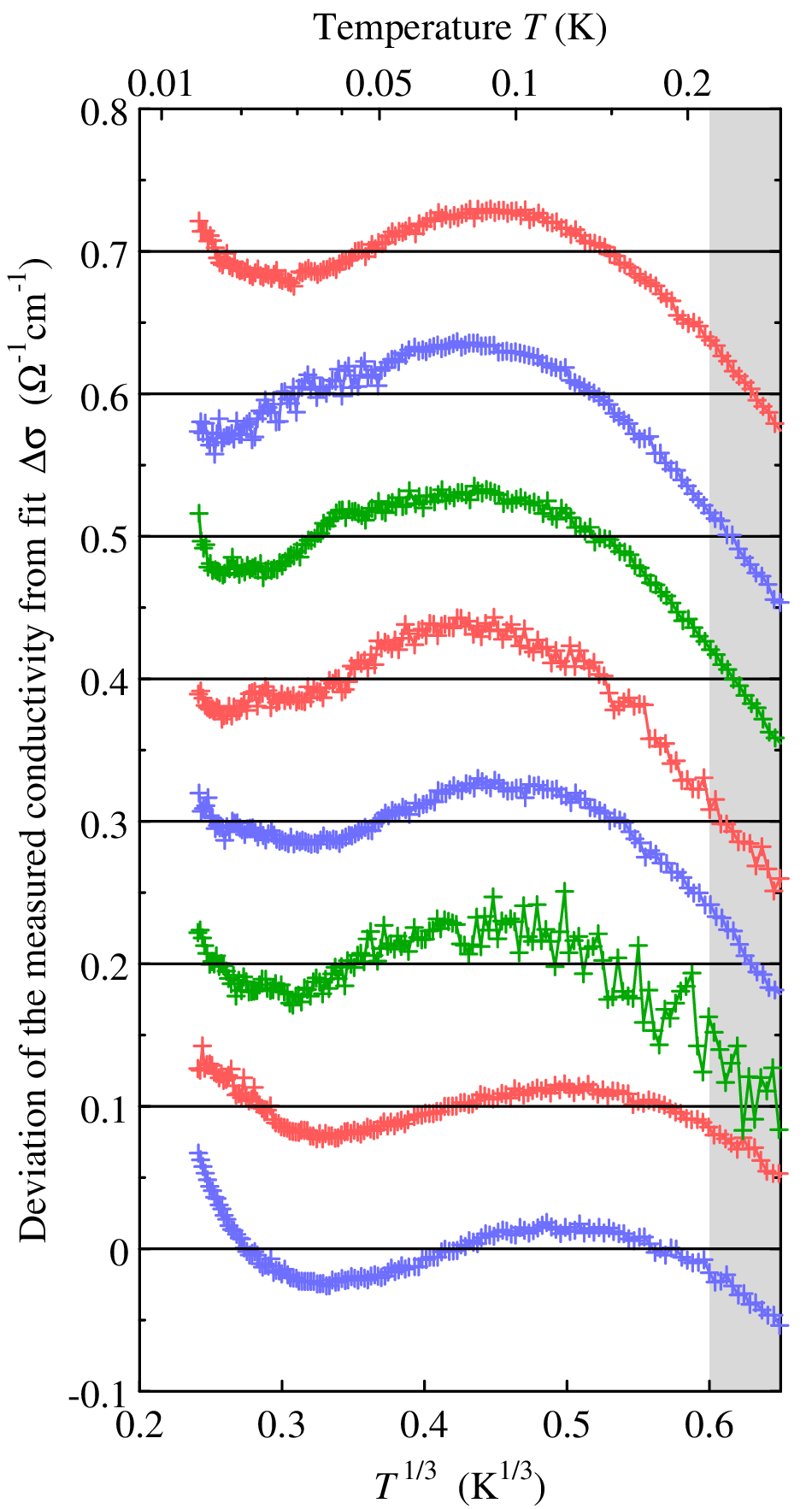}
\caption{(Color online) Temperature dependences of deviations of the
measured $\sigma(T,S = {\rm const.})$ for Si:P, published in Ref.\
\onlinecite{Waf.etal.99}, from the corresponding
$\sigma = a + b\, T^{1/3}$ approximations shown in our Figure
\ref{sip_a}. In order to ensure simultaneously high resolution and
compact presentation, 
$\Delta \sigma(T,S) = \sigma(T,S) - a(S) - b(S)\, T^{1/3} + c(S)$
with appropriately chosen values of $c(S)$ is presented here. From top
to bottom, $S = 2.00$, 1.94, 1.87, 1.82, 1.77, 1.72, 1.66, and
1.61~kbar; the horizontal black lines mark the corresponding $c(S)$.
For further details see caption of Figure \ref{sip_a}.
}
\label{sip_c}
\end{figure}

Nevertheless, the question arises why, as Figure \ref{sip_a} shows,
the $\sigma = a +b\ T^{1/3}$ ansatz can approximate the measured data
over more than one decade of $T$ values reasonably well. To clarify
this point, our Figure \ref{sip_c} displays a magnification of the
weak deviations between measured data and corresponding approximations
to which we already pointed in the above discussion of Figure
\ref{sip_a}.

It is noteworthy that all but one of the curves in Figure \ref{sip_c}
have a pronounced s-shape. This feature indicates that the respective
$\sigma(T,S = {\rm const.})$ exhibit inflection points in the
$\sigma$ versus $T^{1/3}$ plot Figure \ref{sip_a}. Thus, it is not
surprising that $\sigma = a + b\, T^{1/3}$ approximations,
corresponding to straight lines therein, seem to work nicely over a
rather broad $T$ range in the vicinity of these
points.\cite{lin_approx} 

We remark, moreover, that such an s-shape seems not to be consistent
with the used ansatz: If a pure $\sigma = a +b\ T^{1/3}$ behavior were
approached with decreasing $T$, the deviation should have a parabolic
shape at sufficiently low $T$. Thus, also for this reason, the
extrapolations to $T = 0$ given in Figure \ref{sip_a} have to be
called into question.

The problems with the augmented power law analysis described above are
another strong motivation for an as unbiased as possible examination
of the MIT studies on Si:P. Therefore, we now reanalyze these
measurements inspecting the behavior of the logarithmic derivative
$w(T,n = {\rm const.}, S = {\rm const.})$. In doing so, we start with
considering three experiments in which the MIT was tuned by varying
the P concentration, $n$, that is, Refs.\ \onlinecite{Ros.etal.80},
\onlinecite{Ros.etal.83}, and \onlinecite{Stu.etal.93}.

\begin{figure}
\includegraphics[width=0.85\linewidth]{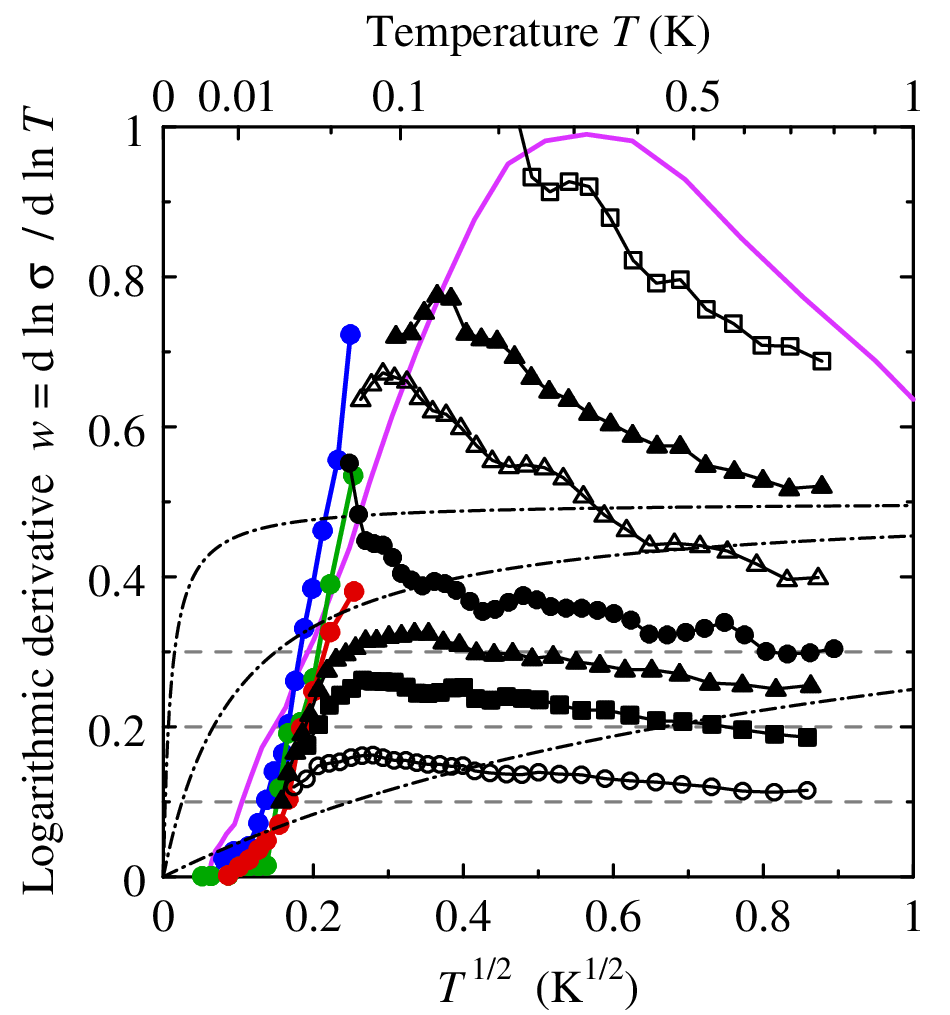}
\caption{(Color online) Comparison of temperature dependences of the
logarithmic derivative of the conductivity, $w$, obtained from three
studies on crystalline Si:P in which the MIT was tuned by varying the
P concentration, $n$: The colored filled circles result from the three
upper curves in Figure 2 of Ref.\ \onlinecite{Ros.etal.80}; we always
considered four neighboring $\ln \sigma (\ln T)$ points in calculating
$w$. The magenta line was derived from the curve for
$n = 3.75 \times 10^{18}\ {\rm cm}^{-3}$ in Figure 1 of Ref.\
\onlinecite{Ros.etal.83}. The black symbols result from Figure 1 of
Ref.\ \onlinecite{Stu.etal.93} when always five neighboring points are
taken into account; from top to bottom, they refer to
$n = 3.38$, 3.45, 3.50, 3.52, 3.56, 3.60, and
$3.67 \times 10^{18}\ {\rm cm}^{-3}$. According to the interpretation
of Ref.\ \onlinecite{Stu.etal.93}, the upper three of these black
curves should correspond to insulating samples, whereas the lower
three curves should originate from metallic conduction. The sample to
which the medium curve ($n = 3.52 \times 10^{18}\ {\rm cm}^{-3}$)
relates is considered in Ref.\ \onlinecite{Stu.etal.93} to be located
very close to the MIT. The dashed-dotted lines represent $w(T)$ which
result from hypothetical relations $\sigma = a + b \, T^{1/2}$ with
$a / b = 0.01$, 0.1, and 1 (from top to bottom).
Dashed gray lines mark constant $w = 0.1$,
0.2, and 0.3.}
\label{sip_1}
\end{figure}

Our Figure \ref{sip_1} contrasts $w(T,n = {\rm const.})$ which we
obtained from the data published in these works with curves resulting
from hypothetical $\sigma(T)$ obeying augmented power laws, for
details see its caption. This diagram shows a more complex behavior
than our corresponding Figure \ref{castner} on Si:As. The following
features of the experimental data are particularly remarkable.

First, there are two $T$ regions with qualitatively different behavior
of $w(T,n = {\rm const.})$. At high $T$, the $w(T,n = {\rm const.})$
have negative slope similar to the Si:As data in Figure \ref{castner},
whereas, at low $T$, the $w(T,n = {\rm const.})$ decrease with $T$ and
seem to tend to 0. The transition temperature between both regimes,
however, is experiment-specific: It amounts to 0.3~K for the data from
Ref.\ \onlinecite{Ros.etal.83} and to roughly 0.1~K for the
measurements in Ref.\ \onlinecite{Stu.etal.93}.

Second, the comparison of the experimental data to hypothetical $w(T)$
curves which were obtained from pure augmented power law behavior of
$\sigma(T)$, Eq.\ (\ref{apl}) with $p = 1/2$, reveals qualitative
discrepancies in both regions: At high $T$, experimental relations and
hypothetical curves have slopes of opposite sign; at low $T$, the
experimental relations vanish far more rapidly with decreasing $T$
than the hypothetical curves. The latter feature corresponds to the
observation in Ref.\ \onlinecite{Ros.etal.83} that the measured
$\sigma(T)$ obeys Eq.\ (\ref{apl}) with $p = 2$, which, however, has
not been followed up in subsequent publications.

Third, focusing on the data obtained from Ref.\
\onlinecite{Stu.etal.93}, marked by black symbols in Figure
\ref{sip_1}, we highlight the strong similarity between the
$w(T,n = {\rm const.})$ of the two groups of samples which were
classified in that work as metallic and insulating, respectively; see
caption of Figure \ref{sip_1}. In particular, we encourage the reader
to compare the three middle curves with each other: They result from
the data for the samples with $n = 3.50$, 3.52, and
$3.56 \times 10^{18}\ {\rm cm}^{-3}$ which were regarded in Ref.\
\onlinecite{Stu.etal.93} as insulating, as very close to the MIT, and
as metallic, respectively.

\begin{figure}
\includegraphics[width=0.85\linewidth]{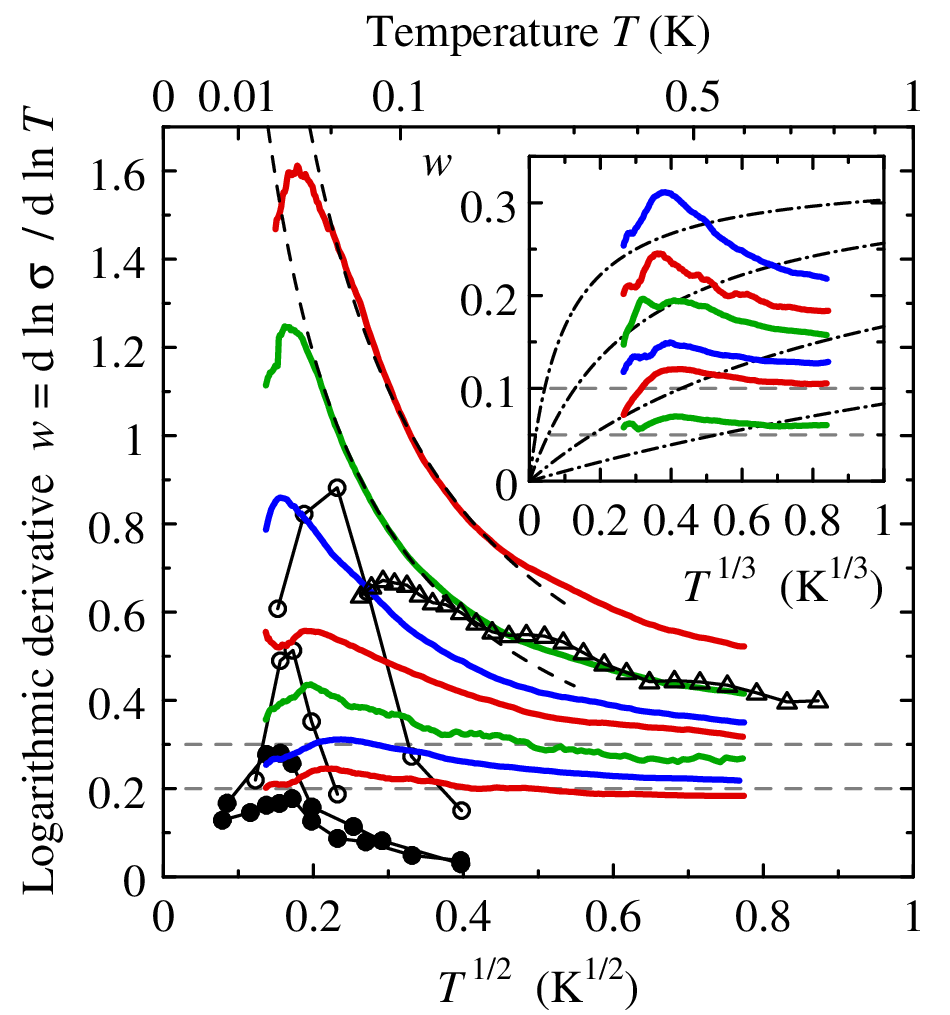}
\caption{(Color online) Comparison of temperature dependences of the
logarithmic derivative of the conductivity, $w$, obtained from two
experiments on crystalline Si:P in which the MIT was tuned by varying
the critical P concentration by means of stress: The points marked by
black circles result from data in Figure 1 of Ref.\ \onlinecite{Tho.etal.83};
three neighboring $\ln \sigma (\ln T)$ points were considered in each
numerical differentiation. From top to bottom, they relate to
$S = 5.73$, 6.33, 6.59, and 6.71~kbar. The authors of Ref.\
\onlinecite{Tho.etal.83} concluded that the first two $S$ values
(empty circles) belong to the insulating range,
whereas the latter two
values (filled circles) belong to the metallic range. The colored
curves were obtained from $\sigma(T,S)$ data presented in Figure 1 of
Ref.\ \onlinecite{Waf.etal.99}; in the calculation of these $w(T)$, we
took always 30 neighboring $\sigma(T)$ points into account. From top
to bottom, $S = 1.50$, 1.56, 1.61, 1.66, 1.72, 1.77, and 1.82~kbar. In
Ref.\ \onlinecite{Waf.etal.99}, according to augmented power law fits,
only the $\sigma(T,S = {\rm const.})$ for the last two stress values
were interpreted as indicating metallic conduction. For comparison,
data for one sample from Ref.\ \onlinecite{Stu.etal.93}, measured
without stress, are included; they are marked by triangles as in our
Figure \ref{sip_1}. The two dashed black curves represent Eq.\
(\ref{derivmer}) for Efros-Shklovskii hopping with $\nu = 1/2$; the
only free parameter, $T_0$, was adjusted at $T = 0.1\ {\rm K}$. With
enhanced resolution concerning $w$, the inset shows
$w(T,S = {\rm const.})$ for $S = 1.77$, 1.82, 1.87, 1.94, 2.00, and
2.17~kbar obtained from $\sigma(T,S)$ data published in Figure 1 of Ref.\ \onlinecite{Waf.etal.99}.
Therein, all these stress values were regarded to belong to the
metallic range. In the inset, the data are plotted versus $T^{1/3}$ to
facilitate a better test of the approximation
$\sigma = a + b \, T^{1/3}$ used in Ref.\ \onlinecite{Waf.etal.99}.
The dashed-dotted lines show hypothetical $w(T)$ for this ansatz with
$a / b = 0.1$, 0.3, 1, and 3. In both plots, gray lines mark constant
$w$ to simplify the judgement of the slope of the curves.}
\label{sip_2}
\end{figure}

Next, we turn to the experiments in which the MIT in Si:P was tuned by
means of stress, that is by varying the critical P concentration. Data
from such publications by Thomas et al.\cite{Tho.etal.83} and by
Waffenschmidt et al.\cite{Waf.etal.99} are reanalyzed in our Figure
\ref{sip_2}. The latter measurements seem to be particularly precise;
therefore, the detailed discussion of the augmented power law approach
in the first part of this subsection focused on them. Here, in Figure
\ref{sip_2}, the colored curves relate to data published in Ref.\
\onlinecite{Waf.etal.99}. -- Also in Ref.\ \onlinecite{phd_waff},
$w(T,S = {\rm const.})$ diagrams were derived therefrom. Similarities
with and differences to our Figure \ref{sip_2} will be discussed at
the end of this subsection. --

In comparing the colored curves with $w(T)$ relations obtained from
previous experiments as well as with results from Coulomb glass
theory, we made the following observations: \linebreak (i) On the one
hand, down to roughly 0.1~K, there is a nice agreement with a $w(T)$
curve which results from the data in Ref.\ \onlinecite{Stu.etal.93},
measured without stress; the corresponding $\sigma$, however, differ
by a factor of about 2.7 at 0.8~K, compare Figure 3 of Ref.\
\onlinecite{Waf.etal.99} and the related discussion in that work. On
the other hand, the $w(T)$ obtained from Ref.\
\onlinecite{Waf.etal.99} agree only qualitatively with the $w(T)$
derived from the previous stress tuning of the MIT, Ref.\
\onlinecite{Tho.etal.83}. \linebreak (ii) In a part of the
non-metallic region, for $w \gtrsim 0.7$ and
$0.05\ {\rm K} < T < 0.1\ {\rm K}$, the
colored $w(T)$ can be well approximated by Eq.\ (\ref{derivmer}) with
$\nu = 1/2$ corresponding to variable-range hopping in the Coulomb
glass; this finding is of high significance because only one parameter
had to be adjusted in each such fit. (iii) As for the measurements
without stress discussed above, the $w(T)$ obtained from Ref.\
\onlinecite{Waf.etal.99} exhibit puzzling maxima at low $T$. In this
case, however, they occur only at roughly 50~mK, that is at a
considerably lower temperature than in the other studies.

Concerning the character of the MIT, two features of the data from
Ref.\ \onlinecite{Waf.etal.99} reanalyzed in our Figure \ref{sip_2}
are particularly important: First, all these colored $w(T)$ curves
show the maxima pointed to above, not only the curves which should be
related to metallic behavior according to Ref.\
\onlinecite{Waf.etal.99}, i.e.\ the curves with $\max(w) < 1/3$. Even
for the stress values for which $\max(w) > 1$, what implies with very
high likelihood non-metallic behavior, such maxima are present.
Therefore, and because the maxima occur at roughly the same $T$ for
all stress values and, moreover, at clearly lower $T$ than in the
studies without stress analyzed in Figure \ref{sip_1}, we think one
should consider the measurements below about 50~mK with caution.

Second, above this temperature, $\mbox{d} w / \mbox{d} T < 0$ holds
for all curves presented. Thus, according to Subsection 2.6, we regard
all these stress values to belong to the insulating side of the MIT.
This interpretation is corroborated by the comparison with the
hypothetical $w(T)$ obtained from $\sigma = a + b \, T^{1/3}$ in the
inset of Figure \ref{sip_2}. In Ref.\ \onlinecite{Waf.etal.99},
however, all the measured $\sigma(T)$ from which we obtained the
curves in the inset were classified as metallic.

The observation that, above 50~mK, $\mbox{d} w / \mbox{d} T < 0$ also
when $w$ is clearly smaller than 1/3, even when $w$ is of the order
0.1, accords with the findings for Si:As discussed above. Thus, the
same reasoning as for Si:As suggests the hypothesis that Si:P
exhibits a discontinuous MIT; but this is in contradiction to the
measurements below 50~mK in Ref.\ \onlinecite{Waf.etal.99}.

Finally, for completeness, we refer to the three
$w(T,S = {\rm const.})$ diagrams in the PhD thesis by 
S.~Waffenschmidt, presented in Figures 4.16 and 4.17 of Ref.\
\onlinecite{phd_waff}; unfortunately, neither of them was included in
the corresponding journal publication, Ref.\ \onlinecite{Waf.etal.99}.
These diagrams resemble our Figure \ref{sip_2}, but they have less
explanatory power for two reasons: The random deviations of the $w(T)$
points in Ref.\ \onlinecite{phd_waff} are considerably larger than
those of the points in our Figure \ref{sip_2} because, in Ref.\
\onlinecite{phd_waff}, seven neighboring $\sigma(T)$ points were
considered in the calculation of the $w$ values whereas we always took
30 of these very dense data points into account. Furthermore, in the
above mentioned diagrams of Ref.\ \onlinecite{phd_waff}, linear $T$
scales are used whereas our Figure \ref{sip_2} presents $w$ versus
$T^{1/2}$ and versus $T^{1/3}$. Thus our plots provide better
compromises between the demands for a wide $T$ range and a high
low-$T$ resolution; moreover, for $p = 1/2$ and $1/3$, they have a
higher significance in testing of hypothetical approximate
$w \propto T^p$ for $w \ll p$, as it follows from
$\sigma = a + b \, T^p$.

Nevertheless, also the lower plot in Figure 4.16 of Ref.\
\onlinecite{phd_waff} shows clearly that, for 
$0.1\ {\rm K} < T < 0.6\ {\rm K}$ and $0.1 < w < 0.2$, the slope of
$w(T)$ is negative, in agreement with our analysis above, but in
contradiction to the interpretation in Ref.\ \onlinecite{Waf.etal.99}.

\subsection{Crystalline Si:B}

\begin{figure}
\includegraphics[width=0.85\linewidth]{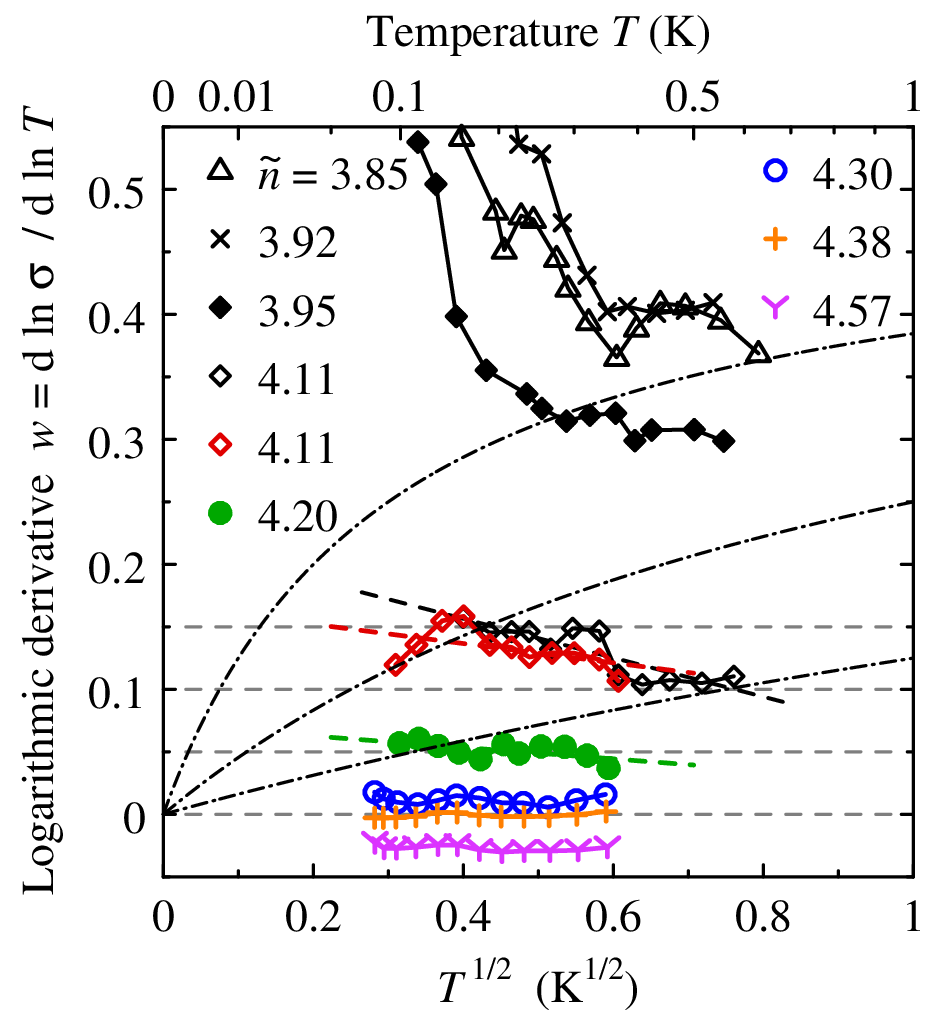}
\caption{(Color online) Temperature dependences of the logarithmic
derivative of the conductivity, $w$, for eight crystalline Si:B samples
investigated in Refs.\ \onlinecite{Sara.Dai.02} and 
\onlinecite{Dai.etal.91}. The acceptor concentration $\tilde{n}$ is
given as multiple of $10^{18}\ {\rm cm}^{-3}$. The data marked black
were obtained by digitizing Figure 1 of Ref.\ \onlinecite{Sara.Dai.02},
while the colored points result from the digitization of Figure 2 of
Ref.\ \onlinecite{Dai.etal.91}. For both data sets, always six
neighboring $\ln \sigma (\ln T)$ points were considered in the
numerical differentiation. The full lines only serve as a guide to the
eye, while the black and colored dashed lines arise from linear
regression of $w(T^{1/2})$. The dashed-dotted lines represent $w(T)$
resulting from the hypothetical relation $\sigma = a + b \, T^{1/2}$
with $a / b = 0.3$, 1, and 3 (from top to bottom). To facilitate
judging the slope for the samples closest to the MIT, dashed gray
lines mark constant $w = 0.$, 0.05, 0.1, and 0.15.}
\label{sarachik}
\end{figure}

The question arises whether the inconsistencies in the augmented power
law approximations of $\sigma(T)$ which have been demonstrated for
Si:As and Si:P in the previous two subsections occur only in n-type
Si. Therefore, we consider now uncompensated crystalline Si:B. More
than two decades ago, this p-type semiconductor was studied by Dai et
al.\ in Ref.\ \onlinecite{Dai.etal.91} in order to obtain the critical
exponent of $\lim_{T \rightarrow 0} \sigma(T,n)$ concerning the B
concentration, $n$, for the hypothetically continuous MIT.

About ten years later, in Ref.\ \onlinecite{Sara.Dai.02}, two of these
authors reported scaling of the $T$ dependences of $\sigma(T,n)$ in
the hopping region, compare Subsection 2.5. Therein, they stressed
that the scaling curve includes data for three samples interpreted as
metallic in their previous work and concluded that additional careful
investigations down to ``as low a temperature as possible'' are
required to solve this puzzle. 

Our Figure \ref{sarachik} presents $w(T,n = {\rm const.})$ relations
obtained from the data published in Refs.\ \onlinecite{Sara.Dai.02}
and \onlinecite{Dai.etal.91}, for details see its caption. This plot
resembles our corresponding graph for Si:As, Figure \ref{castner}, to
a large extent as it is explicated in the following two paragraphs.

Consider first the $w(T)$ for the three samples with $n = 3.85$, 3.92,
and $3.95 \times 10^{18}\ {\rm cm}^{-3}$, classified as insulating in
Ref.\ \onlinecite{Dai.etal.91} as well as in Ref.\
\onlinecite{Sara.Dai.02}. In all three cases, simultaneously, the
smallest value of $w$ falls clearly below 0.5, while, on sliding
average, the respective slope $\mbox{d} w / \mbox{d} T$ is obviously
negative. As discussed in Subsection 2.6, this finding is incompatible
with the hypothesis of a continuous MIT with $\sigma \propto T^{1/2}$
just at the transition.

We now turn to the three samples which were assumed to be metallic in
Ref.\ \onlinecite{Dai.etal.91} but regarded as insulating in Ref.\
\onlinecite{Sara.Dai.02}, that means the samples with $n = 4.11$,
4.20, and $4.30 \times 10^{18}\ {\rm cm}^{-3}$. We emphasize that, on
the average, $\mbox{d} w / \mbox{d} T$ is clearly negative for the
former two of these samples, notwithstanding that $w$ is far smaller
than 0.5, even considerably smaller than 0.2. In consequence, also
from the perspective chosen here, these two samples are very likely
insulating.

Concerning the third of these samples,
$n = 4.30 \times 10^{18}\ {\rm cm}^{-3}$, a definite conclusion is not
possible for us at the current state because the digitized data are
not precise enough to reach a reliable decision on the sign of
$\mbox{d} w / \mbox{d} T$. We remark, however, that also a scaling
analysis as in Ref.\ \onlinecite{Sara.Dai.02} is rather uncertain in
this case: The $T$ dependences of $\sigma$ for $n = 4.20$ and 
$4.30 \times 10^{18}\ {\rm cm}^{-3}$ are so weak that these curves
cannot be made to overlap each other in a mastercurve construction.
Therefore, one has to rely on the implicit assumptions discussed in
detail in Appendix B. In particular, one has to trust in the mean
hopping energy tending to zero as the MIT is approached, that means in
$\mbox{d} \sigma / \mbox{d} T = 0$ marking the MIT at sufficiently low
temperature.

Summarizing this subsection,  we conclude that the inconsistencies in
the augmented power law approximations of $\sigma(T)$ for heavily
doped crystalline Si seem not to be specific to n-type doping. This
interpretation is also supported by observations on partially
compensated Si:(P,B), see Figures 1 in Refs.\ \onlinecite{Moe.89} and
\onlinecite{Hir.etal.89}.

\subsection{Crystalline $^{70}$Ge:Ga}

\begin{figure}
\includegraphics[width=0.85\linewidth]{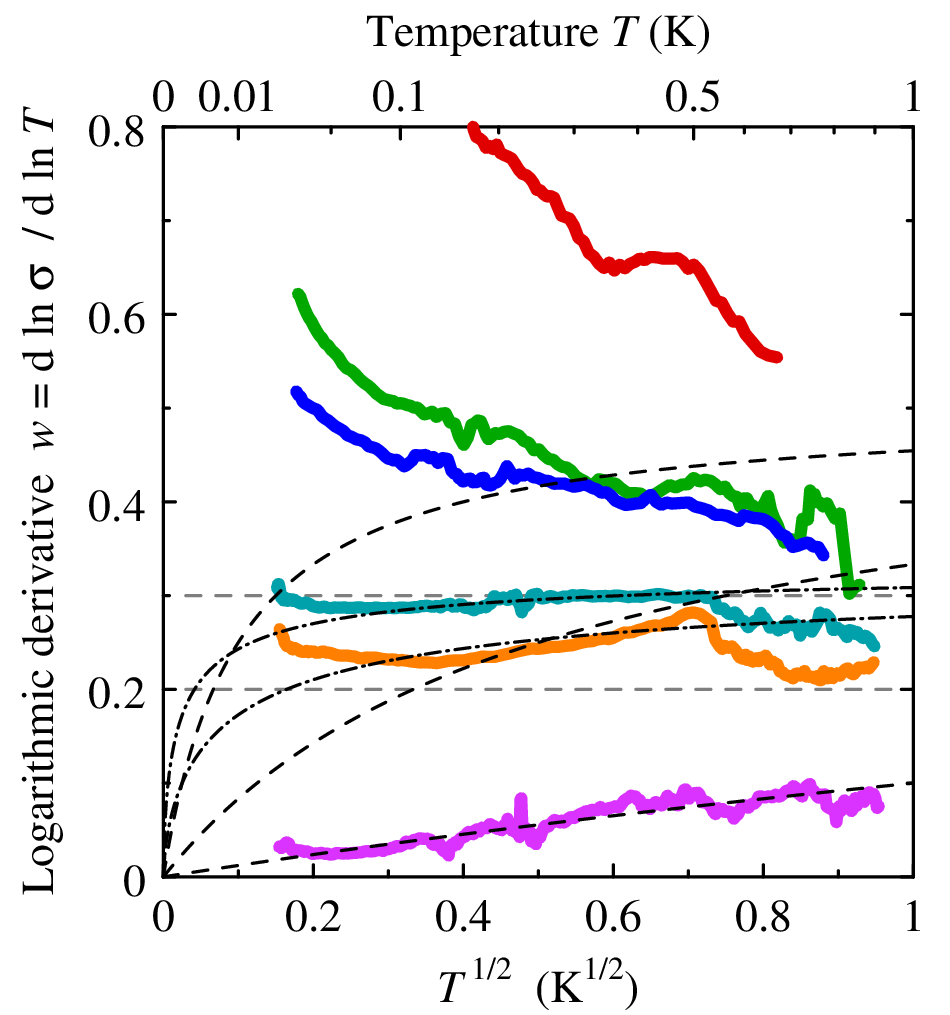}
\caption{(Color online) Temperature dependences of the logarithmic
derivative of the conductivity, $w$, for six crystalline neutron
transmutation doped $^{70}$Ge:Ga samples investigated in Ref.\
\onlinecite{Wata.etal.98}. They are shown as colored curves and relate
to Ga concentrations, $n$, of 1.853, 1.856, 1.858, 1.861, 1.863, and
$1.912 \times 10^{17}\ {\rm cm}^{-3}$, from top to
bottom.\cite{Wata.etal.98} To obtain these relations, we reconstructed
the $\sigma(T, n = {\rm const.})$ from Figure 1 of Ref.\
\onlinecite{Wata.etal.98} and differentiated the individual
$\ln \sigma (\ln T)$ numerically considering groups of 30 neighboring
data points. The dashed and dashed-dotted lines represent $w(T)$
curves resulting from hypothetical $\sigma = a + b \, T^p$ with
$p = 1/2$ and 1/3, respectively. Here, $a / b = 0.1$, 0.5, and 4 for
$p = 1/2$, whereas $a / b = 0.08$ and 0.2 for $p = 1/3$. To facilitate
judging the slope of $w(T)$, dashed gray lines mark constant $w = 0.2$
and 0.3.}
\label{watanabe}
\end{figure}

The next question to answer is: Might the inconsistencies of the
augmented power law approximations of $\sigma(T)$ close to the MIT
which were pointed out in the previous subsections be specific only to
heavily doped Si?

This is not the case as our Figure \ref{watanabe} shows. For six
$^{70}$Ge:Ga samples with different values of the Ga acceptor
concentration, $n$, it presents $w(T, n = {\rm const.})$ curves which
we obtained from data by Watanabe et al.\ published in Ref.\
\onlinecite{Wata.etal.98}; for details see caption of Figure
\ref{watanabe}. These authors used an elaborate sample preparation
technique based on neutron transmutation for doping to achieve a high
degree of homogeneity;\cite{Wata.etal.98} for an earlier MIT study
utilizing this technology, see Ref.\ \onlinecite{Zabr.Zino}. Moreover,
they applied a two-step irradiation process in order to fine-tune the
concentration of Ga acceptors close to the MIT.\cite{Wata.etal.98}

Watanabe et al.\ analyzed their data by means of augmented power law
fits and interpreted them in terms of a continuous
MIT.\cite{Wata.etal.98} This way, they classified the samples to which
the upper three curves in  Figure \ref{watanabe} are related as
insulating, whereas they regarded the samples from which the lower
three curves result as metallic.\cite{Wata.etal.98} Concerning the
latter three samples, the authors concluded that $\sigma(T)$ can be
well approximated by augmented power laws with $p = 1/2$ far from the
MIT and $p = 1/3$ in its immediate vicinity.\cite{Wata.etal.98}

To check this interpretation, our Figure \ref{watanabe} contrasts the
$w(T, n = {\rm const.})$ relations which we obtained by numerical
differentiation of the measured $\sigma(T, n = {\rm const.})$ with
$w(T)$ curves resulting from analytical differentiation of
hypothetical $\sigma = a + b \, T^p$ with $p = 1/2$ and 1/3, evaluated
for several values of $a/b$. This graph strongly resembles the
corresponding plots in the previous subsections. In detail, we
interpret it as follows.

Consider first the sample with the smallest $w$, that is, the sample
with $n = 1.912 \times 10^{17}\ {\rm cm}^{-3}$. In this case, in
agreement with Ref.\ \onlinecite{Wata.etal.98}, analytically
differentiated $\sigma = a + b \, T^{1/2}$ with $a/b = 4$ seems to
well approximate the curve obtained from the measured data by
numerical differentiation. Note, however, that $w < 0.1$ holds for
this sample in the entire $T$ range considered here.

Simultaneously, due to the strong discrepancies between the slopes of
the numerically calculated $w(T)$ curves and the analytically obtained
ones, the hypothesis $\sigma = a + b \, T^{1/2}$ clearly fails for all
other samples. In particular, within the region $0.3 < w < 0.5$,
positive slopes of the $w(T, n = {\rm const.})$ would be expected if
the MIT were continuous and $\sigma$ were proportional to $T^{1/2}$ at
the transition itself. In fact, however, on sliding average,
$\mbox{d} w / \mbox{d} T$ is negative in this $w$ region.  Thus,
again, and in agreement with Ref.\ \onlinecite{Wata.etal.98}, the
approximation $\sigma = a + b \, T^{1/2}$ is not applicable in the
immediate vicinity of the hypothetical MIT.

The question whether or not an augmented power law with $p = 1/3$ may
be valid close to a continuous MIT cannot be finally decided for
$^{70}$Ge:Ga here because the $\sigma(T)$ data are not precise enough.
Nevertheless, at the current stage, there are more cons than pros as
is explained in the following three paragraphs.

In trying to answer this question, we focus on the other two samples
which were regarded as metallic in  Ref.\ \onlinecite{Wata.etal.98},
that is, on the samples with $n = 1.861$ and
$1.863 \times 10^{17}\ {\rm cm}^{-3}$; for them, $0.2 < w < 0.35$
holds within the $T$ interval studied in Figure \ref{watanabe}. In the
medium $T$ range of this graph, at first glance, the $w(T)$ obtained
by numerical differentiation from the measured data and the results of
analytical differentiation of augmented power laws with $p = 1/3$ seem
to be consistent with each other, in accord with Ref.\
\onlinecite{Wata.etal.98}. However, the substantial deviations between
the respective curves which occur at the lower and upper ends of the
$T$ interval considered in Figure \ref{watanabe} clearly conflict with
this interpretation.

Note, moreover, that the $w(T)$ curves for $n = 1.853$, 1.856, and
$1.863 \times 10^{17}\ {\rm cm}^{-3}$ exhibit unusual bumps at about
0.5~K, presumably resulting from some measurement artifacts. These
features raise additional doubts about the plausibility of the seeming
consistency in the medium $T$ range for $n = 1.861$ and
$1.863 \times 10^{17}\ {\rm cm}^{-3}$ to which we pointed in the
previous paragraph.

Finally, we remark that, on average over the whole $T$ range,
$\mbox{d} w / \mbox{d} T$ seems to be slightly negative and
approximately zero for the samples with $n = 1.861$ and
$1.863 \times 10^{17}\ {\rm cm}^{-3}$, respectively. Therefore, at
least the former sample should actually be insulating in contradiction
to the classification in Ref.\ \onlinecite{Wata.etal.98}.

In consequence, since $w(T = {\rm const.}, n)$ decreases with
increasing $n$, the data reconsidered here seem to imply the following
conclusion: If the MIT is continuous and $\sigma(T) \propto T^p$ at
the MIT itself, then $p \lesssim 0.25$; alternatively, the MIT may be
discontinuous.  Hence, because of the above arguments, the
interpretation by Watanabe et al.\ that $\sigma = a + b \, T^{1/3}$ in
the immediate vicinity of the MIT is unlikely to be valid. An
improvement of precision and accuracy of these $\sigma(T)$
measurements should be very promising.

\subsection{Crystalline CdSe:In}

In the previous subsections, we have examined MIT studies of only
elemental semiconductors. Now we ask whether the characteristic
features of $\sigma(T,n)$ and $w(T,n)$ which we have exposed therein
can be identified also in reports on compound semiconductors. First,
we turn to n-type CdSe, a II-VI semiconductor. The (actual or alleged)
hopping conduction in In-doped compensated CdSe was analyzed from
different perspectives in three subsequent publications: In Ref.\
\onlinecite{Zha.etal.90}, Zhang et al.\ probed the Coulomb gap; they
interpreted their measurements in terms of a crossover between
$\sigma(T)$ following Eq.\ (\ref{mer}) with $\nu = 1/2$ at low $T$ and
Eq.\ (\ref{mer}) with $\nu = 1/4$ at high $T$. Later, reinterpreting
and extending these investigations, universality of the crossover and
two-parameter scaling were claimed in Refs.\ \onlinecite{Aha.etal.92}
and \onlinecite{Zha.Sara.95}, respectively. -- Refs.\
\onlinecite{Zha.etal.90} and \onlinecite{Aha.etal.92} consider the
same five samples, but dopant concentration values were redetermined
in Refs.\ \onlinecite{Aha.etal.92}. Out of these publications, only
Ref.\ \onlinecite{Zha.Sara.95} presents $\sigma$ values in absolute
units. --

Remarkably, for the sample with the highest net donor concentration
among the ones investigated in Ref.\ \onlinecite{Zha.Sara.95},
$\sigma(T)$ changes only by a factor of about 2.4 over the lowest
decade of the $T$ range studied, that is, between 0.062 and 0.62~K,
see Figure 1 therein. In terms of simple approximations, this low
value can be interpreted in three different ways: (i) In case, the
sample was located just at a continuous MIT, so that $\sigma(T)$
followed a pure power law, the quotient 2.4 would correspond to an
exponent of roughly 0.4. (ii) If, alternatively, $\sigma(T)$ could be
described by the augmented power law Eq.\ (\ref{apl}) with $p = 1/2$,
then, in consequence of the ratio 2.4, the constant contribution $a$
would be positive and thus indicate metallic conduction. (iii) Of
course, such a low ratio could also result from variable-range hopping
described by Eq.\ (\ref{mer}) with $T_0$ being comparable to the $T$
values considered. For $\nu = 1/2$ and $1/4$, the ratio 2.4 implies
$T_0 = 0.10\ {\rm K}$ and $0.99\ {\rm K}$, respectively; Table I of
Ref.\ \onlinecite{Zha.etal.90} reports $T_0 = 0.10\ {\rm K}$ and
$0.65\ {\rm K}$, respectively, relating to data for higher
temperatures in the latter case.

\begin{figure}
\includegraphics[width=0.85\linewidth]{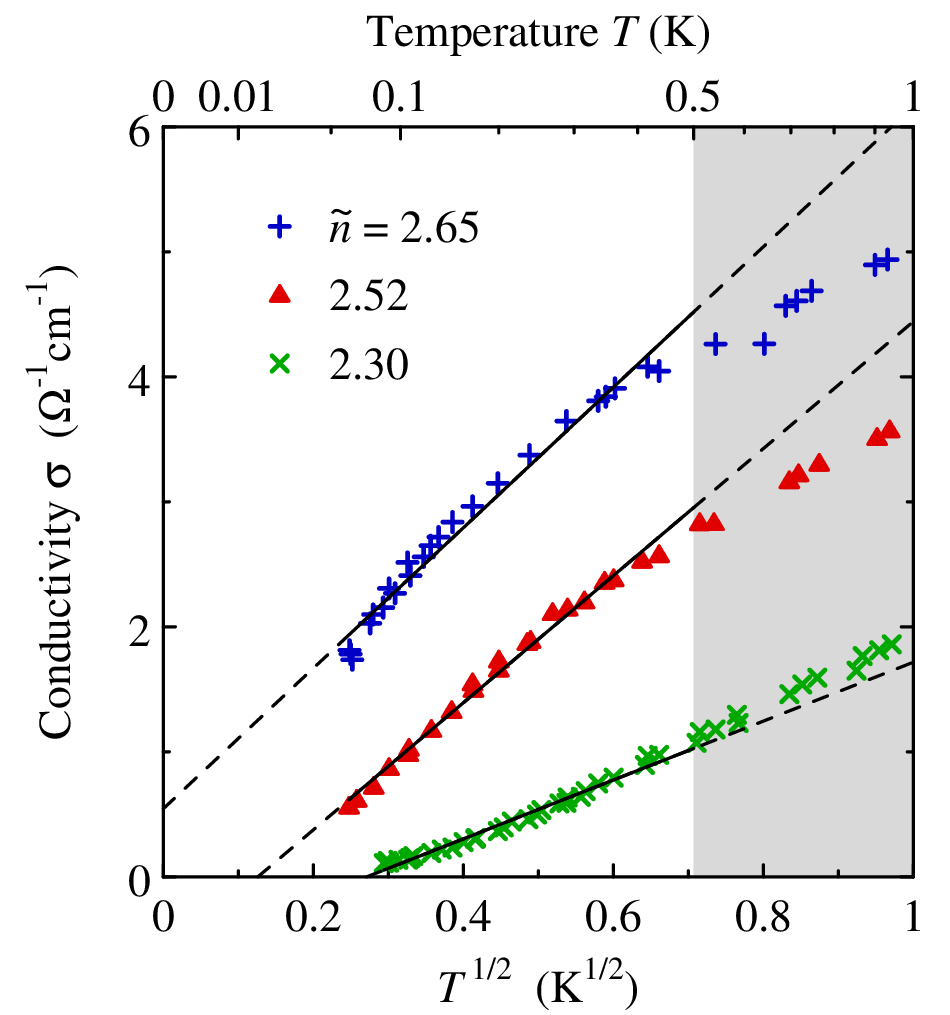}
\caption{(Color online) Temperature dependences of the conductivity of
three CdSe:In samples redrawn from Figure 1 of Ref.\
\onlinecite{Zha.Sara.95}. The values of the net donor concentration, 
$\tilde{n}$, are given as multiple of $10^{17}\ {\rm cm}^{-3}$; they
are taken from Table I of that work. All these samples are regarded as
insulating in Ref.\ \onlinecite{Zha.Sara.95}. The straight lines show
respective $\sigma = a + b \, T^{1/2}$ approximations: They are given
as full lines within the $T$ range considered in the fits,
0.06 -- 0.5~K, and as dashed lines in the extrapolation region outside
of it. The region above 0.5~K is shaded to simplify the discussion.
}
\label{cdse1}
\end{figure}

Because of this ambiguity, it should be very interesting to find out
whether or not $T \rightarrow 0$ extrapolations based on the augmented
power law Eq.\ (\ref{apl}) with $p = 1/2$ can be used for the
classification of the CdSe:In samples, too, and which result they
yield. To clarify these points, we digitized the $\sigma(T)$ data
published in Figure 1 of Ref.\ \onlinecite{Zha.Sara.95} focusing on
the three samples with the highest net donor concentrations. These
data are redrawn in a $\sigma$ versus $T^{1/2}$ plot in Figure
\ref{cdse1} here. 

When covering the shaded part of our graph, one sees that Eq.\
(\ref{apl}) seems to be reasonably well fulfilled between about 0.06
and 0.5~K, that means over almost one temperature decade as also in
our Figure \ref{AA_fig} on GeSb$_2$Te$_4$ films. According to the
corresponding $T \rightarrow 0$ extrapolations, the CdSe:In sample
with the highest net donor concentration,
$n = 2.65 \times 10^{17}\ {\rm cm}^{-3}$, should be metallic in
contradiction to the interpretation in Ref.\ \onlinecite{Zha.Sara.95},
while the two others should be insulating in agreement with that work.

Additionally taking the data points between 0.5 and 1~K into account,
however, totally changes the situation: The previously obtained
augmented power law fits now cease to be good approximations of the
experimental data, so that the $T \rightarrow 0$ extrapolation for
$n = 2.65 \times 10^{17}\ {\rm cm}^{-3}$ and the therefrom concluded
sample classification are called into question.

The here demonstrated interpretational ambiguity of augmented power
law fits to $\sigma(T)$ data sheds bright light on how questionable
this approach is for CdSe:In, too. Thus, our Figure \ref{cdse1} shows
a further example of the decisive influence of the choice of the MIT
criterion on the sample classification obtained.

Note the resemblance between this graph and corresponding plots for
Si:P, Figure 1 of Ref.\ \onlinecite{Stu.etal.93} and our Figures
\ref{sip_a} and \ref{sip_b}. From all these graphs, it is obvious how
strongly the usual localization theory motivated augmented power law
extrapolation results depend on which $T$ interval is chosen to fit
Eq.\ (\ref{apl}) to the experimental data. Nevertheless, together,
these figures support the idea that the character of the MIT should be
the same in CdSe:In and Si:P. 

\begin{figure}
\includegraphics[width=0.85\linewidth]{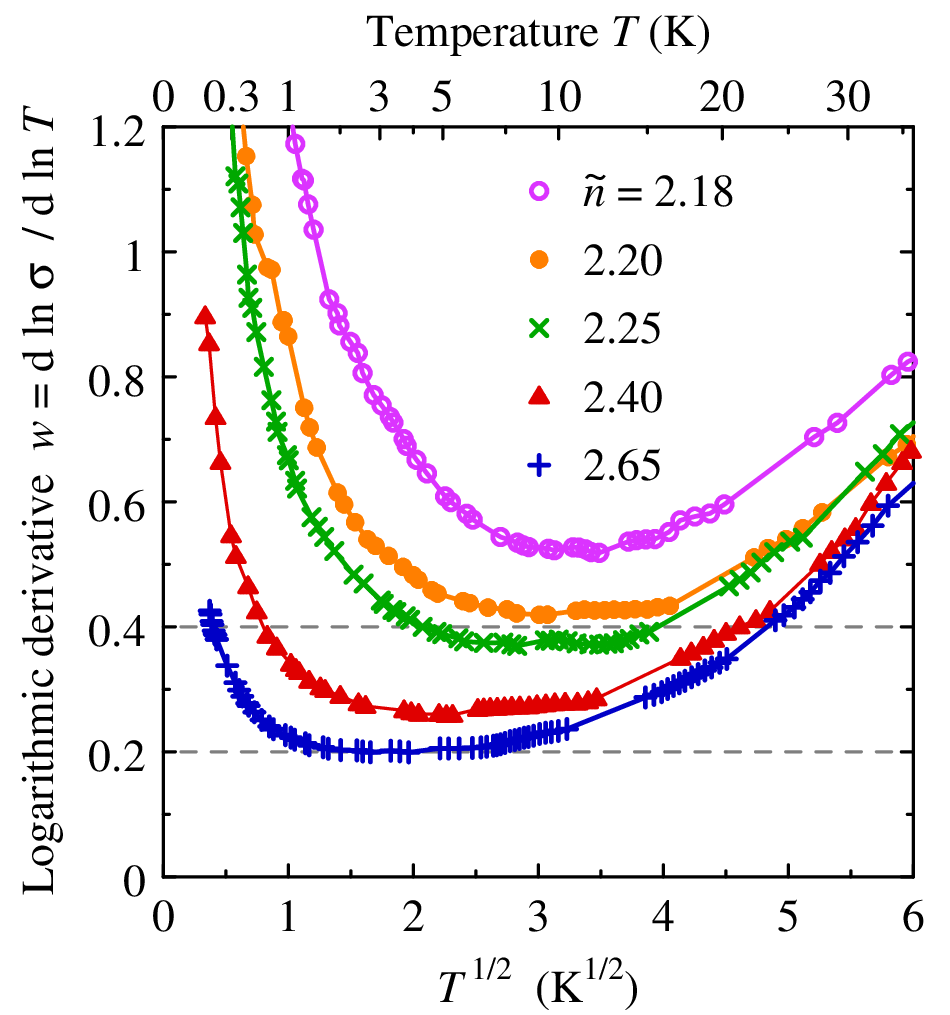}
\caption{(Color online) Temperature dependences of the logarithmic
derivative of the conductivity, $w$, for the five crystalline CdSe:In
samples investigated in Refs.\ \onlinecite{Zha.etal.90} and
\onlinecite{Aha.etal.92}.
The values of
$w(T,n)$ were obtained from Figure 1 of Ref.\ \onlinecite{Aha.etal.92}
as described in the text. The net dopant concentrations $\tilde{n}$,
given as multiple of $10^{17}\ {\rm cm}^{-3}$, are taken from Table I
of Ref.\ \onlinecite{Aha.etal.92}.
}
\label{cdse2}
\end{figure}

The next question, of course, concerns the behavior of a set of
$w(T, n = {\rm const.})$ curves for CdSe:In. To answer it, we 
digitized the $\sigma(T)$ data for five samples published in Figure 1
of Ref.\ \onlinecite{Aha.etal.92}. This graph includes values from a
very wide $T$ range, from about 60~mK up to roughly 100~K. Utilizing
the numerical differentiation method explained in Appendix C, we 
calculated $1 / w(T, n = {\rm const.})$ sliding a $\ln \sigma$ window
of width 1.0 along the $\ln T (\ln \sigma)$ curves, analogously to our
approach in Subsection 3.1. To guarantee that at least 7 data points
are taken into account in each slope calculation, the $\ln \sigma$
window was correspondingly expanded if necessary. This procedure
ensures that the random errors of the $w$ values are kept small in the
medium temperature range, where $w(T)$ seems to vary only slowly.

The resulting $w(T, n = {\rm const.})$ data are presented in Figure
\ref{cdse2}.  Three features of this graph deserve particular
attention. First, for $T \lesssim 1\ {\rm K}$, all
$w(T, n = {\rm const.})$ exhibit negative slope, in agreement with
Figure 3b of Ref.\ \onlinecite{Zha.etal.90}. Therefore, we support
the sample classification in Refs.\ 
\onlinecite{Zha.etal.90,Aha.etal.92,Zha.Sara.95} where all these
samples are considered as insulating. Moreover, we point out that the
considerable negative slope of $w(T)$ is the origin of the problems
with the augmented power law approximations of $\sigma(T)$ discussed
above. 

Second, in contrast, at high $T$, that means above roughly 15~K, $w$
increases with $T$ for all samples considered here. This behavior very
likely arises from a second conduction mechanism yielding a
substantial contribution to $\sigma(T)$, see the discussion in
Subsection 2.6, compare also Figure \ref{w_fig}. 

Third, in between the low- and high-$T$ regimes, all
$w(T, n = {\rm const.})$ exhibit pronounced minima. The corresponding
temperature value, $T_{\rm min}(n)$, decreases as $n$ increases, that
means as the MIT is approached, and so does
$w_{\rm min}(n) = w(T_{\rm min}(n),n)$. -- This feature could already
be foreshadowed from inspection of Figure 3b of Ref.\
\onlinecite{Zha.etal.90}, restricted to $T \lesssim 10\ {\rm K}$. -- 
Such characteristic minima are not special to CdSe:In, they were
observed already in amorphous alloys, see Figure 7 in Ref.\
\onlinecite{Moe.Adk} and Figures 7a/b in Ref.\
\onlinecite{Moe.etal.99} for a-Si$_{1-x}$Cr$_x$ and
a-Si$_{1-x}$Ni$_x$, respectively; see also our Figure \ref{w_fig_cond}
on GeSb$_2$Te$_4$. Finally, we stress that, for the CdSe:In sample
closest to the MIT, $w_{\rm min} \approx 0.2$. This is incompatible
with the idea of a continuous MIT at which $\sigma(T) \propto T^p$
with $p = 1/2$ or $1/3$; for the reasoning, see Subsection 2.6. 

Concluding this subsection, we encourage the reader to compare Figure
\ref{cdse2} to the diagrams in our Section 5. They show the behavior
of $w(T,x)$ in the vicinity of the MIT for four simple, qualitatively
different phenomenological models of $\sigma(T,x)$, where $x$ stands
for an arbitrary control parameter. In our opinion, Figure \ref{cdse2}
closest resembles Figure \ref{pri_discont_sca_2m}. The latter diagram
presents a set of $w(T, x = {\rm const.})$ curves for a discontinuous
MIT which is superimposed by an additional high-temperature conduction
mechanism and for which $\mbox{d} \sigma / \mbox{d} T = 0$ indicates
the transition point in the low-$T$ limit. This resemblance is in
accord with the interpretation in Ref.\ \onlinecite{Zha.Sara.95},
where the existence of a finite minimum metallic conductivity was
concluded from two-parameter scaling of the $T$ dependences of
$\sigma$.

\subsection{Crystalline n-Cd$_{0.95}$Mn$_{0.05}$Se}

\begin{figure}
\includegraphics[width=0.85\linewidth]{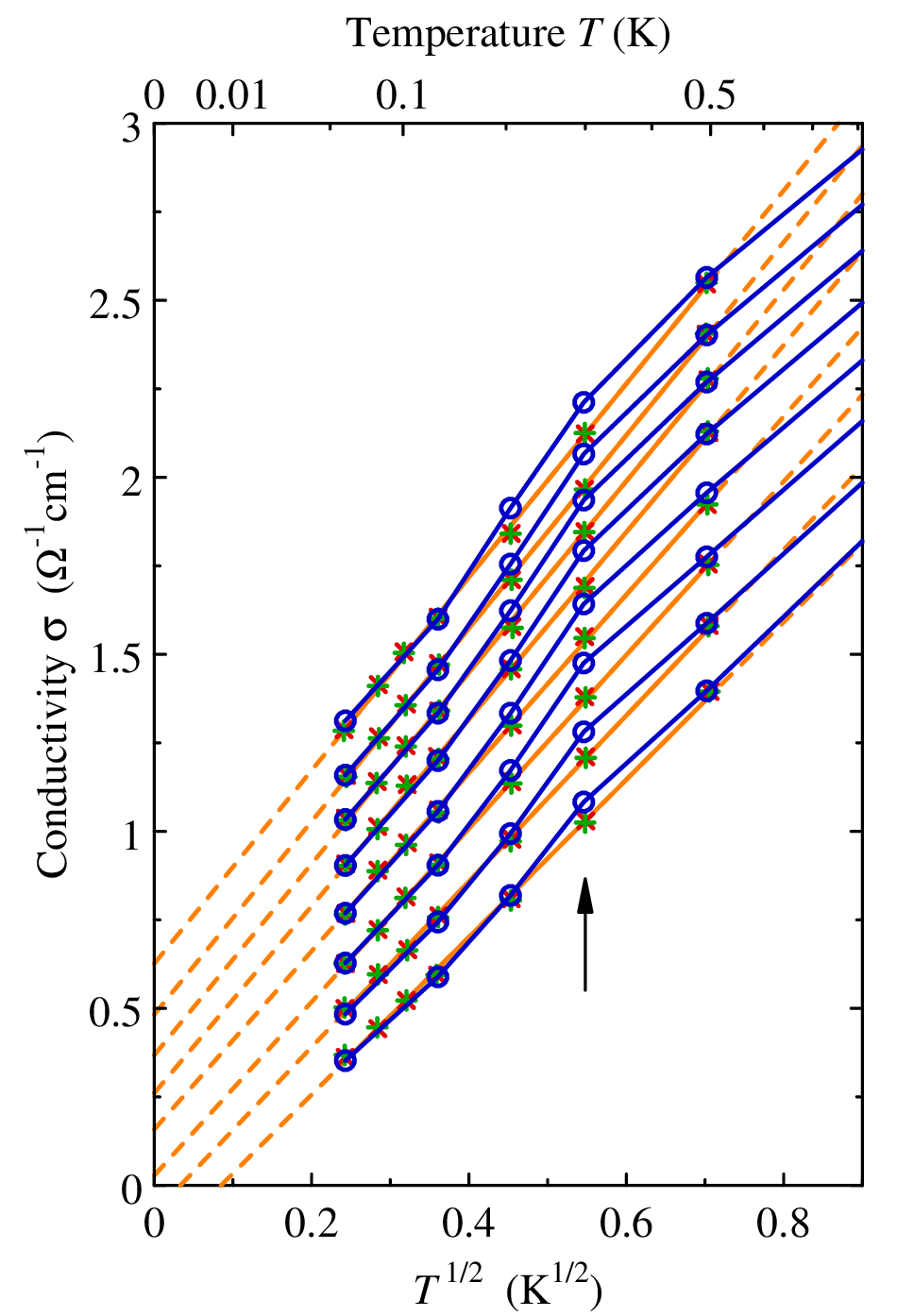}
\caption{(Color online) Temperature dependences of the conductivity
of an n-Cd$_{0.95}$Mn$_{0.05}$Se sample with net In-donor
concentration $4 \times 10^{17}\ {\rm cm}^{-3}$ under various magnetic
fields published in
Refs.\ \onlinecite{Wojt.etal.86} and
\onlinecite{Dietl.etal.86}; from top to bottom, $H = 28.5$, 25.2,
22.7, 20.2, 17.7, 15.2, 12.7, and 10.3~kOe. The data marked by red
$\times$ and green $+$ are redrawn from the apparently identical plots in
Figure 1a of Ref.\ \onlinecite{Wojt.etal.86} and Figure 6 of Ref.\
\onlinecite{Dietl.etal.86}; the data given by blue circles were
obtained by digitizing the $\rho(H,T = {\rm const.})$ curves in Figure
4 of Ref.\ \onlinecite{Dietl.etal.86}. The orange straight lines refer
to $\sigma = a + b \, T^{1/2}$ approximations of the data from Figure 6
of Ref.\ \onlinecite{Dietl.etal.86} for the interval 0.059 -- 0.493~K;
they are given as full lines within this range and as dashed lines in
the extrapolation region outside of it. 
}
\label{dietl_a}
\end{figure}

\begin{figure}
\includegraphics[width=0.85\linewidth]{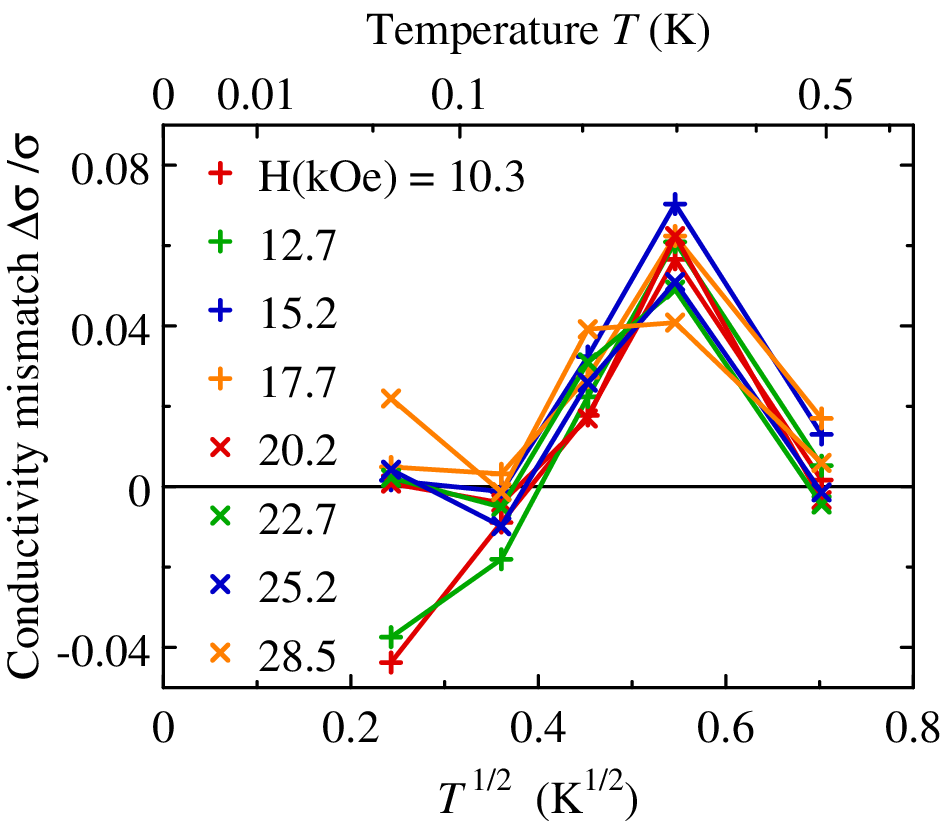}
\caption{(Color online) Mismatch between the digitized values of the
data sets published in Figures 4 and 6 of Ref.\
\onlinecite{Dietl.etal.86}. This diagram presents \protect\linebreak
$(\sigma({\rm Figure\ 4}) - \sigma({\rm Figure\ 6})) /
\sigma({\rm Figure\ 6})|_{(T,H_i)}$ versus $T^{1/2}$ for the $H_i$
values listed in the caption of Figure \ref{dietl_a}. 
}
\label{dietl_b}
\end{figure}

In certain cases, alternatively to utilizing stress, the MIT can also
be tuned by applying a magnetic field. Thereby, the variation of the
critical dopant concentration can originate from two effects acting in
opposite directions: On the one hand, the field squeezes the dopant
wave functions and thus increases the critical dopant concentration.
On the other hand, the field reduces the effective disorder by
partially aligning d-spins which, in turn, affect the current carrying
shallow impurity levels via s-d exchange interaction; this way, the
field reduces the critical dopant concentration. For details see
Refs.\ \onlinecite{Efros.Shkl.84} and \onlinecite{Dietl.etal.86},
respectively.

As an example of employing the second of these effects in the study of
the MIT, we now evaluate the investigation of the semimagnetic
crystalline semiconductor \linebreak n-Cd$_{0.95}$Mn$_{0.05}$Se described in
Refs.\ \onlinecite{Wojt.etal.86} and \onlinecite{Dietl.etal.86}. In
these measurements, the magnetic field was directed perpendicular to
the current. Thus, interpretation complications can arise from the
tensor character of the conductivity. However, the authors of Ref.\
\onlinecite{Wojt.etal.86} claim the magnetoresistance to be isotropic
in this case, so that the corresponding corrections should be
negligible in the data evaluation. In the following, we take this
assumption for granted.

\begin{figure}[!]
\includegraphics[width=0.85\linewidth]{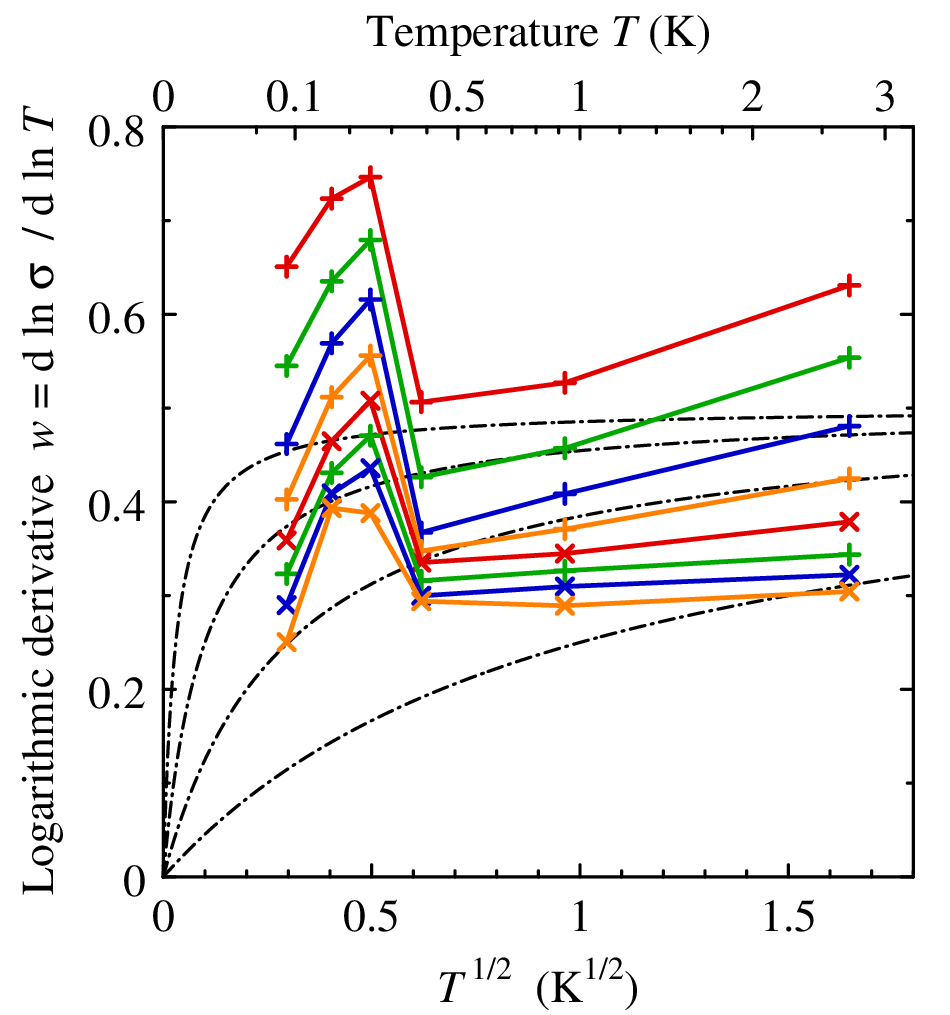}
\caption{(Color online) Temperature dependences of the logarithmic
derivative of the conductivity, $w$, for an \protect\linebreak
n-Cd$_{0.95}$Mn$_{0.05}$Se
sample studied in Ref.\ \onlinecite{Dietl.etal.86}, in which the MIT
was tuned by varying the critical dopant concentration by means of
magnetic field; compare caption of Figure \ref{dietl_a}. The data
presented here were obtained from Figure 4 of Ref.\
\onlinecite{Dietl.etal.86}; the symbols have the same meaning as in
our Figure \ref{dietl_b}. For comparison, the dashed-dotted lines
represent $w(T)$ resulting from the hypothetical relation
$\sigma = a + b \, T^{1/2}$ with $a / b = 0.03$, 0.1, 0.3, and 1
(from top to bottom). 
}
\label{dietl_c}
\end{figure}

To inspect an as broad as possible $T$ range, to ask for the limit of
the linear range of $\sigma(T^{1/2},H = {\rm const.})$ analogously to
our approach in Subsection 4.2, and, moreover, to check our
digitization, we took into consideration all available published data.
Thus we digitized not only the apparently identical plots of
$\sigma(T,H = {\rm const.})$ versus $T^{1/2}$ in Figure 1a of Ref.\
\onlinecite{Wojt.etal.86} and Figure 6 of Ref.\
\onlinecite{Dietl.etal.86}, but also the
$\log_{10} \rho(T= {\rm const.},H)$ versus $H^{1/2}$ diagram Figure 4
of Ref.\ \onlinecite{Dietl.etal.86}, which includes data from a wider
$T$ range.

Surprisingly, the digitization precision check uncovered a systematic
mismatch between these data sets: Our Figure \ref{dietl_a} shows that,
in particular, the presumably ``more original'' data points in Figure
4 of Ref.\ \onlinecite{Dietl.etal.86} mark significantly larger values
of $\sigma(0.3\ {\rm K},H)$ than the corresponding data points in
Figure 1a of Ref.\ \onlinecite{Wojt.etal.86} and Figure 6 of Ref.\
\onlinecite{Dietl.etal.86}; this temperature is indicated by an arrow.
The first impression is confirmed by the detailed presentation of the
mismatch in our Figure \ref{dietl_b}, showing that this deviation
considerably exceeds the random digitization errors. -- Of course, we
double-checked this finding very carefully. --

Two strange effects may help to evaluate this discrepancy: First, all
$\sigma(T^{1/2},H = {\rm const.})$ given in Figure 6 of Ref.\
\onlinecite{Dietl.etal.86} can be described almost exactly by
$\sigma = a(H) + b(H) \, T^{1/2}$, compare our Figure \ref{dietl_a},
whereas the $\sigma(T^{1/2},H = {\rm const.})$ obtained from Figure 4
of that work exhibit a slight s-shape, similarly to our finding for
Si:P in Subsection 4.2. Second, if, instead of the actually presented
values, the data from Figure 4 of Ref.\ \onlinecite{Dietl.etal.86}
were shown in Figure 6 of that work (or in Figure 1a of Ref.\
\onlinecite{Wojt.etal.86}), then the quality of the fits to
Finkel'stein's renormalization group equations therein would be
considerably damaged. Of course, thirty years after publication, it
may be impossible to solve this puzzle. However, in future work,
caution is advised in drawing conclusions from Figures 1a of Ref.\
\onlinecite{Wojt.etal.86} and Figure 6 of Ref.\
\onlinecite{Dietl.etal.86}.

What does the logarithmic derivative $w(T,H = {\rm const.})$ tell us
about this MIT? To answer the question, we calculated $w$ starting
from $\ln \sigma(\ln T)$ relations obtained from Figure 4 of Ref.\
\onlinecite{Dietl.etal.86}; in this case, we considered all pairs of
neighboring $T$ values. The results are presented in our Figure
\ref{dietl_c}: Again, the numerically obtained relations differ
qualitatively from the $w(T)$ expected according to the hypothetical
$\sigma = a + b \, T^{1/2}$. As for Si:P, all $w(T, H = {\rm const.})$
exhibit maxima at roughly the same temperature, at about 0.2~K in the
present case. Such a maximum occurs even for the field strength
$10.3\ {\rm kOe}$ which, according to the corresponding maximum value
of $w$, as well as to Ref.\ \onlinecite{Dietl.etal.86}, should clearly
fall into the insulating region. Furthermore, above this temperature,
up to 0.4~K, a considerable decrease of $w$ with increasing $T$ is
present also for the $H$ values interpreted as belonging to the
metallic region in Ref.\ \onlinecite{Dietl.etal.86}, even for the
largest field, for which $w$ falls down to about 0.3. Therefore,
although we consider these data with caution due to the discrepancies
uncovered above, we conclude that they can definitely not be
interpreted to indicate a continuous MIT at which
$\sigma \propto T^{1/2}$.

\subsection{Crystalline 
Cd$_{0.95}$Mn$_{0.05}$Te$_{0.97}$Se$_{0.03}$:In}

Persistent photoconductivity is another elegant way to fine-tune the
MIT.\cite{Katsu.87,Glod.etal.93,Lei.etal.98} It seems to be based on
the existence of deep donors (or acceptors) with the capture rate
being very small at low temperatures.\cite{Katsu.87} Thus, at low $T$,
the excitation of electrons (or holes) from such levels to shallow
states by illumination can cause a very long-lasting conductivity
increase.

G\l\'{o}d et al.\ used this effect in Ref.\ \onlinecite{Glod.etal.93}
to study the MIT in the diluted magnetic semiconductor
Cd$_{0.95}$Mn$_{0.05}$Te$_{0.97}$Se$_{0.03}$:In. They aimed to compare
with the transition tuned in compound semiconductors by magnetic
field, see previous subsection.

In Ref.\ \onlinecite{Glod.etal.93}, the authors investigated four
samples with different degrees of doping; the room-temperature
electron concentrations were reported to range from
$1.8 \times 10^{17}$ to $3.1 \times 10^{17}\ {\rm cm}^{-3}$. Before
and after illumination, $\sigma(T)$ was measured between about 20 and
500~mK. At the end, this report concludes: ``The obtained data
contradict suggestions that the polaron formation could result in a
discontinuous transition.''

\begin{figure}[!]
\includegraphics[width=0.85\linewidth]{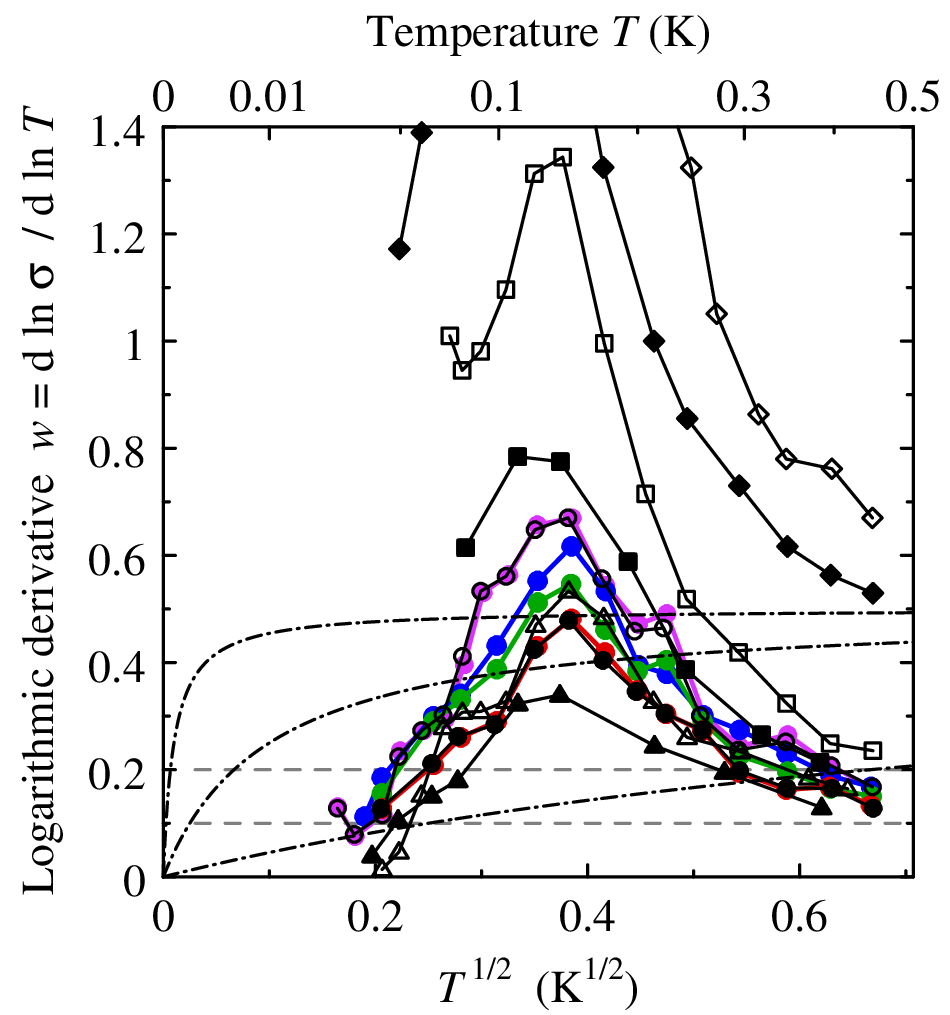}
\caption{(Color online) Temperature dependences of the logarithmic
derivative of the conductivity, $w$, for light-controlled transport in
crystalline Cd$_{0.95}$Mn$_{0.05}$Te$_{0.97}$Se$_{0.03}$:In studied in
Ref.\ \onlinecite{Glod.etal.93}. The presented $w(T)$ data were
obtained from the $\sigma(T)$ data published in Ref.\
\onlinecite{Glod.etal.93} as described in the text. Here, we use the
same geometric symbols as in the original $\sigma(T)$ plots: All
colored symbols relate to data from Figure 1 of Ref.\
\onlinecite{Glod.etal.93}, which reports on measurements of
one sample
after different illumination times (increasing from top to bottom
here); black empty and solid symbols denote data obtained from the
$\sigma(T)$ of four samples shown in Figure 2 of Ref.\
\onlinecite{Glod.etal.93} which were measured before and after long
illumination, respectively. For comparison, dashed-dotted lines
represent $w(T)$ resulting from the hypothetical relation
$\sigma = a + b \, T^{1/2}$ with $a / b = 0.01$, 0.1, and 1 (from top
to bottom). To facilitate judging the slope of the $w(T)$, dashed gray
lines mark constant $w = 0.2$ and 0.1.
}
\label{dietl3}
\end{figure}

To check this statement, we digitized the $\sigma(T)$ shown in Figures
1 and 2 of Ref.\ \onlinecite{Glod.etal.93} and obtained $w(T)$
therefrom by numerical differentiation considering always sets of
three neighboring points. The results of this data evaluation are
presented in our Figure \ref{dietl3}. In several aspects, as detailed
below, this graph strikingly resembles our Figures \ref{sip_1} and
\ref{sip_2}, which concern Si:P. 

Note, in particular, that all $w(T)$ curves in Figure \ref{dietl3}
have negative slope above about 170~mK, even when $w < 0.2$. This
feature conflicts with the interpretation by G\l\'{o}d et al., who
claimed that part of their $\sigma(T)$ originate from metallic
transport. These authors, however, left open which of the individual
$\sigma(T)$ they considered to indicate metallic transport and which
not; in other words, they did not define a specific MIT point.
Furthermore, as already stressed above for other disordered solids,
this feature is incompatible with the assumption of a continuous MIT
with $\sigma$ being proportional to $T^{1/2}$ or $T^{1/3}$ at the
transition.

Note, moreover, that all $w(T)$ exhibit clear maxima at roughly the
same temperature, at about 140~mK, independently of the respective
maximum value of $w(T)$. In this context, we mention the investigation
of the light-induced MIT in a related compound semiconductor,
crystalline Cd$_{1-x}$Mn$_x$Te:In, in Ref.\ \onlinecite{Lei.etal.98}.
Examining that study uncovered similar strange behavior of $w(T)$, but
in that case at higher temperatures: All $w(T)$ obtained from the data
in Ref.\ \onlinecite{Lei.etal.98} exhibit maxima at roughly
0.5~K.\cite{Moe.etal.98}

Finally, concerning Figure \ref{dietl3}, we point to the rapid
decrease of all $w(T)$ with decreasing $T$ below about 110~mK, far
more rapid than expected according to Eq.\ (\ref{apl}) with $p = 1/2$.
Qualitatively, it resembles the rapid decrease of the $w(T)$ curves
for Si:P in Figure \ref{sip_1} below about 50~mK. In the case of
Cd$_{0.95}$Mn$_{0.05}$Te$_{0.97}$Se$_{0.03}$:In, this feature results
from the saturation tendency of the individual $\sigma(T)$ below
roughly 100~mK apparent in Figures 1 and 2 of Ref.\
\onlinecite{Glod.etal.93}. The authors speculated that this saturation
may be related to spin glass freezing, but they seem not to have
followed up on this idea in later publications. To us, this feature is
more likely due to some experimental artifacts.

\subsection{Disordered Gd}

We turn now to non-crystalline solids. For several such substances,
already more than a decade ago, the augmented power law approach to
locating the MIT was shown to fail, see Ref.\ \onlinecite{Moe.Adk} and
citations therein. Here, from this group of solids, we first examine
the recent investigation of disordered Gd films by Misra et al., Ref.\
\onlinecite{Mis.etal.11}. By means of rf magnetron sputtering, the
authors grew thin Gd films on substrates held at
130~K.\cite{Mis.etal.11} These films were found to be stable below
77~K. Their sheet resistance could be tuned by annealing at the
deposition temperature.\cite{Mis.etal.11} 

Misra et al.\ evaluated their data by means of a scaling analysis
taking for granted continuity of the MIT; they considered the $T$
region from about 5 up to about 50~K. At first glance, their
interpretation seems to work nicely, see Figure 3 in Ref.\ 
\onlinecite{Mis.etal.11}. Several parameters, however, have to be 
adjusted in that approach, which is always associated with a certain
risk of misinterpretation. Therefore, we examine here the
justification of the augmented power law fits with adjustable
exponents to the measured $\sigma(T)$ data, the first step of that
scaling analysis.

In their Figure 2, Misra et al.\ present a plot of $w(T)$ for a series
of samples. It exhibits a $T$-independent separatrix between metallic
and insulating samples, a feature which seems to corroborate the
assumed continuity of the MIT, compare Figure \ref{pri_cont} in our
Section 5. This conclusion, however, is not justified for a simple
reason: The authors present only the analytical differentiations of
their augmented power law fits to $\sigma(T)$ but no numerically
differentiated experimental data. Thus, concerning the continuity of
the MIT, first and foremost, Misra et al.\ only got out what they put
into their data analysis.

\begin{figure}
\includegraphics[width=0.85\linewidth]{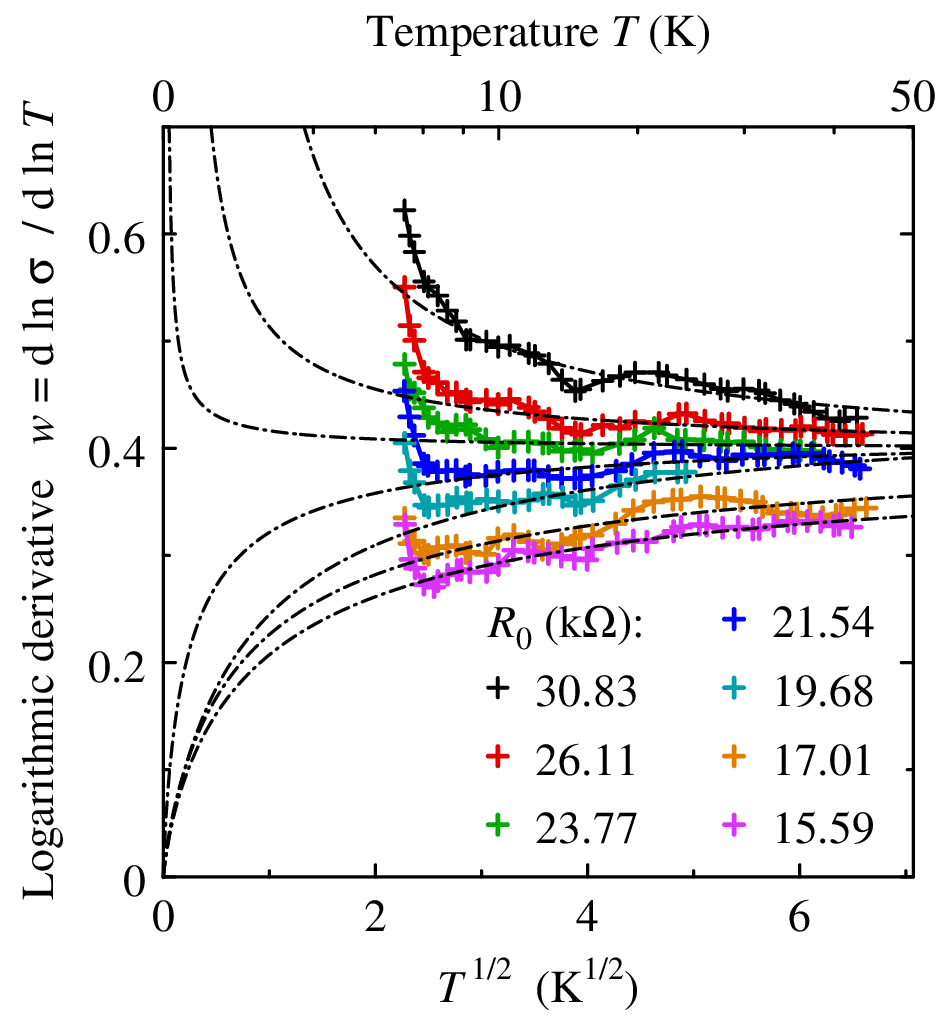}
\caption{(Color online) Temperature dependences of the logarithmic
derivative of the conductivity, $w$, for seven samples of disordered
Gd investigated in Ref.\ \onlinecite{Mis.etal.11}. For this
reanalysis, we digitized a part of the data sets published in Figure 1
of Ref.\ \onlinecite{Mis.etal.11}, displaying $T$ dependences of the
normalized conductivity $\sigma_{\rm n}$. The samples are labeled
by their sheet resistance at 5~K, $R_0$. We obtained the values of
$w(T)$ in two ways: Crosses mark results of numerical differentiation
obtained sliding a window of variable width along the
$\ln \sigma_{\rm n} (\ln T)$ curves, for details see text;
dashed-dotted lines result from analytical differentiation of
augmented power law fits with freely adjustable exponents to the
experimental $\sigma_{\rm n}(T;R_0)$ above 8~K.}
\label{misra}
\end{figure}

Our Figure \ref{misra} compares $w(T)$ relations resulting from the
numerical differentiation of the original data with estimates obtained
by the analytical differentiation of augmented power law fits. In the
former approach, it was non-trivial to find a good compromise between
two contradictory demands on the window width: On the one hand, the
window shifted along the $\ln \sigma (\ln T)$ curves has to be
sufficiently wide to damp the random fluctuations; on the other hand,
it has to be kept sufficiently small so that not too much of the
low-$T$ part of $w(T)$ is lost. Therefore, we use here a window of
variable width: At low $T$, we consider the first two data points,
then the first three points, and so on until the $\ln T$ window
reaches the width 0.3; after this, we hold the width constant and
shift the window along the curve. This approach is not ideal, but it
seems to be the best one can do in this case. Certainly, it might
overvalue random fluctuations at low $T$.

At about 8~K, our Figure \ref{misra} shows a qualitative change in the
behavior of the $w(T)$ curves obtained by numerical differentiation.
Below this threshold, they exhibit pronounced upturn-like deviations
from the analytically differentiated augmented power law fits to the
$\sigma(T)$ data points between 8 and about 50~K. Remarkably, these
upturns have qualitatively the same shape for all the samples
considered here, independently of $R_0$, the sheet resistance at 5~K,
and independently of whether $w(T)$ decreases of increases with $T$
above 10~K. Hence, it is unlikely that they occur only by chance.

We stress that these upturns are not only present in the $w(T)$ of the
samples with $R_0 \ge 23.77\ \rm{k \Omega}$, classified as insulating
in Ref.\ \onlinecite{Mis.etal.11}, but also in the $w(T)$ of the
samples with $R_0 \le 21.54\ \rm{k \Omega}$, considered as metallic
therein. Because this feature is not compatible with the
hypothetically metallic $\sigma(T)$ being described by an augmented
power law, its presence questions the sample classification in Ref.\
\onlinecite{Mis.etal.11}.

The amplitude of the upturns decreases with $R_0$. This might be
related to the onset of this feature being shifted to lower $T$ with
decreasing $R_0$. To some extent, this finding resembles results for
insulating samples of amorphous Si$_{1-x}$Ni$_x$; in that case, the
$w(T,x = {\rm const.})$ exhibit pronounced upturns which are shifted
to lower $T$ with increasing $x$, see Figure 7 of Ref.\
\onlinecite{Moe.etal.99}. Compare also Figures \ref{w_fig_cond} and
\ref{cdse2} in our review, regarding GeSb$_2$Te$_4$ and CdSe:In,
respectively.

Anyway, the presented comparison of both approaches to obtaining
$w(T)$ shows that the interpretation in Ref.\ \onlinecite{Mis.etal.11}
is not conclusive. More precise measurements taking into account also
moderately lower temperatures are required to reach convincing
results.

\subsection{Nanogranular Pt-C}

As further recent examples of studies on non-crystalline systems, we
consider now the investigations of the MIT vicinity of nanogranular
W-C and Pt-C published in Refs.\ \onlinecite{Huth.etal.09} and
\onlinecite{Sac.etal.11}, respectively. These experiments investigated
the electrical transport in mesoscopic samples of granular films which
were produced by focused electron beam induced deposition of a
metal-organic precursor and, in the case of Pt-C, subsequent electron
beam irradiation. Due to the very small size of the samples, precise
such electrical measurements are demanding. 

\begin{figure}
\includegraphics[width=0.85\linewidth]{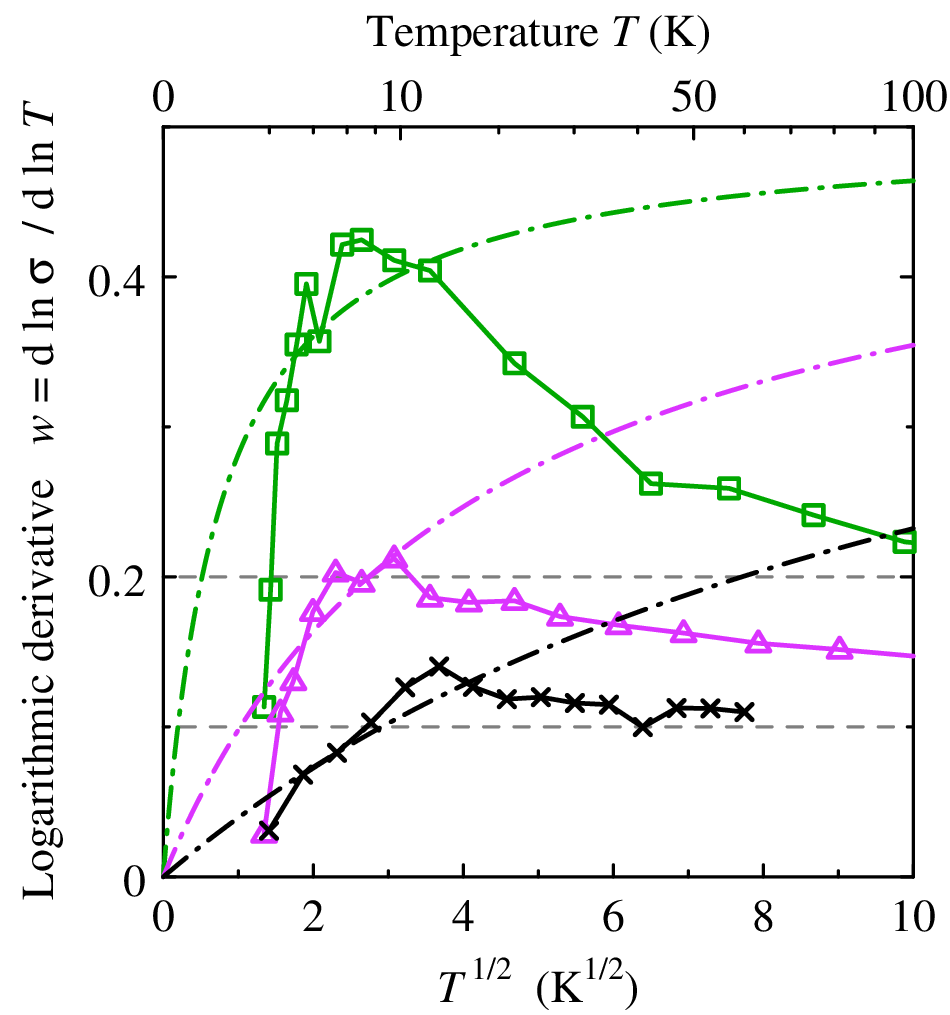}
\caption{(Color online) Comparison of temperature dependences of the
logarithmic derivative of the conductivity, $w$, obtained in two
different ways for three nanogranular Pt-C samples from Ref.\
\onlinecite{Sac.etal.11}. These samples were prepared by focused
electron beam induced deposition and subsequent electron beam
irradiation. Data points marked by green squares and magenta triangles
are redrawn from Figure 1b of Ref.\ \onlinecite{Sac.etal.11}. They
result from numerical differentiation of the measured $\sigma(T)$
values and refer to two samples irradiated with doses of 0.48 and
$0.64\ {\rm \mu C / \mu m^2}$, respectively. In addition, data for a
sample irradiated with $0.80\ {\rm \mu C / \mu m^2}$ are included,
labeled by black crosses. They were obtained by digitizing Figure 2a
of Ref.\ \onlinecite{Sac.etal.11} and subsequent numerical
differentiation, in which pairs of neighboring $\ln \sigma(\ln T)$
points were considered. The dashed-dotted curves, relating to the same
measurements of the three samples, were determined by analytical
differentiation of the $\sigma = a + b \, T^{1/2}$ fits presented in
Figure 2a of Ref.\ \onlinecite{Sac.etal.11}. The dashed gray lines mark
marks 
constant $w = 0.1$ and 0.2.}
\label{ptc}
\end{figure}

For six W-based granular films, Figure 5 of Ref.\
\onlinecite{Huth.etal.09} shows $w(T^{1/2})$ data obtained by
numerical differentiation. Its inset compares these data with
analytically differentiated augmented power law approximations for
three of the samples. Two problems are obvious here: Below roughly
30~K, $w$ depends on the W content in a nonsystematic manner.
Furthermore, the curves obtained by numerical differentiation of the
measured data and by analytical differentiation of the augmented power
law approximations, respectively, deviate considerably from each other
in this $T$ region.

The latter problem is even more striking in the study of Pt-C, Ref.\
\onlinecite{Sac.etal.11}, as we will show now. According to Figure 2a
of this publication, $\sigma = a + b \, T^{1/2}$ fits seem to well
approximate the measured data between about 1.5 and about 20~K. To
check this conclusion, our Figure \ref{ptc} compares data obtained by
numerical differentiation of $\ln \sigma(\ln T)$ with analytic
differentiations of $\sigma = a + b \, T^{1/2}$ approximations for
three of the investigated samples. We interpret it as follows.

For the two samples labeled by squares and triangles, which were
irradiated with doses of 0.48 and $0.64\ {\rm \mu C / \mu m^2}$,
respectively, the results of both approaches clearly contradict each
other: The numerical differentiation yields $w(T)$ increasing with
decreasing $T$ down to roughly 8~K and then rapidly decreasing below
this temperature. These data suggest that $w(T)$ may either vanish
proportionally to a high power of $T^{1/2}$ as $T \rightarrow 0$ or
reach 0 already between 1 and 2~K.\cite{rem_gr_pt} -- Note the
resemblance to our Figure \ref{sip_1} on Si:P, in particular to the
behavior of the colored curves below 50~mK therein, as well as to our
Figure \ref{dietl3} regarding persistent photoconductivity in
Cd$_{0.95}$Mn$_{0.05}$Te$_{0.97}$Se$_{0.03}$:In. -- On the contrary,
the analytically differentiated augmented power law approximations
decrease monotonically with $T$ everywhere and reach 0 only at
$T = 0$. The sample labeled by crosses, which was irradiated with a
dose of $0.80\ {\rm \mu C / \mu m^2}$, exhibits similar behavior of
$w(T)$ but less pronounced.

Thus, the $w(T)$ obtained by analytical differentiation of
$\sigma = a + b \, T^{1/2}$ fits to measured $\sigma(T)$ substantially
differ from the results of numerical differentiations not only above
10~K but also below 4~K. Hence, the goodness of the fits shown in
Figure 2a of Ref.\ \onlinecite{Sac.etal.11} arises only from the
restriction to the transition region between two qualitatively
different regimes which both cannot be described by
$\sigma = a + b \, T^{1/2}$. We remark that this interpretation is
already suggested by careful inspection of the $\sigma$ versus
$T^{1/2}$ plots for 0.80 and $1.28 {\rm \mu C / \mu m^2}$ in Figure
2a of Ref.\ \onlinecite{Sac.etal.11}.

The discrepancies uncovered here may be overlooked on first reading of
Ref.\ \onlinecite{Sac.etal.11} for two strange reasons: (i) For the
sample irradiated with a dose of $0.64\ {\rm \mu C / \mu m^2}$, Figure
2a of Ref.\ \onlinecite{Sac.etal.11} does not include any $\sigma(T)$
points below 4~K, whereas Figure 1b thereof presents $w(T)$ data for
this sample down to 2.3~K. (ii) For the sample obtained by irradiation
with a dose of $0.48\ {\rm \mu C / \mu m^2}$, the $w(T)$ data in
Figure 1b of Ref.\ \onlinecite{Sac.etal.11} are not consistent with
the $\sigma(T)$ values given in Figure 2a thereof: If these $w(T)$
data are correct, then the $\sigma$ versus $T^{1/2}$ plot must exhibit
a saturation tail at low $T$ as it is present in Figure 2a of Ref.\
\onlinecite{Sac.etal.11} for the two samples irradiated with doses of
0.80 and $1.28\ {\rm \mu C / \mu m^2}$, respectively, but it does not.

In this way, the comparison presented in our Figure \ref{ptc}
disproves one of the central conclusions drawn in Ref.\
\onlinecite{Sac.etal.11} by Sachser et al.: It is not justified to
claim the presence of a universal $T^{1/2}$ contribution to
$\sigma(T)$ at low temperatures as the authors did.

\subsection{Common features of all $w(T,x)$ diagrams}

Concluding this part of our review, we highlight the destructive
result common to all the examinations of various experiments from the
literature presented in the current section: For none of the samples
with $w > 0.1$, the behavior of $w(T)$ can be understood in terms of a
metallic conduction mechanism which causes $\sigma(T)$ to obey an
augmented power law, Eq.\ (\ref{apl}), with $p = 1/2$ or 1/3 over a
wide $T$ range.

Simultaneously, we emphasize the constructive result of these
analyses: In all examined experiments, there is a wide $T$ range
within which the following correlation exists. The logarithmic
derivative $w(T)$ seems to always increase with decreasing $T$ when
$w > 0.1$, that means not only when $w > 1/2$, as for (almost)
exponential $\sigma(T)$, but also even when $0.1 < w \le 1/2$ arising
from comparably weak $T$ dependences of $\sigma$. Only one
$^{70}$Ge:Ga sample with $w \approx 0.25$ considered in Subsection 4.4
might form a slight exception; on broad average, $w(T)$ is roughly
constant in this case.

This correlation conflicts with the classification into metallic and
insulating samples according to augmented power law extrapolations to
$T = 0$ used in most of the studies examined here. Furthermore, since,
at fixed $T$, $w$ seems to decrease monotonically when the MIT is
approached from the insulating side, for example with increasing
dopant concentration, this correlation is incompatible with the
idea of continuous MIT at which $\sigma \propto T^p$ with $p > 0.1$.

Additionally, in some but not all studies, our examinations have
uncovered a qualitatively different behavior at the lower end of the
$T$ range investigated: Here, $w(T)$ decreases with $T$ and vanishes
far more rapidly than proportionally to $T^{1/2}$. At the crossover
between both scenarios, these $w(T)$ exhibit pronounced maxima with
the unusual characteristics described in the following two paragraphs.

Concerning the corresponding temperature value, $T_{\rm max}$, we
emphasize two findings: On the one hand, in each of the individual
experiments, $T_{\rm max}$ seems to be almost independent of the
control parameter. On the other hand, for Si:P, $T_{\rm max}$ differs
considerably from experiment to experiment, see Subsection 4.2.

Concerning the peak value of $w(T,x = {\rm const.})$, note the
following: For Si:P and
Cd$_{0.95}$Mn$_{0.05}$Te$_{0.97}$Se$_{0.03}$:In, such extrema were
found even in the region $w \gtrsim 1$, that means for apparently
clearly insulating samples, see Subsections 4.2 and 4.7, respectively.

Together, the findings described in the previous two paragraphs give
rise to serious doubts about the soundness of the maxima and thus of
the very rapid decreases of $w(T)$ with decreasing $T$ below
$T_{\rm max}$.

Because of the observations listed above, none of the experimental
studies analyzed here can be considered as conclusive support for a
continuous MIT. Instead, as we will show in Section 5, contrasting
flow diagrams of $w(T,x = {\rm const.})$ obtained from four
qualitatively different phenomenological models for the dependence of
$\sigma$ on $T$ and control parameter $x$ with the
``high-temperature'' parts of our Figures \ref{castner}, \ref{sip_1},
\ref{sip_2}, \ref{sarachik}, \ref{cdse2}, \ref{dietl3}, and \ref{ptc}
favors the opposite interpretation: This comparison supports the
hypothesis that $\lim_{T \rightarrow 0} \sigma(T,x)$ exhibits a
discontinuity at the MIT. Nevertheless, since the experimental data
are not completely consistent, further and more precise measurements
are encouraged.

\section{Four possible scenarios of $w(T,x)$}

In the previous two sections, considering various solids, we have
presented numerous diagrams of $T$ dependences of the logarithmic
derivative of the conductivity, $w(T,x = {\rm const.})$, which were
obtained by numerical differentiation of experimental data from the
literature. In most cases, independently of the nature of the control
parameter, $x$, these results strikingly conflict with the
interpretation in the respective publication. 

However, as summarized in Subsection 4.10, our diagrams exhibit
obvious similarities with each other. These features are intriguing
because, as already pointed to in Subsection 2.6 and Ref.\
\onlinecite{Moe.etal.99}, such plots of $w(T,x = {\rm const.})$ for
several samples with different values of $x$ can be an informative
fingerprint of the character of the MIT. 

The direct quantitative evaluation of $w(T,x)$ is hindered by
experimental uncertainties and by the necessity of assumptions on
$\sigma(T,x)$. Therefore, we here take the opposite approach and
analyze the consequences of qualitatively different phenomenological
hypotheses: For four very simple phenomenological models describing
$\sigma(T,x)$ close to the MIT, we obtain families of
$w(T,x = {\rm const.})$ curves. In this way, we demonstrate how the
character of the MIT determines the qualitative features of such flow
diagrams and thus evaluate alternative interpretations of
$\sigma(T,x)$ measurements.

First, we assume the MIT occurring at $x_{\rm c}$ to be continuous.
That means, we suppose $\sigma(T = {\rm const.}, x)$ to be continuous
there not only at any finite $T$ but also in the limit
$T \rightarrow 0$. To construct a simple but sufficiently flexible
model of $\sigma(T,x)$ with these characteristics, we start from Eq.\
(10) of Ref.\ \onlinecite{Moe.etal.99}: We combine modified stretched
exponential and augmented power law dependences for the insulating and
metallic sides of the MIT, respectively, so that
\begin{eqnarray}
&& \sigma(T,x) = \nonumber\\
&& \nonumber\\
&& \left\{ \begin{array}{l}
T^p \exp(-(T_0(x)/T)^p)\\ \\ a(x) + b(x) \, T^p
\end{array} \mbox{\ for\ } 
\begin{array}{l} x < x_{\rm c} \\ \\ x \ge x_{\rm c} \end{array} ,
\right.
\label{m_cont_mit}
\end{eqnarray}
where $T_0(x)$, $a(x)$, and $b(x)$ are continuous functions, and where
$T_0(x)$ decreases monotonically with $x$, whereas $a(x)$ increases
monotonically with $x$. Furthermore, in order to ensure continuity at
$x_{\rm c}$, we presume \linebreak $T_0(x \rightarrow x_{\rm c} - 0) = 0$,
$a(x \rightarrow x_{\rm c} + 0) = a(x_{\rm c}) = 0$, and
$b(x \rightarrow x_{\rm c} + 0) = b(x_{\rm c}) = 1$. In Eq.\
(\ref{m_cont_mit}), for simplicity, all quantities are given in
dimensionless form. Moreover, the exponents of the transport
mechanisms involved are assumed to have the same value, $p$;
certainly, this guess is a gross simplification, but abandoning it
would not modify the qualitative features discussed below. -- Our \linebreak
continuity and monotonicity presumptions on $T_0(x)$, $a(x)$, and
$b(x)$ as well as our simplifying assumption on the exponents also
apply to the three models considered below. --

\begin{figure}
\includegraphics[width=0.85\linewidth]{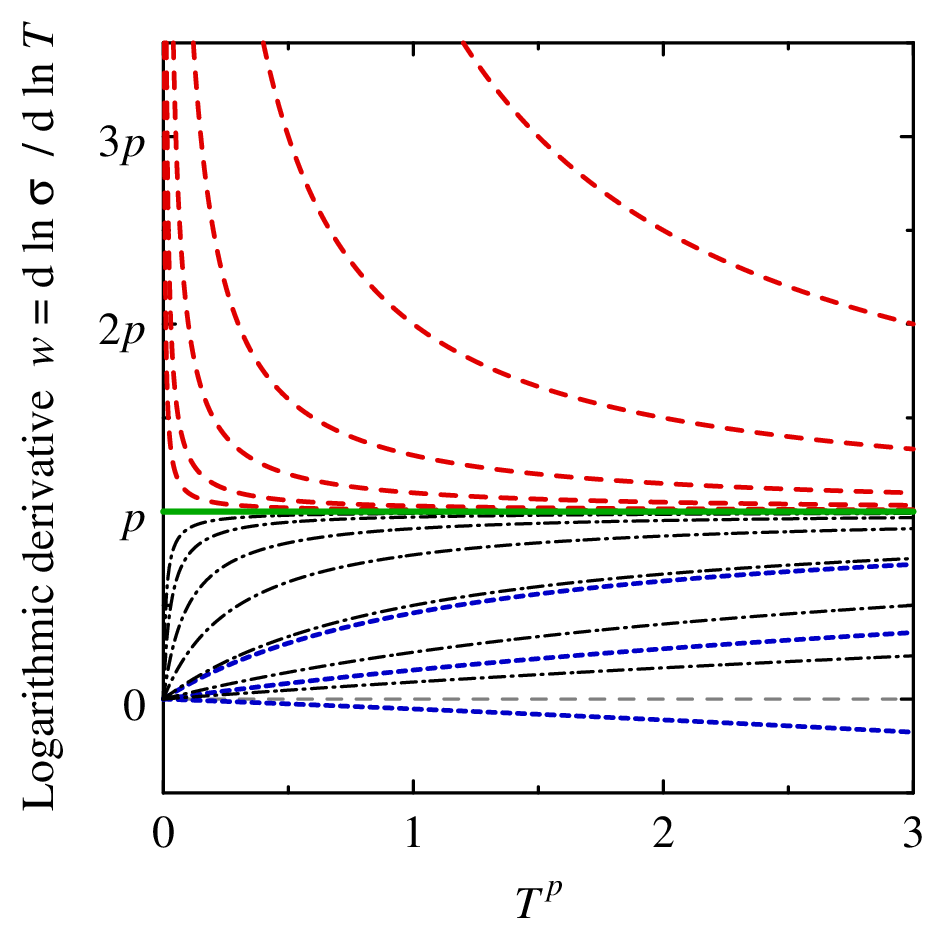}
\caption{(Color online) Behavior of the logarithmic derivative of the
conductivity, $w(T,x = {\rm const.})$, for the
continuous MIT modeled by
Eq.\ (\ref{m_cont_mit}). Dashed (red): non-metallic with
$T_0^{\ p} = 3$, 1, 0.3, 0.1, 0.03, and 0.01 (from top to bottom);
dashed-dotted (black): metallic with $a = 0.01$, 0.03, 0.1, 0.3, 1, 3,
10 and $b \equiv 1$ (from top to bottom); short-dashed (blue):
metallic with $a = 1$, 3, 10 and $b = 1 - 0.15\,a$ (for smaller values
of $a$, such curves would almost coincide with the corresponding
relations for $b \equiv 1$); full (green): separatrix, i.e.\
$w(T,x = x_{\rm c})$. The horizontal dashed gray line marks $w = 0$.}
\label{pri_cont}
\end{figure}

Figure \ref{pri_cont} shows the flow diagram of 
$w(T,x = {\rm const.})$ for two versions of Eq.\ (\ref{m_cont_mit}),
$b(x) \equiv 1$ and $b(x) = 1 - 0.15 \, a(x)$. In both cases, the
regions with metallic and non-metallic behavior do not overlap each
other; they are separated by a horizontal line, $w(T,x_{\rm c}) = p$.
Above this separatrix, that is for non-metallic behavior, the slope,
$\mbox{d} w / \mbox{d} T$, is always negative, while, below the
separatrix, that is in the metallic region, $\mbox{d} w / \mbox{d} T$
is positive whenever $w > 0$; compare Subsection 2.6.

\vspace{0.20cm}

On the metallic side of the MIT, when $b(x) \equiv 1$ as in Eq.\ (10)
of Ref.\ \onlinecite{Moe.etal.99}, our Eq.\ (\ref{m_cont_mit}) yields
only functions $w(T,x = {\rm const.})$ which are positive everywhere
except at $T = 0$. Hence, for realizing a sign change of 
$\mbox{d} \sigma / \mbox{d} T$ and thus of $w(T = {\rm const.},x)$,
the parameter $b$ must depend on the control parameter $x$. In Figure
\ref{pri_cont}, we model such a situation by means of the assumption
$b(x) = 1 - 0.15 \, a(x)$. (This relation between $a(x)$ and $b(x)$
leads to a good resolution of the family of curves; moderately
modifying it does not alter the qualitative features of Figure
\ref{pri_cont} provided that $b(x_{\rm c}) = 1$ remains valid.)

\vspace{0.20cm}

One property of this model is particularly noteworthy: The MIT and the
sign change of $\mbox{d} \sigma / \mbox{d} T$ occur at different
values of $a$ and therefore at different values of $x$; for Eq.\
(\ref{m_cont_mit}) with $b(x) = 1 - 0.15 \, a(x)$, when $a = 0$ and
when $a = 6.7$, respectively. Hence, according to the model considered
here, these phenomena are different in nature, which also follows from
the mathematical consideration in Subsection 2.2 by logical
contraposition.

\vspace{0.20cm}

Furthermore, just at the MIT described by Eq.\ (\ref{m_cont_mit}),
$\mbox{d} \sigma / \mbox{d} T = p \, T^{p-1}$. Hence, this derivative
is positive at any finite $T$; it diverges as $T \rightarrow 0$ if
$p < 1$.

\vspace{0.20cm}

Second, for studying how a discontinuous MIT is reflected in the
$w(T,x = {\rm const.})$ flow diagram, the above model of a continuous
MIT has to be modified only slightly. We incorporate an additional
constant into the prefactor for the insulating side and change the
limiting value of the constant part for the metallic side. Thus we
obtain the following generalization of Eq.\ (11) of Ref.\
\onlinecite{Moe.etal.99}:
\begin{eqnarray}
&& \sigma(T,x) = \nonumber\\
&& \nonumber\\
&& \left\{ \begin{array}{l}
(1 + T^p) \exp(-(T_0(x)/T)^p)\\ \\ a(x) + b(x) \, T^p
\end{array} \mbox{\ for\ } 
\begin{array}{l} x < x_{\rm c} \\ \\ x \ge x_{\rm c} \end{array} 
\right.
\label{m_discont_mit}
\end{eqnarray}
with $T_0(x \rightarrow x_{\rm c} - 0) = 0$, 
$a(x \rightarrow x_{\rm c} + 0) = a(x_{\rm c}) = 1$, and
$b(x \rightarrow x_{\rm c} + 0) = b(x_{\rm c}) = 1$. This model
exhibits a discontinuity in $\lim_{T \rightarrow 0} \sigma(T, x)$,
whereas $\sigma(T = {\rm const.}, x)$ remains continuous for any
$T > 0$; compare Figure 11 of Ref.\ \onlinecite{Moe.etal.99}.

\vspace{0.20cm}

\begin{figure}
\includegraphics[width=0.85\linewidth]{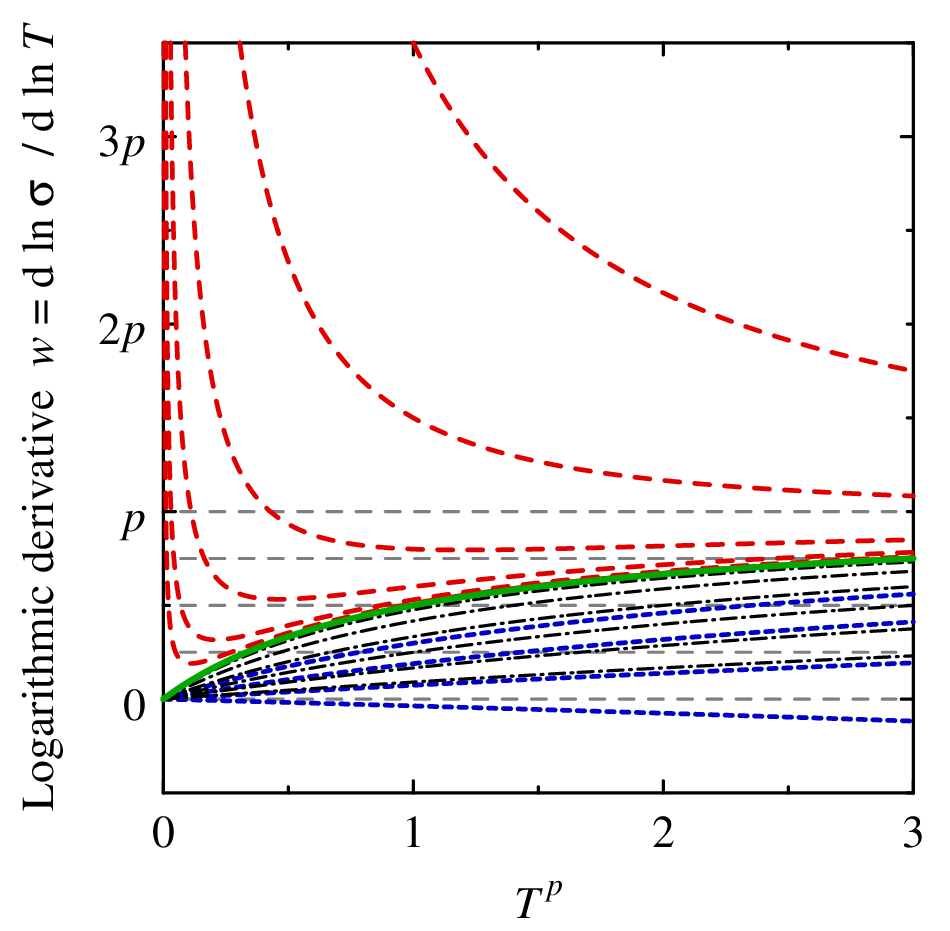}
\caption{(Color online) Behavior of $w(T,x = {\rm const.})$ for the
discontinuous MIT modeled by Eq.\ (\ref{m_discont_mit}), at which
$\mbox{d} \sigma / \mbox{d} T^p \, = 1$. Dashed (red): non-metallic with
$T_0^{\ p} = 3$, 1, 0.3, 0.1, 0.03, and 0.01 (from top to bottom);
dashed-dotted (black): metallic with $a = 1.1$, 1.4, 2, 3, 5, 10 and
$b \equiv 1$ (from top to bottom); short-dashed (blue): metallic with
$a = 2$, 3, 5, 10 and $b = 1 - 0.15\,(a - 1)$; full (green):
separatrix, $w(T,x = x_{\rm c})$. To facilitate judging the slope of
the curves, dashed gray lines mark constant $w = 0$, $0.25\,p$,
$0.5\,p$, $0.75\,p$, and $p$.}
\label{pri_discont}
\end{figure}

Figure \ref{pri_discont} presents the flow diagram of 
$w(T,x = {\rm const.})$ for two versions of Eq.\
(\ref{m_discont_mit}), that is for $b(x) \equiv 1$ and for
$b(x) = 1 - 0.15 \, (a(x) - 1)$. Again, in both cases, the metallic
and non-metallic regions do not overlap each other and, within the
metallic region, $w > 0$ is correlated with positive slope of
$w(T, x = {\rm const.})$.

\vspace{0.20cm}

However, in the non-metallic region, $x < x_{\rm c}$, the
$w(T, x = {\rm const.})$ now have negative slopes for all $x$ values
only in the limit $T \rightarrow 0$, in contrast to the curve set in
Figure \ref{pri_cont}. At finite $T$, the sign of the slope varies:
Four of the curves exhibit a minimum; it occurs if and only if
$T_0 < 1$. The corresponding temperature, $T_{\rm min}(x)$, can take
any positive value, whereas $0 < w(T_{\rm min},x) < p$. 

\vspace{0.20cm}

Note, when the MIT is approached from the insulating side, that means
when \linebreak $T_0 \rightarrow 0$, both $T_{\rm min}(x)$ and
$w(T_{\rm min},x)$ decrease and tend to zero; this can also be easily
derived analytically. In consequence, the separatrix
$w(T, x = x_{\rm c})$ tends to zero as $T$ vanishes, unlike the
separatrix in Figure \ref{pri_cont}. 

\vspace{0.20cm}

Because of the minima in $w(T, x = {\rm const.})$, for this model,
simultaneously exhibiting positive $w$ and negative slope of
$w(T, x = {\rm const.})$ at a certain measuring temperature is
sufficient but not necessary for a sample to be insulating.
Furthermore, the identification of insulating samples is hindered by
the following restriction: The closer $x$ to $x_{\rm c}$, the lower
the temperatures which have to be considered in order to rule out the
possibility of metallic conduction by means of detecting
$\mbox{d} w / \mbox{d} T < 0$. 

\vspace{0.20cm}

As in the case of Eq.\ (\ref{m_cont_mit}), the parameter $b$ must
depend on $x$ when a sign change of $w(T = {\rm const.},x)$ is to be
described.  In Figure \ref{pri_discont}, we emulate such a situation
assuming $b(x) = 1 - 0.15 \, (a(x) - 1)$ in analogy to our above
consideration of a continuous MIT. 

\vspace{0.20cm}

Again, the MIT and the sign change of $\mbox{d} \sigma / \mbox{d} T$,
now occurring when $a = 1$ and $a = 7.7$, respectively, do not
coincide. Moreover, as for the model of a continuous MIT considered
above, just at the MIT, $\mbox{d} \sigma / \mbox{d} T = p \, T^{p-1}$
is positive at any finite $T$, and diverges as $T \rightarrow 0$ if
$p < 1$.

\vspace{0.20cm}

Third, the question arises how to construct a model which yields a 
sign change of $\mbox{d} \sigma / \mbox{d} T$ just at the MIT itself.
According to Subsection 2.2, such an MIT must be discontinuous. Thus
we start from Eq.\ (\ref{m_discont_mit}). Small modifications are
sufficient to reach our aim: We only omit the $T$ dependence of the
preexponential factor for $x < x_{\rm c},$ appropriately change the
limit of $b(x)$, and obtain
\begin{eqnarray}
&& \sigma(T,x) = \nonumber\\
&& \nonumber\\
&& \left\{ \begin{array}{l}
\exp(-(T_0(x)/T)^p)\\ \\ a(x) + b(x) \, T^p
\end{array} \mbox{\ for\ }  
\begin{array}{l} x < x_{\rm c} \\ \\ x \ge x_{\rm c} \end{array} 
\right.
\label{m_discont_mit_sca}
\end{eqnarray}
with $T_0(x \rightarrow x_{\rm c} - 0) = 0$, 
$a(x \rightarrow x_{\rm c} + 0) =$ \linebreak $a(x_{\rm c}) = 1$,
$b(x \ne x_{\rm c}) < 0$ and
$b(x \rightarrow x_{\rm c} + 0) = b(x_{\rm c}) = 0$. This way, the
discontinuity of $\lim_{T \rightarrow 0} \sigma(T, x)$ as well as the
continuity of $\sigma(T = {\rm const.},x)$ for $T > 0$ are maintained.
Now, for any $T > 0$, $\mbox{d} \sigma / \mbox{d} T = 0$ holds at and
only at $x_{\rm c}$. 

\begin{figure}
\includegraphics[width=0.85\linewidth]{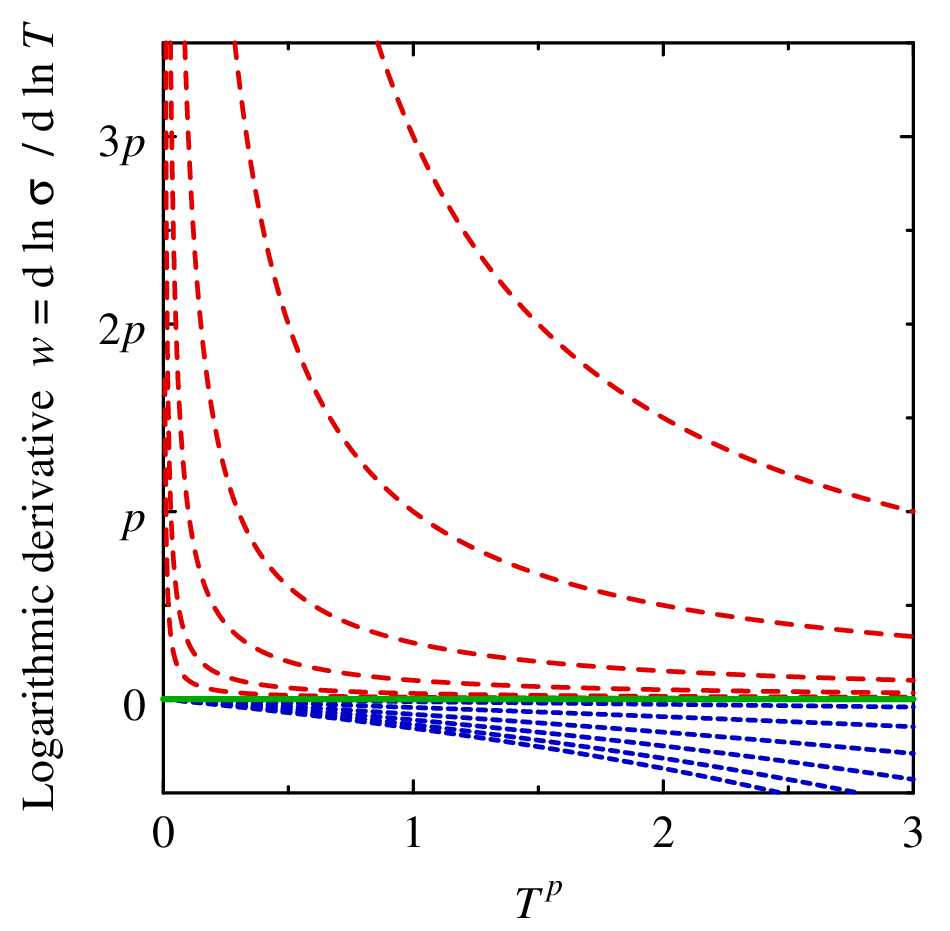}
\caption{(Color online) Behavior of $w(T,x = {\rm const.})$ for the
discontinuous MIT modeled by Eq.\ (\ref{m_discont_mit_sca}), at which
$\mbox{d} \sigma / \mbox{d} T$ changes sign. Dashed (red):
non-metallic with $T_0^{\ p} = 3$, 1, 0.3, 0.1, 0.03, and 0.01 (from
top to bottom); short-dashed (blue): metallic with $a = 1.1$, 1.4, 2,
3, 5, 10 and $b = -0.15\,(a - 1)$ (from top to bottom); full (green):
separatrix, $w(T,x = x_{\rm c})$.}
\label{pri_discont_sca}
\end{figure}

The flow diagram of $w(T,x = {\rm const.})$ for this model is given in
Figure \ref{pri_discont_sca}, where \linebreak
$b(x) = -0.15 \, (a(x) - 1)$ is
assumed in analogy to the cases previously considered. Again, as in
Figure \ref{pri_cont} illustrating Eq.\ (\ref{m_cont_mit}),
non-metallic and metallic regions are separated by a horizontal line,
but here by $w(T, x_{\rm c}) = 0$. Within the metallic region, in
contrast to Figure \ref{pri_cont}, $w(T,x)$ is now negative for any
$x > x_{\rm c}$ and any $T > 0$. Furthermore, we point out that, in
Figure \ref{pri_discont_sca}, the slope $\mbox{d} w / \mbox{d} T$ is
always negative on both sides of the MIT.

One specific feature of Eq.\ (\ref{m_discont_mit_sca}) has to be
stressed: On the insulating side of the MIT, the $T$ dependences of
$\sigma$ can be scaled according to Eq.\ (\ref{scaling}). As discussed
in Appendix B, this universality implies that, at the MIT, for any
$T > 0$, $\mbox{d} \sigma / \mbox{d} T = 0$, so that
$\mbox{d} w / \mbox{d} T = 0$; in the present case, this follows also
directly from Eq.\ (\ref{m_discont_mit_sca}) and is confirmed by
Figure \ref{pri_discont_sca}.

Fourth, in experiments, however, this inference on
$\mbox{d} \sigma / \mbox{d} T$ may only be valid as low-temperature
approximation, see Subsections 2.1 and 2.6. Therefore, we now
incorporate into Eq.\ (\ref{m_discont_mit_sca}) the influence of an
additional high-temperature conduction mechanism which enlarges
$\sigma(T,x)$ on both sides of the MIT, the more, the higher $T$. To
that end, we follow the data analysis approach of a multiplicative
decomposition of $\sigma(T)$ in the hopping region proposed in Ref.\
\onlinecite{Moe.etal.85} and extrapolate the contribution of the
high-temperature mechanism from the non-metallic into the metallic
region similarly as in Ref.\ \onlinecite{Moe.90b}. This results in
\begin{eqnarray}
&& \sigma(T,x) = \nonumber\\
&& \nonumber\\
&& \left\{ \begin{array}{l}
(1 + h(T)) \exp(-(T_0(x)/T)^p)\\ \\ a(x) + b(x) \, T^p + h(T)
\end{array}  \mbox{\ for\ } 
\begin{array}{l} x < x_{\rm c} \\ \\ x \ge x_{\rm c} \end{array} 
\right.
\label{m_discont_mit_sca_2m}
\end{eqnarray}
with $T_0(x)$, $a(x)$, and $b(x)$ fulfilling the same conditions as
for Eq.\ (\ref{m_discont_mit_sca}). In line with Ref.\
\onlinecite{Moe.90b}, we assume the high-temperature contribution,
$h(T)$, to be independent of $x$. Furthermore, in order to keep the
low-temperature behavior described by Eq.\ (\ref{m_discont_mit_sca}),
we demand that $h(T) / T^p \rightarrow 0$ as $T \rightarrow 0$. Under
this condition, the qualitative influence of $h(T)$ on the flow
diagram of $w(T,x = {\rm const.})$ does not depend on the specific
form of $h(T)$. As an example, we consider here
$h(T) = 0.01 \, T^{4 p}$, describing a quadratic $T$ dependence in
case $p = 1/2$.

\begin{figure}
\includegraphics[width=0.85\linewidth]{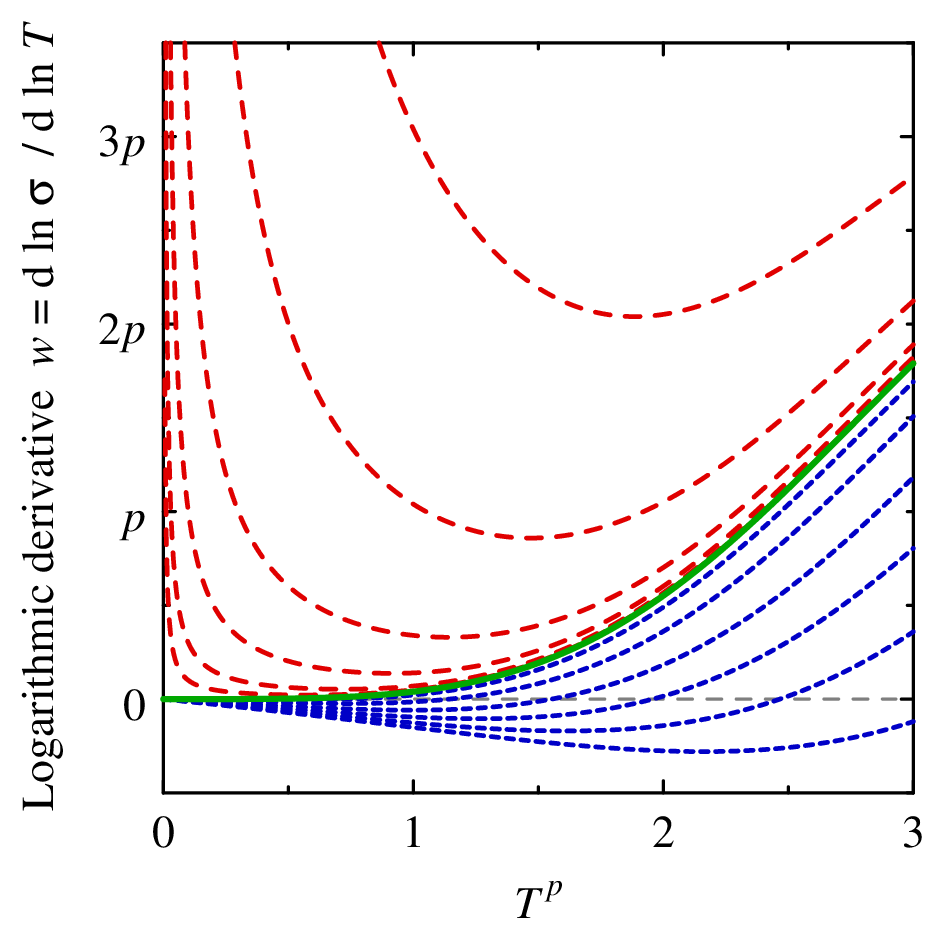}
\caption{(Color online) Behavior of $w(T,x = {\rm const.})$ for a
discontinuous MIT which is superimposed by a high-$T$
mechanism contributing to the electrical conduction on both sides of the MIT.
This transition is modeled by Eq.\ (\ref{m_discont_mit_sca_2m}) with
$h(T) = 0.01 \, T^{4 p}$. Here, just at the MIT,
$\mbox{d} \sigma / \mbox{d} T = 0$ holds only in the limit as $T \rightarrow 0$.
For the meaning of the dashed red, short-dashed
blue, and full green lines see caption of Figure \ref{pri_discont_sca}.
The dashed gray line indicates $w = 0$.}
\label{pri_discont_sca_2m}
\end{figure}

The flow diagram of $w(T,x = {\rm const.})$ for Eq.\
(\ref{m_discont_mit_sca_2m}) with $b(x) = -0.15 \, (a(x) - 1)$ is
shown in Figure \ref{pri_discont_sca_2m}. It resembles Figure
\ref{pri_discont}, which illustrates Eq.\ (\ref{m_discont_mit}), in
two aspects: On the insulating side of the MIT, there are minima, but
now in all curves. Furthermore, the separatrix, $w(T,x_{\rm c})$, is
$T$-dependent and, again, tends to zero as $T \rightarrow 0$.

In the present case, however, $w(T,x_{\rm c})$ vanishes as
$T \rightarrow 0$ far more rapidly than in Figure \ref{pri_discont}
obtained from Eq.\ (\ref{m_discont_mit}). For the immediate vicinity
of the MIT, more precisely, for $T_0^p \ll 0.3$ and $T_{\rm lea} < 1$,
this difference has important consequences:  When curves relating to
the same value of $T_0$ are compared, the minimum of
$w(T,x = {\rm const.})$ is reached in Figure \ref{pri_discont_sca_2m}
at a considerably higher temperature than in Figure \ref{pri_discont}.
Therefore, the width of the $x$ interval within which
$w(T,x = {\rm const.})$ exhibits positive slope at given $T_{\rm lea}$
despite $x < x_{\rm c}$ is far smaller for Eq.\
(\ref{m_discont_mit_sca_2m}) than it is for Eq.\
(\ref{m_discont_mit}). Furthermore, at $T_{\rm lea}$, the smallest
positive value of $w$ which is correlated with negative
$\mbox{d} w / \mbox{d} T$ is now far smaller as well. For these
reasons, Eq.\ (\ref{m_discont_mit_sca_2m}) may be considered to be
more ``experimentalist-friendly'' than Eq.\ (\ref{m_discont_mit}).

On the metallic side of the MIT, two new features arise from the
high-temperature contribution $h(T)$: With increasing $T$, all these
$w(T,x = {\rm const.})$  first pass through a minimum and then change
sign at some larger $T$ value. The latter feature indicates the
existence of minima in $\sigma(T,x = {\rm const.})$ as they have been
observed in experiments on crystalline n-Ge:(As,Ga) and Si:As as well
as on amorphous Si$_{1-x}$Cr$_x$ and Si$_{1-x}$Ni$_x$; we refer here
to Figure 1 of Ref.\ \onlinecite{Zabr.Zino}, Figure 4a of Ref.\
\onlinecite{Koo.Cas.90}, Figure 7 of Ref.\ \onlinecite{Moe.etal.85},
and Figure 8a of Ref.\ \onlinecite{Moe.etal.99}, respectively. The
corresponding $T$ values are the lower the closer $x$ to $x_{\rm c}$,
see Ref.\ \onlinecite{Moe.etal.85}.

Finally, we highlight a noteworthy relation between the flow diagrams
presented in this section: Qualitatively, Figures \ref{pri_discont}
and \ref{pri_discont_sca} can be understood as resulting from Figure
\ref{pri_discont_sca_2m} by zooming the $T$ scale out or in,
respectively.

In the current section, we have presented four simple phenomenological
models of $\sigma(T,x)$ close to the MIT yielding flow diagrams of
$w(T,x = {\rm const.})$ which qualitatively differ from each other.
The comparison of experimental graphs to these diagrams can be very
helpful in evaluating different hypotheses on the MIT: Note the
qualitative similarities between our Figures \ref{castner} and
\ref{sarachik}, on the one hand, and our Figure \ref{pri_discont_sca},
on the other hand, as well as the resemblance of Figure \ref{cdse2} of
the present work and Figure 7 of Ref.\ \onlinecite{Moe.Adk} to our
Figure \ref{pri_discont_sca_2m}. All these similarities support the
hypothesis of a discontinuous MIT. Note, furthermore, the qualitative
differences between our Figures \ref{castner}, \ref{sarachik}, and
\ref{cdse2}, on the one hand, and Figure \ref{pri_cont}, on the other
hand. These discrepancies disprove the phenomenological model of a
continuous MIT considered here, Eq. (\ref{m_cont_mit}), for any value
of $p$ above 0.1.

\section{Summarizing discussion}

\subsection{Analysis of the available MIT criteria}

In the present work, motivated by the recent, seemingly very
surprising publication on phase-change materials Ref.\
\onlinecite{Sie.etal.11}, we have examined and critically reviewed the
measurement interpretations in numerous experimental studies of the
MIT in disordered solids. To find out whether or not the results of
Ref.\ \onlinecite{Sie.etal.11} are striking indeed, we started in
Section 2 with elucidating the fundamentals of the corresponding data
analyses. For this aim, we discussed in detail the available
approaches to locate the MIT, that means to discriminate between
metallic and insulating samples. In doing so, we showed that making
this decision is by far not as simple as it might seem at first
glance: Our investigation highlighted substantial biases inherent to
the diverse available methods. Although these biases predetermine the
result on the character of the MIT to a large extent, their influence
has been overlooked in many publications. The related problems are
summarized in the following paragraphs.

The MIT in disordered solids is primarily a zero-temperature
phenomenon. Therefore, in contrast to early studies as well as to the
recent Ref.\ \onlinecite{Sie.etal.11}, conclusions concerning this
matter can be drawn only by means of $T \rightarrow 0$ extrapolations.
As an approximate substitute, the value of some physical observable at
the lowest measuring temperature may be identified with its value at
$T = 0$. Such an approximation, however, seems to be meaningful only
if the considered observable changes sign at the MIT and if,
simultaneously, the corresponding control parameter value is
sufficiently weakly $T$-dependent.

In this sense, the empirical criterion ``sign change of
$\mbox{d} \rho / \mbox{d} T$ at the lowest experimentally accessible
temperature'' has often been used in the literature. However, too
little attention has been paid to the point that the {\it supposed}
validity of this criterion necessarily implies that
$\lim_{T \rightarrow 0} \sigma(T,x)$ is a discontinuous function of
the control parameter $x$ and jumps from zero to some finite value at
the MIT. In a number of cases, this conductivity value was found to
roughly equal Mott's estimate of the hypothetical minimum metallic
conductivity. However, as shown in Appendix A, such a correlation is
natural already for dimensional reasons.

Consequently, as long as claiming all samples with negative
$\mbox{d} \rho / \mbox{d} T$ to be insulating is only an assumption in
the data analysis, this correlation alone must not be considered a
confirmation of Mott's theory. Thus, the criterion ``sign change of
$\mbox{d} \rho / \mbox{d} T$ at the lowest experimentally accessible
temperature'' exhibits a substantial interpretation bias toward a
discontinuous MIT with a finite minimum metallic conductivity.

We stress that this criterion seems to be incompatible with the
currently available microscopic theories: First, since it implies the
existence of a finite minimum metallic conductivity, it conflicts with
all the theories which yield continuity of the MIT, in  particular
with the scaling theory of localization. Second, since it presumes
$\mbox{d} \sigma / \mbox{d} T = 0$ to hold just at the MIT, it is
incompatible with the idea of an Anderson transition, occurring when
the Fermi energy crosses a mobility edge. The latter interpretation,
however, neglects electron-electron interaction. Therefore, this
discrepancy is not a valid disproof of the hypothesis that, in the
limit $T \rightarrow 0$, the MIT is connected with
$\mbox{d} \rho / \mbox{d} T$ changing sign.

An alternative approach which has been frequently applied for the last
three decades tries to find out when the metallic region is left while
the control parameter is varied. In this case, the breakdown of the
augmented power law approximation of $\sigma(T)$ is taken as MIT
criterion. \linebreak However, the usual restriction of the breakdown
identification to only ask whether or not the adjusted parameter
values are physically meaningful implies a substantial analysis bias.
The reason is that the possibility of an approximation breakdown
indicated by small but systematic deviations between measured data and
adjusted function is ignored this way. 

In a similar manner, $\sigma(T)$ ceasing to be describable by a
stretched Arrhenius law may be interpreted as indication of leaving
the insulating region. Such a parameter adjustment, however, is not
meaningful when the mean hopping energy is smaller than the lowest
experimentally accessible temperature. Moreover, this classification
method has the drawback that, in these data analyses, samples with
very weak $T$ dependences of $\sigma$ are often classified as metallic
without further checks. 

Thus, both the two approaches based on augmented power law fits and
on stretched Arrhenius law fits, respectively, tend to misclassify
weakly insulating samples as metallic. In consequence, the continuity
of the MIT concluded in such data evaluations may be only an artifact,
and not real.

We stress the significance of just these weakly insulating samples,
which exhibit merely nonexponential $\sigma(T)$ within the accessible
$T$ range. Their reliable discrimination from metallic ones is of
central importance to the trusty characterization of the MIT: An
incorrect classification of one or of a very few samples can easily
change the answer to the question whether
$\lim_{T \rightarrow 0} \sigma(T,x)$ is a continuous function of the
control parameter $x$ or whether it exhibits a discontinuity at the
MIT. Therefore, the mentioned tendency to misinterpret weakly
insulating samples as metallic causes an interpretation bias toward
continuity of the MIT. Furthermore, in case the MIT were indeed
continuous, such a misclassification could considerably modify the
obtained value of the critical exponent of
$\lim_{T \rightarrow 0} \sigma(T,x)$.

One cannot escape this problem: Since the mean hopping energy very
likely continuously tends to zero when the MIT is approached from the
insulating side, going to lower and lower temperatures is of limited
value in the characterization of the MIT. In each real experiment,
there is a control parameter region of finite width in which only
nonexponential $\sigma(T)$ can be observed although the corresponding
samples are insulating. For this reason, every success in
unambiguously classifying a given set of samples which is reached by
diminishing the lowest accessible $T$ value is ruined again when the
density of the considered values of the control parameter is
increased.

In other words: The relevant temperature scale is set by the mean
hopping energy. Hence, very likely in each experiment, when
continuously varying the control parameter, one passes the MIT at
infinitely high temperature, even if the measurements are performed at
a few mK.

We remark that related severe experimental difficulties arise also for
another reason: The critical exponent of the characteristic
temperature seems to be rather large; it was estimated to amount to
$2.1 \pm 0.1$ for crystalline n-Ge:(As,Ga) and to $3.0 \pm 0.4$ for
amorphous Si$_{1-x}$Cr$_x$, see Refs.\ \onlinecite{Zabr.Zino} and
\onlinecite{Moe.90b}, respectively. Therefore, we expect that, to
reduce the width of the control parameter region within which only
nonexponential $\sigma(T)$ can be observed merely by a factor of 2,
the lowest measuring temperature has to be diminished at least by a
factor of 4, possibly even by one order of magnitude.

In this context, the study of the logarithmic temperature derivative
of the conductivity, $w(T) = \mbox {d} \ln \sigma / \mbox {d} \ln T$,
has turned out to be very helpful in classifying individual samples:
Assume, on the metallic side of the MIT, $\sigma(T,x)$ follows an
augmented power law, $\sigma(T,x) = a(x) + b(x) \, T^p$. Then, for any
metallic sample with negative $\mbox{d} \rho / \mbox{d} T$,
corresponding to positive $w$, on the one hand, $w(T)$ tends to zero
as $T \rightarrow 0$ and, on the other hand, $w(T)$ cannot exceed $p$.
It is particularly important that, in this case, the slope of the
logarithmic derivative is positive, $\mbox{d} w / \mbox{d} T > 0$.
Consequently, under the assumption made, all samples for which, at low
$T$, the inequalities $w > 0$ and $\mbox{d} w / \mbox{d} T < 0$
simultaneously hold must be insulating.

This implication has the big advantage to unambiguously identify also
a large part of the weakly insulating samples with nonexponential
$\sigma(T)$. It should be noted, however, that it is based on an
assumption, although a plausible one: For each metallic sample with
$w > 0$, the deviation of $\sigma(T)$ from exact augmented power law
behavior is expected to be sufficiently small, that means at least so
small that ${\mbox d} w / {\mbox d} T$ retains the positive sign.

Simultaneously, the evaluation of the logarithmic derivative $w$
yields valuable information also on two other points: First, it is
very sensitive to experimental inaccuracies which, despite being
small, may qualitatively alter the judgement on the nature of
conduction of individual samples; see Sections 3 and 4 for several
examples. Second, even more importantly, flow diagrams of $w(T)$ for
sets of control parameter values enable conclusions about the
character of the MIT. Such diagrams for simple phenomenological models
of continuous and discontinuous transitions differ substantially from
each other; this was demonstrated in some detail in our Section 5. In
particular, if the MIT is continuous and $\sigma = a + b \, T^p$ holds
on its metallic side, then $0 < w < p$ implies
$\mbox{d} w / \mbox{d} T > 0$. Conflicts between this implication and
experimental findings turned out to be central to our study.

\subsection{The case GeSb$_2$Te$_4$}

The above summarized analysis of the available data evaluation
approaches provided the basis for critically examining the
interpretations in various experimental studies of the MIT in our
Sections 3 and 4. In the former part, we scrutinized the seemingly
surprising conclusions of the recent work by Siegrist et al.\ on the
MIT in phase-change materials, Ref.\ \onlinecite{Sie.etal.11}, which
motivated our work. These authors claimed that GeSb$_2$Te$_4$ differs
from other disordered systems in two features: the character of the
MIT and the strong deviation of the critical charge carrier
concentration from the Mott criterion estimate. Our reanalysis of data
from Ref.\ \onlinecite{Sie.etal.11} in Section 3 disproved both these
claims of differences by means of the following two arguments.

First, as demonstrated in Subsection 3.1, the $\sigma(T)$ curves of
GeSb$_2$Te$_4$ obtained from Ref.\ \onlinecite{Sie.etal.11} resemble
data from other disordered solids in the following sense: There are
two GeSb$_2$Te$_4$ samples for which, between about 5 and 100~K, 
$\sigma(T)$ can be well approximated by the ansatz
$\sigma = a + b \, T^{1/2}$. For one of these samples, the parameter
$a$ has a positive value being by a factor of 100 smaller than the
minimum metallic conductivity estimate in Ref.\
\onlinecite{Sie.etal.11}, whereas, for the other, $a$ is approximately
zero. -- For the $T$ range from 0.35 to about 2~K, an analogous
finding was reported in a very recent subsequent study of the MIT in
GeSb$_2$Te$_4$, Ref.\ \onlinecite{Vol.etal.15}, published by three of
the authors of Ref.\ \onlinecite{Sie.etal.11}. -- In many previous
publications of experiments on various disordered solids, such
situations were interpreted as indicating continuity of the MIT, where
all samples with positive values of $a$ were regarded as metallic. In
Ref.\ \onlinecite{Sie.etal.11}, however, Siegrist et al.\ classified
the sample for which $a$ has a small positive value as clearly
insulating and concluded the existence of a finite minimum metallic
conductivity. The obvious contradiction between both these
interpretations supports the sceptical perspective taken in our
introduction: Ref.\ \onlinecite{Sie.etal.11} is a notable example of
how the choice of the data evaluation method may predetermine the
conclusion on the character of the MIT.

Second, the asserted deviation of the critical charge carrier
concentration from the Mott criterion value is unfounded since the
latter was obtained in Ref.\ \onlinecite{Sie.etal.11} from an
unrealistic guess of the effective Bohr radius of the participating
states: In Subsection 3.2, several arguments were given for the
electronic transport in the insulating region close to the MIT
proceeding via deep defect states presumably originating from
vacancies instead of via shallow impurity states as presupposed in
Ref.\ \onlinecite{Sie.etal.11}. The latter states can only be crucial
in high-quality crystalline semiconductors, additionally provided that
the cores of host and impurity atoms resemble each other.

In consequence, there is no reason to agree with Siegrist et al.\ with
respect to ascribing an unusual quantum state of matter to
GeSb$_2$Te$_4$ and related phase-change materials. 

Nonetheless, in Subsection 3.1, also the widely-used analysis by means
of $\sigma = a + b \, T^{1/2}$ fits was found to be problematic. We
checked the validity of this approximation by considering the
logarithmic derivative $w(T)$. From this perspective, the
GeSb$_2$Te$_4$ sample from Ref.\ \onlinecite{Sie.etal.11} which was
prepared by annealing at $T_{\rm ann} =\  175\ {\rm ^oC}$ is
particularly puzzling: According to the $\sigma = a + b \, T^{1/2}$
fit, it should be clearly insulating, in agreement with the
classification in Ref.\ \onlinecite{Sie.etal.11}. However, although
$w > 1/2$ within a wide $T$ range reaching down to roughly 10~K,
$w(T)$ seems to vanish as $T \rightarrow 0$. Such a behavior may
indicate metallic transport with an augmented power law exponent
considerably exceeding $1/2$ or, alternatively, a superposition of at
least two $T$-dependent mechanisms. Moreover, this behavior of $w(T)$
may originate from thermal decoupling or from sample inhomogeneities.
In all these cases, the \linebreak $\sigma = a + b \, T^{1/2}$
analysis as a whole cannot be trusted.

Thus, the data evaluation by Siegrist et al., the extrapolations of
the $\sigma = a + b \, T^{1/2}$ approximations to $T = 0$, and the
analysis of the low-$T$ behavior of $w(T)$ yield classifications of
the samples into insulating and metallic ones which pairwise
contradict each other. Therefore, based on the data in Ref.\
\onlinecite{Sie.etal.11}, a precise determination of the transition
point between metallic and insulating behavior is impossible. In
consequence, in analyzing these data, the answer to the question
whether the MIT in GeSb$_2$Te$_4$ is continuous or discontinuous
depends on the perspective taken.

In part, the contradictions discussed above may originate from
specific features of the $\sigma(T)$ data sets which arise from
experimental imperfections. This hypothesis is suggested by our
comparison of the low-$T$ part of the experimental $w(T)$ curves of
the two most insulating GeSb$_2$Te$_4$ samples from Ref.\
\onlinecite{Sie.etal.11} with the theoretical expectations for three
different mechanisms of activated conduction. 

Further, strong support for this hypothesis comes from the results of
the subsequent GeSb$_2$Te$_4$ study Ref.\ \onlinecite{Vol.etal.15}.
In this work, the investigated $T$ range was extended by one order of
magnitude down to 0.35 K. However, apparently, the samples were
prepared in the same way as in Ref.\ \onlinecite{Sie.etal.11}.
According to the measurements in Ref.\ \onlinecite{Vol.etal.15}, also
for the samples annealed at 175 and  $200\ {\rm ^oC}$, $w(T)$ clearly
increases with decreasing $T$ at the lower end of the $T$ range
considered; for the latter sample, $w(T)$ has a minimum value of 0.35.
Moreover, for $T_{\rm ann} = 225\ {\rm ^oC}$, $w(T)$ is roughly
constant between 0.6 and 20~K and amounts to 0.14 there, in clear
contradiction to the behavior of analytically differentiated
hypothetical $\sigma = a + b \, T^{1/2}$. Thus, all these three
samples are very likely insulating.

Beyond the classification of the individual samples, these findings
contain valuable information on the character of the MIT: Because of
the low values taken by $w$, the $w(T)$ for the samples annealed at
200 and $225\ {\rm ^oC}$ clearly conflict with the interpretation in
terms of a continuous MIT with $\sigma \propto T^{1/2}$ at the
transition, in contrast to the interpretation in Ref.\
\onlinecite{Vol.etal.15}, but in agreement with the results for many
other disordered solids in our Section 4.

Nevertheless, Refs.\ \onlinecite{Sie.etal.11} and
\onlinecite{Vol.etal.15} demonstrated the possibility to fine-tune
localization in GeSb$_2$Te$_4$ by annealing. In this approach, various
preparation parameters, in particular the composition, can be kept
constant. Thus, future such experiments with enhanced precision and
accuracy of the $\sigma(T)$ measurements will very likely yield
further valuable information on the MIT in disordered systems. At the
current stage, however, one related question on GeSb$_2$Te$_4$ is
still completely open: To what extent do percolation effects arising
from inhomogeneities on the grain size scale caused by segregation
mask the generic behavior of homogeneous disordered solids?

\subsection{Comparison with other solids}

Naturally, while analyzing the MIT in phase-change materials, we were
confronted with the question of how trustworthy the interpretations in
publications on the MIT in other disordered solids are. Therefore, in
Section 4, we examined a large number of such studies, in particular
frequently cited key publications. The results are alarming: The
interpretation problems described above are not specific to the
reanalyzed GeSb$_2$Te$_4$ measurements. Instead, the inspection of
$w(T)$ curve sets for further nine different disordered solids
uncovered serious inconsistencies of the augmented power law
interpretation in all these cases. Our observations concerning this
matter are summarized in the following paragraphs.

(a) The evaluation of data for crystalline Si:As from Ref.\
\onlinecite{Shaf.etal.89} lead to an important conclusion: When, at
fixed $T$, in consequence of increasing donor concentration $n$, the
MIT is approached from the insulating side, the slope of
$w(T, n = {\rm const.})$ at low $T$ stays negative while $w$ decreases
down to $w$ values below 0.1. This is incompatible with the assumption
of a continuous MIT at which $\sigma(T) \propto T^{1/2}$ or
$\sigma(T) \propto T^{1/3}$, in contrast to the interpretation in the
original work, Ref.\ \onlinecite{Shaf.etal.89}.

(b) The MIT in crystalline Si:P was studied in particularly great
detail in the literature. Therefore, exemplarily, we here reproduced
and extended the augmented power law analysis from Ref.\
\onlinecite{Waf.etal.99}.  Our corresponding diagrams demonstrate the
considerable ambiguity of such sample classifications: The obtained
transition point depends on the $T$ range considered and in particular
on the exponent of $T$ assumed; moreover, the optimum exponent value
varies substantially with temperature and distance to the MIT.
Furthermore, our check of the augmented power law approach showed that
the rather wide linear range in the $\sigma$ versus $T^{1/3}$ plot
presented in Ref.\ \onlinecite{Waf.etal.99} does not result from
convergence to asymptotic behavior, but that it is implied by
inflection points of the measured $\sigma(T^{1/3})$. In order to gain
deeper insight, we turned to inspecting $w(T)$ and observed the
following. Contrary to the case of Si:As, the $w(T)$ of Si:P exhibit
maxima, not only for possibly metallic but also for clearly insulating
samples. However, because the corresponding temperature,
$T_{\rm max}$, seems to be almost independent of the distance to the
MIT and because, moreover, its value is experiment-specific, serious
doubts about the reliability of the measurements close to and below
the respective $T_{\rm max}$ are suggested. Nevertheless, remarkably,
in the temperature range between the puzzling maxima and about 0.6~K,
a similar behavior is present as in the case of crystalline Si:As: At
fixed $T$, while $w$ decreases with increasing P concentration or
increasing stress, $\mbox{d} w / \mbox{d} T$ stays negative at least
until $w \approx 0.1$. Hence, extrapolations based on
$\sigma = a + b \, T^{1/2}$ or $\sigma = a + b \, T^{1/3}$ are not
justified, so that the conclusions about the continuity of the MIT in
Refs.\ \onlinecite{Tho.etal.83}, \onlinecite{Waf.etal.99},
\onlinecite{Ros.etal.80}, \onlinecite{Ros.etal.83}, and
\onlinecite{Stu.etal.93} are called into question.

(c) The results on $w(T, n = {\rm const.})$ for the p-type
semiconductor Si:B, for which the original publications, Refs.\
\onlinecite{Sara.Dai.02} and \onlinecite{Dai.etal.91}, present
different interpretations, resemble the findings on the n-type
semiconductor Si:As to a large extent. Again, continuity of the MIT
with $\sigma(T) \propto T^{1/2}$ or $\sigma(T) \propto T^{1/3}$ just
at the transition can be excluded.

(d) Also the $w(T, n = {\rm const.})$ curves for
neutron-transmutation-doped $^{70}$Ge:Ga obtained from the data in
Ref.\ \onlinecite{Wata.etal.98} qualitatively resemble the
corresponding results for Si:As. Our inspection of the
$w(T, n = {\rm const.})$ for $^{70}$Ge:Ga lead to the conclusion that
$\sigma(T) \propto T^{1/2}$ at the MIT can definitely be ruled out,
while $\sigma(T) \propto T^{1/3}$ at the MIT, as concluded in Ref.\
\onlinecite{Wata.etal.98}, is very unlikely but not impossible.
Improving the precision of the $\sigma(T)$ measurements should enable
a final decision on the latter hypothesis.

(e) In contrast to all other experiments examined in Section 4, the
publications on crystalline CdSe:In which we considered there, Refs.\
\onlinecite{Zha.etal.90}, \onlinecite{Aha.etal.92}, and
\onlinecite{Zha.Sara.95}, focus merely on allegedly insulating
samples. Because, however, these studies of various aspects of the
hopping conduction took into account also samples with only rather
weak $T$ dependences of $\sigma$, trying to approximate these
$\sigma(T, n = {\rm const.})$ by augmented power laws is suggested.
Indeed, a $\sigma$ versus $T^{1/2}$ plot shows that, when the analysis
is restricted to the $T$ range 0.06 -- 0.5~K, one of the CdSe:In
samples could be regarded as metallic as well; this resembles the
situation for GeSb$_2$Te$_4$ discussed in Section 3. When, however, a
wider $T$ range is considered, a substantial curvature of
$\sigma(T^{1/2})$ becomes obvious and questions the $T \rightarrow 0$
extrapolation according to the ansatz $\sigma = a + b \, T^{1/2}$,
just as in the case Si:P. Our subsequent inspection of
$w(T, n = {\rm const.})$ uncovered interesting similarities to three
doped elemental semiconductors: Below about 1~K, these curves strongly
resemble the corresponding relations for crystalline Si:As, Si:B, and
$^{70}$Ge:Ga; $\mbox{d} w / \mbox{d} T$ is always negative, even if
$w \approx 0.2$.  On the contrary, for CdSe:In, above about 15~K,
$\mbox{d} w / \mbox{d} T$ is always positive, very likely due to a
second conduction mechanism substantially contributing to $\sigma(T)$.
The temperature value at which, between the two regions,
$w(T, n = {\rm const.})$ has its minimum decreases as the MIT is
approached, similarly as it was previously observed in studies of
a-Si$_{1-x}$Cr$_x$ and a-Si$_{1-x}$Ni$_x$; this feature seems to occur
in the case of GeSb$_2$Te$_4$, too, see Section 3.

(f) As an example of tuning the MIT by applying a magnetic field of
variable field strength $H$, we considered the study of the
semimagnetic crystalline semiconductor n-Cd$_{0.95}$Mn$_{0.05}$Se
published in Refs.\ \onlinecite{Wojt.etal.86} and
\onlinecite{Dietl.etal.86}. In this case, our data analysis uncovered
significant inconsistencies between the values published, on the one
hand, in the almost identical $\sigma$ versus $T^{1/2}$ diagrams in
Figures 1a and 6 of Refs.\ \onlinecite{Wojt.etal.86} and
\onlinecite{Dietl.etal.86}, respectively, and, on the other hand, in
the $\log_{10} \rho$ versus $H^{1/2}$ diagram Figure 4 of Ref.\
\onlinecite{Dietl.etal.86}, presenting detailed data on a wider
$(T,H)$-range. Thus, substantial doubts about the reliability of the
data basis for the theoretical interpretation in these publications
arise. Furthermore, our analysis showed that the behavior of the
$w(T, H = {\rm const.})$ obtained numerically from the data presented
in Figure 4 of Ref.\ \onlinecite{Dietl.etal.86} clearly conflicts with
the $w(T)$ resulting from hypothetical $\sigma = a + b \, T^{1/2}$:
Again, pronounced maxima of $w(T)$ occur for all considered values of
$H$ at roughly the same temperature, here at roughly 0.2~K, even if
the transport is very likely non-metallic. Moreover, from 0.2 to
0.4~K, $w$ decreases considerably in all cases, even for the largest
field, and thus highest conductivity, for which $w$ falls down to
about 0.3. Hence, these data cannot be understood in terms of a
continuous MIT at which $\sigma \propto T^{1/2}$.

(g) Evaluating the $\sigma(T)$ data from the attempt in Ref.\
\onlinecite{Glod.etal.93} to fine-tune the MIT in the persistent
photoconductor crystalline
Cd$_{0.95}$Mn$_{0.05}$Te$_{0.97}$Se$_{0.03}$:In by illumination, we
obtained a set of $w(T)$ curves which qualitatively resemble the
results for Si:P: All $w(T)$ exhibit maxima at roughly 0.14~K,
although in several cases the behavior of $w(T)$ above 0.14~K seems to
indicate clearly exponential character of $\sigma(T)$. Below this
temperature, the $w(T)$ decline far more rapidly with decreasing $T$
than expected in consequence of $\sigma = a + b \, T^{1/2}$. Above it,
in all cases, $\mbox{d} w / \mbox{d} T$ stays negative at least up to
0.4~K, even if $w$ is only a little larger than 0.1. Hence, the
interpretation in Ref.\ \onlinecite{Glod.etal.93} that a part of the
$\sigma(T)$ shown therein exhibits metallic behavior seems unfounded.

(h) The annealing-induced MIT in thin films of disordered Gd was
studied in Ref.\ \onlinecite{Mis.etal.11} by means of a scaling
analysis relying upon $\sigma = a + b \, T^p$ approximations with
adjusted $p$ and $w(T)$ relations obtained analytically therefrom.
However, the $w(T)$ which we obtained by numerical differentiation of
the $\sigma(T)$ from Ref.\ \onlinecite{Mis.etal.11} exhibit
substantial systematic low-temperature deviations from the analytical
differentiation of respective $\sigma = a + b \, T^p$ approximations.
Thus, our observation questions the basic assumptions of that work
and, in consequence, the validity of the scaling analysis therein. 

(i) Finally, for nanogranular Pt-C studied in Ref.\
\onlinecite{Sac.etal.11}, the numerically obtained $w(T)$ data
(partially from Ref.\ \onlinecite{Sac.etal.11}) clearly conflict with
$w(T)$ curves which we derived analytically from the
$\sigma = a + b \, T^{1/2}$ fits therein: Again, all numerically
obtained $w(T)$ have maxima at about the same $T_{\rm max}$. Below
$T_{\rm max}$, they decrease far more rapidly with $T$ than expected
according to $\sigma = a + b \, T^{1/2}$; above $T_{\rm max}$, the
slopes of numerically and analytically obtained $w(T)$ have opposite
signs. These findings disprove one of the main results of Ref.\
\onlinecite{Sac.etal.11}, that is the presence of a $T^{1/2}$
contribution to $\sigma(T)$ at low $T$. Furthermore, it is noteworthy
that, concerning important low-$T$ details, Figures 1b and 2a of Ref.\
\onlinecite{Sac.etal.11} presenting $w$ versus $T^{1/2}$ and
normalized $\sigma$ versus $T^{1/2}$, respectively, are clearly not
consistent with each other.

\subsection{Conclusions}

As a whole, our reanalyses in Sections 3 and 4 provide overwhelming
evidence against the usual, localization theory motivated
interpretations of the conductivity data in terms of a continuous MIT
with $\sigma(T) \propto T^p$ and $p = 1/2$ or $1/3$ at the transition.

Let us leave aside for a moment the experiment-dependent part of the
inconsistencies between augmented power law approximations of
$\sigma(T)$ on the one hand and numerically derived $w(T)$ curve sets
on the other hand, that is the $T$ region below and close to the
puzzling maxima of $w(T)$. Then one feature is common to most of the
$w(T)$ sets which we obtained in examining numerous studies on various
disordered solids in Section 4: When approaching the MIT at fixed $T$
from the insulating side, while $\mbox{d} w / \mbox{d} T$ stays
negative, \linebreak $w$ decreases down to $w$ values below 0.2, in
the cases of Si:As, Si:P, and Si:B even down to $w$ values below 0.1.
Because of its apparent universality, this feature cannot be explained
in terms of sample-specific inhomogeneities. Thus, it conflicts with
all existing and future theories which yield continuity of the MIT and
$\sigma(T) \propto T^p$ with $p > 0.1$ at the transition itself.

The following possibilities remain: (i) Even at the lowest $T$
considered, several mechanisms might superimpose each other, for all
disordered solids considered in similar ways. \linebreak (ii) The MIT
might be continuous but with the exponent $p$ being very small,
$p < 0.1$. \linebreak (iii) The MIT might be discontinuous, so that
the minimum metallic conductivity would be finite. 

To us, the discontinuity hypothesis (iii) seems to have the greatest
likelihood to be valid for two reasons. First, it is supported by the
comparison of the $w(T)$ flow diagrams obtained from experiments on
many solids in Section 4 with the results for the phenomenological
models considered in Section 5. According to this comparison, the MIT
seems to be indicated by $\mbox{d} \sigma / \mbox{d} T$ changing sign
at infinitely small $T$. Second, as we discussed in Subsection 2.5,
the discontinuity hypothesis is consistent with the scaling of the $T$
dependences of $\sigma$ in the hopping region which had been observed
in several MIT studies. As detailed in Appendix B, it is very likely
that, even for any $T$ within the temperature validity range of such a
scaling relation, the MIT occurs precisely at that control parameter
value at which $\mbox{d} \sigma / \mbox{d} T = 0$. Nevertheless, at
the current stage, because of the experiment-dependent features of
$w(T)$ left aside for the moment, we consider our outcome for the
character of the MIT still only as speculation, although as a
well-founded one.

By questioning the $\sigma(T \rightarrow 0)$ extrapolations based on
augmented power laws and thus the resulting sample classifications,
our findings also shed new light on the long-standing critical
exponent puzzle of the
MIT;\cite{Hir.etal.88, Dai.etal.91, Stu.etal.93} see in particular
Table I in Ref.\ \onlinecite{Hir.etal.88}. They suggest a surprisingly
simple solution: All the different critical exponent values for the
control parameter dependence of $\lim_{T \rightarrow 0} \sigma(T)$ in
the metallic region which have been reported for various experiments
in the literature may not have any physical meaning.

In future experiments, to achieve a compelling phenomenological
description of $\sigma(T)$ close to the MIT, the transition has to be
considered from different perspectives. In particular, in each such
study, consistent $w(T)$ data sets for a series of samples have to be
obtained from experiment by numerical differentiation. Because $w(T)$
is very sensitive to small errors of $\sigma(T)$, high precision and
accuracy of the measurements are most important. Hence, additionally
to ensuring optimum preparation reproducibility and a high degree of
sample homogeneity, the following aspects have to be taken care of.

For each individual data point, as in Ref.\ \onlinecite{Moe.etal.99},
the resistance measurement should best be performed only after
temperature equilibration has been completed, instead of taking data
while $T$ is slowly floating. Since the $\sigma(T)$ curves are only
weakly structured, the therefore necessary reduction in the number of
data points is not an issue. Furthermore, the linearity of the
current-voltage relation has to be carefully checked to eliminate the
influence of Joule heating as well as that of non-Ohmic effects. Last
but not least, thermal decoupling problems have to be eliminated very
thoroughly, even if they are so small that they are almost not
noticeable in a conventional plausibility inspection of $\sigma(T)$.
For corresponding tests, $w(T)$ data from insulating samples are very
useful: They are expected to show systematic control parameter
dependence and $\mbox{d} w / \mbox{d} T \le 0$ down to the lowest
measuring $T$. To meet all the listed prerequisites is absolutely
necessary for the reliability of experimental investigations of the
MIT.

Such a strategy focusing on high-quality $w(T)$ diagrams for sets of
samples should enable sensitive checks of the individual data analysis
approaches. Thus, it should be very helpful in resolving discrepancies
between data evaluations from different perspectives as they have been
uncovered in the present work. Only if precision and/or accuracy
improvements fail to yield an unambiguous sample classification on the
basis of $w(T)$, the intricate extension of the temperature range
toward lower $T$ will be indispensable to locate the MIT and to reach
a conclusion about its character.

Furthermore, in consequence of the experience gained in our data
analyses, we suggest for future experiments to also attach particular
importance to the detailed study of the control parameter region
where, at the lowest experimentally accessible temperature,
$\mbox{d} \sigma / \mbox{d} T$ changes sign. A careful, unbiased
analysis of the corresponding weak $T$ dependences of $\sigma$,
incorporating samples with positive $\mbox{d} \sigma / \mbox{d} T$ as
well as ones with negative $\mbox{d} \sigma / \mbox{d} T$, may be very
interesting. It could yield hints whether the hypothesis about the
existence of a line of continuous phase transitions in the finite-$T$
part of the $(x,T)$-plane proposed in Subsection 2.5 may have a chance
to be true. -- If this idea turns out to be true indeed, it will
considerably simplify the unambiguous identification of the MIT. -- In
such an analysis, the relation between the values of
$\mbox{d} \sigma / \mbox{d} T$ and of the optimal exponent of a
corresponding augmented power law approximation, in which all three
parameters are adjusted, may yield valuable information. For a first
detailed investigation of this matter see Ref.\ \onlinecite{Moe.90a},
in particular Figure 5 therein. Moreover, searching for a
non-analyticity in the relations between the parameters of
phenomenological models of $\sigma(T,x  = {\rm const.})$ which is
correlated with $\mbox{d} \sigma / \mbox{d} T = 0$ in the low-$T$
limit may be a promising strategy, compare Refs.\
\onlinecite{Moe.etal.85} and \onlinecite{Moe.90b}.

The MIT in disordered solids has fascinated physicists for more than
fifty years. Nevertheless, for numerous publications of experiments on
various such substances, the reanalyses of the reported data from
different perspectives in this review have uncovered serious
interpretation inconsistencies. According to our comparison with
qualitatively different phenomenological models for temperature and
control parameter dependences of the conductivity, the hypothesis of a
discontinuous MIT occurring precisely when
$\mbox{d} \sigma / \mbox{d} T$ changes sign at infinitely small $T$
has a considerably greater likelihood to correctly describe nature
than current localization theory claiming continuity of the MIT. It is
time to solve this puzzle. For experimentalists, improving the
measurement precision should be the most promising strategy. For
theoreticians, constructing a microscopic model which is capable to
explain the seemingly generic features of the behavior of
$\mbox{d} \ln \sigma / \mbox{d} \ln T$ illuminated in this review
should be a very rewarding challenge.

\begin{acknowledgments}
I am much obliged to C.J.~Adkins, T.G.~Castner, W.~ L\"oser,
W.~M\"obius, M.~Richter, and J.C.~Sch\"on for many detailed
discussions and numerous critical remarks on this manuscript.
Moreover, I am indebted to M.~Schreiber for drawing my attention to
the publication by Siegrist et al.\ and for a stimulating
exchange of views on it. Last not least, I am grateful to one of the
referees for his knowledgeable critical remarks, in particular
concerning the need of a completion in Subsection 2.2 and of adding 
Subsection 2.7.
\end{acknowledgments}

\newpage

\begin{center}
{\bf APPENDICES}
\end{center}

\appendix

\section{Dimensional analysis of relations between characteristic
charge carrier concentrations and minimum metallic conductivity}

When it is known which input quantities are relevant, dimensional
analysis can yield valuable restrictions on the possible outcome of
detailed theories. Here, we apply this approach to studying the MIT in
heavily doped crystalline semiconductors. In the field of impurity
conduction, all theories with the smallest possible material
specificity have four dimensioned parameters: the universal constants
electron charge, $e$ [A\,s], and reduced Planck constant, $\hbar$
[kg\,m$^2$\,/\,s], as well as the material parameters effective mass,
$m^*$ [kg], and permittivity, $\varepsilon$
[(A$^2$\,s$^4$)\,/\,(kg\,m$^3$)].

For determining the critical charge carrier concentration of the MIT,
one expresses the edge length of the cube containing one charge
carrier, in the following denoted as characteristic length,
$l_{\rm c}$ [m], as product of powers of the input parameters,
\begin{equation}
l_{\rm c} = C_1 e^{\alpha_1} \hbar^{\beta_1} m^{*\,\gamma_1}
\varepsilon^{\delta_1} \,,
\end{equation}
where $C_1$ is a dimensionless number. Dimensional analysis yields the
following system of equations for the powers of the units kg, m, s,
and A.
\begin{eqnarray}
\beta_1 + \gamma_1 - \delta_1 = 0 \nonumber \\
2 \, \beta_1 - 3 \, \delta_1 = 1 \nonumber \\ 
\alpha_1 - \beta_1 + 4 \, \delta_1 = 0 \nonumber \\
\alpha_1 + 2 \, \delta_1 = 0 \nonumber
\end{eqnarray}
Thus we obtain $\alpha_1 = -2$, $\beta_1 = 2$, $\gamma_1 = -1$, and 
$\delta_1 =1$ what results in
\begin{equation}
l_{\rm c} = C_1 \frac{\hbar^2 \varepsilon}{e^2 m^*}\,.
\label{dim_length}
\end{equation}
Since the coefficient matrix of the above system of linear equations
is nonsingular, the dimensionless number $C_1$ is independent of all 
dimensioned parameters. Hence, the constant $C_1$ is universal, and
Eq.\ (\ref{dim_length}) describes a proportionality.

Not surprisingly, $l_{\rm c}$ agrees with the effective Bohr radius up
to a dimensionless constant. Therefore, for three-dimensional systems,
the corresponding critical concentration,
$n_{\rm c} = 1/ l_{\rm c}^{\ 3}$, must satisfy a universal relation with
the structure of Eq.\ (\ref{mott_crit}), where only the constant at
the right-hand side remains undetermined.

The condition $\mbox{d} \rho / \mbox{d} T = 0$ in the limit as
$T \rightarrow 0$ defines a second characteristic length,
$\widetilde{l_{\rm c}}$, a corresponding density
$\widetilde{n_{\rm c}}$, and a characteristic conductivity value,
$\sigma_{\rm c}$. In analogy to Eq.\ (\ref{dim_length}), the
characteristic length $\widetilde{l_{\rm c}}$ must fulfill the
relation
\begin{equation}
\widetilde{l_{\rm c}} = \widetilde{C_1}
\frac{\hbar^2 \varepsilon}{e^2 m^*}\,,
\label{dim_length_tilde}
\end{equation}
where $\widetilde{C_1}$ is likewise a universal constant. Therefore,
the ratio $\widetilde{l_{\rm c}} / l_{\rm c}$ must be universal and
the ratio of the corresponding charge carrier concentrations,
$\widetilde{n_{\rm c}} / n_{\rm c}$, as well. According to experiment,
the latter quotient should be only slightly larger than 1 in case the
MIT is continuous, while, by definition, it is identical to 1 for a
discontinuous MIT at which $\mbox{d} \rho / \mbox{d} T$ changes sign
in the limit as $T \rightarrow 0$.

Consider now the characteristic conductivity, $\sigma_{\rm c}$.
Expressed in SI base units, it is a multiple of
(A$^2$\,s$^3$)\,/\,(kg\,m$^3$) in the three-dimensional case. This
quantity, too, must equal a product of powers of the input parameters,
\begin{equation}
\sigma_{\rm c} = C_2 e^{\alpha_2} \hbar^{\beta_2} m^{*\,\gamma_2}
\varepsilon^{\delta_2} \,,
\end{equation}
where $C_2$ is dimensionless. We get the following system of equations
for the powers of the units kg, m, s, and A.
\begin{eqnarray}
\beta_2 + \gamma_2 - \delta_2 = -1 \nonumber \\
2 \, \beta_2 - 3 \, \delta_2 = -3 \nonumber \\ 
\alpha_2 - \beta_2 + 4 \, \delta_2 = 3 \nonumber \\
\alpha_2 + 2 \, \delta_2 = 2 \nonumber
\end{eqnarray}
Solving yields  $\alpha_2 = 4$, $\beta_2 = -3$, $\gamma_2 = 1$, and
$\delta_2 = -1$ such that
\begin{equation}
\sigma_{\rm c} = C_2 \frac{e^4 m^*}{\hbar^3 \varepsilon} \,.
\label{dim_cond}
\end{equation}
Since the coefficient matrix of the above system of linear equations
is nonsingular, $C_2$ is a universal constant, too.

Note that, on the right-hand side of this equation, all the material
dependence is contained in the quotient $m^* / \varepsilon$. This is
also the case on the right-hand side of Eq.\ (\ref{dim_length_tilde})
determining $\widetilde{l_{\rm c}}$. Therefore, $\sigma_{\rm c}$,
$\widetilde{l_{\rm c}}$, and $\widetilde{n_{\rm c}}$ are linked by a
universal relation, that is, by
\begin{equation}
\sigma_{\rm c}
= \widetilde{C_1} C_2 \frac{e^2}{\hbar \widetilde{l_{\rm c}}}
= \widetilde{C_1} C_2 \frac{e^2}{\hbar} \widetilde{n_{\rm c}}^{1/3}\,.
\label{dim_cond_length}
\end{equation}

As a further characteristic conductivity value, the minimum metallic
conductivity, $\sigma_{\rm mm}$, be it finite or zero, can be
determined in the same way as $\sigma_{\rm c}$. Thus it must be
related to the critical charge carrier concentration, $n_{\rm c}$,
by an equation of the same form as Eq.\ (\ref{dim_cond_length}),
\begin{equation}
\sigma_{\rm mm} = C_3 \frac{e^2}{\hbar l_{\rm c}}
= C_3 \frac{e^2}{\hbar} n_{\rm c}^{\ 1/3}\,,
\label{dim_mm_cond}
\end{equation} 
where $C_3$ is universal, too. We point out that, up to the
dimensionless prefactor, which remains unknown in such a
consideration, Eq.\ (\ref{dim_mm_cond}) agrees with Eq.\
(\ref{mmc}).

In conclusion, considering the three-dimensional case, we have shown
that the structures of Eqs.\ (\ref{mmc}) and (\ref{mott_crit}) are
robust against various theoretical approximations. In particular,
these equations must also result from a theory, which takes
electron-electron interaction perfectly into account. 

For the two-dimensional case, it can be shown in an analogous manner
that $\sigma_{\rm c}$ is a universal quantity being independent of
$m^*$ and $\varepsilon$.

\section{Mathematical aspects of scaling analyses relating
$\sigma(T, x = {\rm const.})$ data sets of several samples with each
other}

Suppose we have found that $\sigma(T,x)$ exhibits the following
features: (i) For $T \in [T_{\rm lea},T_{\rm utt}]$ and 
$\sigma \in [\sigma_{\rm a},\sigma_{\rm b}]$, the $\sigma(T,x)$ data
of all samples investigated satisfy the scaling relation Eq.\
(\ref{scaling}), repeated here for better readability,
\begin{equation}
\sigma(T,x) = \sigma_{\rm scal}(T/T_0(x))\,,
\label{scaling_a}
\end{equation}
where the control parameter values span the interval
$[x_{\rm a},x_{\rm b}]$; in the following, we abbreviate the quotient
$T/T_0(x)$ by $t$. (ii) At the lower end of
$[\sigma_{\rm a},\sigma_{\rm b}]$, \linebreak $\sigma(T,x  = {\rm const.})$
decreases exponentially with $T$ according to Eq.\ (\ref{mer}),
\begin{equation}
\sigma(T,x) = \sigma_0(x) \, \exp(-(T_0(x)/T)^\nu)\, .
\label{mer_a}
\end{equation}
The validity of the scaling relation Eq.\ (\ref{scaling_a}) implies
that $\sigma_0(x)$ is an $x$-independent constant.

As discussed in Subsection 2.5, we infer from (i) and (ii) that
$[x_{\rm a},x_{\rm b}]$ belongs to the insulating side of the MIT.
Strictly speaking, this conclusion relies on two plausible
suppositions: \linebreak (iii) For $T \in [T_{\rm lea},T_{\rm utt}]$,
Eq.\ (\ref{scaling_a}) is valid for any $x \in [x_{\rm a},x_{\rm b}]$.
(iv) $\lim_{T \rightarrow 0} \sigma(T,x) = 0$ holds as consequence of
(ii) and (iii) for any $x \in [x_{\rm a},x_{\rm b}]$.

In our further analysis, considering the $T$ interval
$(0,T_{\rm utt}]$, we rely on three plausible presumptions about the 
vicinity of the MIT: (v) The conductivity $\sigma(T,x)$ is a
continuous function of both $T$ and $x$. (vi) Without loss of
generality, for any $T > 0$, \linebreak $\sigma(T = {\rm const.},x)$
increases strictly monotonically with $x$. (vii) On the insulating
side of the MIT, $\sigma(T,x = {\rm const.})$ increases strictly
monotonically with $T$ within the scaling domain.

In consequence of (vi), (vii), Eq.\ (\ref{scaling_a}), and since
$\partial \sigma / \partial x = \mbox{d} \sigma_{\rm scal}(t) / \mbox{d} t \  \partial t / \partial x
= \mbox{d} \sigma_{\rm scal}(t) / \mbox{d} t \  ( - T / T_0(x)^2) \  \mbox{d} T_0(x) / \mbox{d} x 
= \mbox{d} \sigma_{\rm scal}(t) / \mbox{d} t \  (1 / T_0(x)) \ ( - T / T_0(x)) \  \mbox{d} T_0(x) / \mbox{d} x 
= \mbox{d} \sigma_{\rm scal}(t) / \mbox{d} t \  \partial t / \partial T \  ( - T / T_0(x)) \  \mbox{d} T_0(x) / \mbox{d} x 
= \partial \sigma / \partial T \  ( - T / T_0(x)) \  \mbox{d} T_0(x) / \mbox{d} x$ so that \linebreak
$\mbox{d} T_0(x) / \mbox{d} x = - (T_0(x) / T) \  \partial \sigma / \partial x \  \Big/ \ \partial \sigma / \partial T$, 
the function $T_0(x)$ must decrease strictly monotonically with
increasing $x$ and can thus be inverted. Hence, there is a function
$x_0(T)$ with $x_0(T_0(x)) = x$. Furthermore, because of (vi), the
critical control parameter value of the MIT, $x_{\rm c}$, must exceed
$x_{\rm b}$.

Now, as key to the physical interpretation, we make a first
generalizing hypothesis: (viii) We suppose that the scaling relation
Eq.\ (\ref{scaling_a}) holds whenever $T \in (0,T_{\rm utt}]$,
$\sigma \in (0,\sigma_{\rm b}]$, as well as $x \ge x_{\rm a}$ and that
$\lim_{t \rightarrow 0} \sigma_{\rm scal}(t) = 0$.

The generalizing hypothesis (viii) has two immediate consequences: It
excludes metallic conduction within the whole considered scaling
domain and thus implies the existence of a finite minimum metallic
conductivity, $\sigma_{\rm mm}$, related to $T \in (0,T_{\rm utt}]$,
where $\sigma_{\rm mm} \ge \sigma_{\rm b}$. Furthermore, hypothesis
(viii) means that, for $x \ge x_{\rm a}$, there is no non-metallic
conduction without $\sigma(T,x)$ satisfying Eq.\ (\ref{scaling_a}) at
sufficiently low $T$. Thus, this way, the corresponding control
parameter interval is extended beyond $x_{\rm b}$ up to the MIT at
$x_{\rm c}$.

We reach an important conclusion on $T_0(x)$: Without loss of
generality, be \linebreak $\sigma_{\rm scal}(T/T_0(x) = 1) = 
\sigma^\star \in [\sigma_{\rm a},\sigma_{\rm b}]$. For 
$T \in (0,T_{\rm lea}]$, we have $\sigma(T,x_{\rm a}) \le 
\sigma(T_{\rm lea},x_{\rm a}) \le \sigma_{\rm a}$, whereas, for
approaching the MIT at $x_{\rm c}$ from the metallic side,
$\sigma(T,x_{\rm c} + 0) \ge \sigma_{\rm mm} \ge \sigma_{\rm b}$.
Thus, according to (v), (vi), and the intermediate value theorem,
for any arbitrarily small $\tilde{T}$, there is one and only one
$\tilde{x}$ with $\sigma(\tilde{T},\tilde{x}) = \sigma^\star$. Since,
because of (viii) and the definition of $\sigma^\star$,
$T_0(\tilde{x})$ equals $\tilde{T}$, we conclude that $T_0(x)$ can
take arbitrarily small values. Therefore, due to $T_0(x)$ decreasing
strictly monotonically with increasing $x$ and due to the physical
condition $T_0(x) > 0$, the $x$ domain of
$\sigma_{\rm scal}(T/T_0(x))$ ends at 
$\lim_{\tilde{T} \rightarrow 0} x_0(\tilde{T})$, and so does the
validity region of the scaling relation Eq.\ (\ref{scaling_a}).

This control parameter value marks the MIT,
$x_{\rm c} = \lim_{\tilde{T} \rightarrow 0} x_0(\tilde{T})$, because,
as concluded above, for $x \ge x_{\rm a}$, there is no non-metallic
conduction without $\sigma(T,x)$ satisfying Eq.\ (\ref{scaling_a}) at
sufficiently low $T$. In other words, the generalizing hypothesis
(viii) together with the continuity and the strict monotonicity of
$\sigma(T = {\rm const.},x)$ imply $T_0(x) \rightarrow 0$ as
$x \rightarrow x_{\rm c} - 0$. In this sense, under the above
suppositions, scaling of the $\sigma(T, x = {\rm const.})$ curves for
various $x$ answers the question about the limiting behavior of the
mean hopping energy, posed in Subsection 2.2. 

Up to now, the condition $\sigma(T,x) \le \sigma_{\rm b}$ restricts
the domain of the argument $t = T/T_0(x)$ of the scaling function
$\sigma_{\rm scal}(t)$ to the finite interval $(0,t_{\rm b}]$ where
$t_{\rm b}$ is given by
$\sigma_{\rm scal}(t_{\rm b}) = \sigma_{\rm b}$. Thus, we are
confronted with the following problem: According to experimental
experience, close to the MIT, the scaling verification is hindered by
the increasingly weak $T$ dependence of $\sigma$ as
$x \rightarrow x_{\rm c}$, see, e.g., Figures 4 and 1 of Refs.\
\onlinecite{Moe.etal.83} and \onlinecite{Liu.etal.92}, respectively.
This is the natural consequence of $T_{\rm utt}/T_0(x)$ diverging as
$x \rightarrow x_{\rm c}$. 

Note, because of this singularity, to verify Eq.\ (\ref{scaling_a}) by
a mastercurve construction also for the region \linebreak
$\{T \le T_{\rm utt}, x < x_{\rm c}, \sigma > \sigma_{\rm b}\}$, one
would need an infinite number of samples. In consequence, there always
is a finite $x$ interval adjacent to the MIT where one cannot reliably
verify Eq.\ (\ref{scaling_a}) by experiment. Since it would seem
unphysical, however, that scaling would break down just when $x$
approaches $x_{\rm c}$, we rely on a second generalizing hypothesis:
(ix) We suppose Eq.\ (\ref{scaling_a}) to be valid whenever
$T \in (0,T_{\rm utt}]$ and $x \in [x_{\rm a},x_{\rm c})$. Within this
domain, $\sigma(T,x) \le \sigma(T_{\rm utt},x) <
\sigma(T_{\rm utt},x_{\rm c})$, so that $\sigma_{\rm c} =
\lim_{x \rightarrow x_{\rm c} - 0} \sigma(T_{\rm utt},x) =
\sigma(T_{\rm utt},x_{\rm c})$ can be interpreted as maximum
non-metallic conductivity.

Because of supposition (v), generalizing hypothesis (ix), and
$\lim_{x \rightarrow x_{\rm c} - 0} T_0(x) = 0$, we  finally obtain
that $\sigma(T,x_{\rm c}) =
\lim_{x \rightarrow x_{\rm c} - 0} \sigma(T,x) = 
\lim_{T_0 \rightarrow 0} \sigma_{\rm scal}(T/T_0) = 
\lim_{t \rightarrow \infty} \sigma_{\rm scal}(t) = 
\lim_{T_0 \rightarrow 0} \sigma_{\rm scal}(T_{\rm utt}/T_0) = 
\sigma_{\rm c}$  holds for any $T \in (0,T_{\rm utt}]$. -- In the
insulating region, the limits for $x$ and $T$ do not commute,
$\sigma_{\rm c} = 
\lim_{T \rightarrow 0} \lim_{x \rightarrow x_{\rm c}} \sigma(T,x) \ne
\lim_{x \rightarrow x_{\rm c}} \lim_{T \rightarrow 0} \sigma(T,x) = 
0$. -- 

Furthermore, we now see that, due to the monotonicity supposition
(vi), provided \linebreak $T \le T_{\rm utt}$, the relation
$\sigma(T,x) < \sigma_{\rm c}$ is satisfied if and only if
$x < x_{\rm c}$. Hence, because of the continuity supposition (v), the
maximum non-metallic conductivity coincides with the minimum metallic
conductivity, $\sigma_{\rm c} = \sigma_{\rm mm}$.

For the MIT itself, $x = x_{\rm c}$, we showed above that $\sigma$ is
independent of $T$, $\partial \sigma(T,x_{\rm c}) / \partial T = 0$.
We now ask for the behavior of $\partial \sigma(T,x) / \partial T$ in
the insulating vicinity of the MIT: Due to Eq.\ (\ref{scaling_a}) and
suppositions (vii) and (ix), $\sigma_{\rm scal}(t)$ increases
monotonically with $t$.  Therefore, the finiteness of
$\sigma_{\rm c} = \lim_{t \rightarrow \infty} \sigma_{\rm scal}(t)$
implies that
$\mbox{d} \sigma_{\rm scal}(t) / \mbox{d} \ln t \rightarrow 0$ as
$t \rightarrow \infty$. Because, moreover, 
$\mbox{d} \sigma_{\rm scal}(t) / \mbox{d} \ln t =
\mbox{d} \sigma_{\rm scal}(T/T_0) / \mbox{d} \ln T = 
T \mbox{d} \sigma_{\rm scal}(T/T_0) / \mbox{d} T = 
T \partial \sigma(T,x) / \partial T$, we thus obtain that, for any
$T \in (0,T_{\rm utt}]$, $\partial \sigma(T,x) / \partial T$
continuously tends to zero as $x \rightarrow x_{\rm c} - 0$.

This way, the scaling relation Eq.\ (\ref{scaling_a}) together with a
few plausible assumptions imply the condition
$\mbox{d} \rho / \mbox{d} T = 0$ to mark the MIT and to yield the
value of the minimum metallic conductivity. Since scaling according to
Eq.\ (\ref{scaling_a}) is a low-$T$ phenomenon, our mathematical
consideration supports the empirical MIT criterion
``$\mbox{d} \rho / \mbox{d} T = 0$ in the limit as
$T \rightarrow 0$'', discussed in detail in Subsection 2.2.

One may now ask, what would happen if any of the above suppositions
were violated in some respect, e.g., (v) by a small positive step in 
$\sigma(T = {\rm const.}, x)$ at $x_{\rm c}$, whereas, as
$x \rightarrow x_{\rm c} - 0$, Eq.\ (\ref{scaling_a}) and the limiting
behavior $T_0(x) \rightarrow 0$ were maintained for
$T \in (0,T_{\rm utt}]$. In this case, the answer is that, together
with the monotonicity supposition (vi) on the $x$ dependence of
$\sigma$, these relations would be sufficient to imply the existence
of a finite minimum metallic conductivity. This can also be easily
shown by an indirect proof: Suppose the MIT is continuous and 
$\lim_{x \rightarrow x_{\rm c} + 0} \sigma(T, x_{\rm c}) = b \, T^p$.
Then, for $T < (\sigma_{\rm c}/b)^{1/p}$, we get 
$\lim_{x \rightarrow x_{\rm c} - 0} \sigma(T, x) = \sigma_{\rm c} > 
\lim_{x \rightarrow x_{\rm c} + 0} \sigma(T, x)$ in contradiction to 
our monotonicity supposition.

The situation would be totally different if $T_0$ tended to a finite
value as $x \rightarrow x_{\rm c}$. Then 
$\lim_{x \rightarrow x_{\rm c} - 0} \sigma(T,x)$ would be
$T$-dependent, in contrast to our conclusion 
$\lim_{x \rightarrow x_{\rm c} - 0} \sigma(T,x) = \sigma_{\rm c}\, =
{\rm const.}$\ drawn above. The limit would fall off exponentially as 
$T \rightarrow 0$. Therefore, in consequence of the very likely
continuity of $\sigma(T = {\rm const.}, x)$, 
$\lim_{x \rightarrow x_{\rm c} + 0} \sigma(T,x)$ would fall off
exponentially as well, so that the MIT would be continuous. -- Note
that this limiting behavior would be incompatible with the
generalizing supposition (viii). -- Moreover, just on the metallic
side of the MIT, $\sigma(T, x = {\rm const.})$ had to follow a
stretched Arrhenius law augmented by a small $x$-dependent constant
contribution instead of an augmented power law. To the best of our
knowledge, however, such a situation has not been reported up to now.

\section{Method for numerical differentiation of functions given by
noisy values at non-equidistant points}

The present study relies to a large extent on the as accurate as
possible numerical differentiation of functions given by sets of
experimental data points. For explaining our approach to this
numerical task, we first remind that uncertainties of derivative
values approximated by difference quotients originate both from random
errors of the data and from nonlinearities of the considered function.
To reduce the influence of random errors, the argument intervals used
to calculate the difference quotients should be as wide as possible.
However, the wider the interval, the stronger is the influence of
nonlinearities. In consequence, there is an optimum interval width;
its size depends on which differentiation formula is used.

When not only two, but many data points are taken into account, the
corresponding averaging further reduces the influence of their random
errors. Thus, we aim to slide a window along the curve and to estimate
the local slope by means of linear regression on all data points
within the window. In case of equidistant argument values, this can be
realized by applying a special Savitzky-Golay filter; in this
approach, the slope value is related to the average of the smallest
and largest argument values within the window.\cite{Sav.Gol}

For the present study, however, we need a procedure which is
applicable to the general case of non-equidistant arguments. The
approach which we use in this work has proved to be very effective in
our research for many years; in particular, Eqs.\ (8) and (9) of Ref.\
\onlinecite{Moe.etal.99} were obtained this way. However, this
approach seems to be widely unknown. Therefore, we present here a
short derivation in general notation.

The idea of our procedure is the following: In each regression, we
additionally determine that argument value for which the linear
regression slope is expected to be the best approximation of the
derivative. In other words, we ask: For which argument choice do 
nonlinearities have the smallest influence? 

To answer this question, we need an adequate quantification of these
deviations. The truncation error, resulting from the truncation of the
corresponding Taylor series, is best suited for this aim. We remind
here that, in considering difference quotients obtained from two data
points with the arguments differing by $h$, the midpoint formula is
far superior to forward and backward differentiation formulas. The
reason is the following: In the former case, the leading term of the
truncation error arising from nonlinearities is proportional to $h^2$,
while in the latter cases it is proportional to $h$ and thus usually
far larger; the random error is always the same and proportional to
$h^{-1}$. Therefore, for the midpoint formula, the total error is
considerably smaller and the optimum interval width is much wider than
for the forward or backward formulas.

Consider now the physical quantities $x$ and $y$ which are linked by
the function $y = f(x)$. Assume, we performed $k$ measurements and
obtained the data points $(x_i,y_i)$, where at least two of the $x_i$
values differ from each other. Assume, furthermore, that the
uncertainties of the $x_i$ are negligible and that the random errors
of all $y_i$ have the same standard deviation. (The generalization of
the approach described here to the case of different standard
deviations of the random errors of the individual $y_i$ is
straightforward.)

In the first step, capturing only lowest-order nonlinearities, we
approximate $f(x)$ by the quadratic ansatz
\begin{equation}
\varphi(x) = a + b\,x + c\,x^2\,,
\label{dif1}
\end{equation}
where the values of the parameters $a$ and $b$ have to be adjusted,
and where, for the moment, we assume that we know the value of $c$. 

To determine the values of $a$ and $b$, we minimize the sum of the
squared deviations,
\begin{equation}
\sum_i (y_i - a - b\,x_i - c\,{x_i}^2)^2\ \rightarrow\ \min\,.
\end{equation}
Differentiation with respect to $a$ and $b$ yields the normal
equations,
\begin{equation}
a \, k + b  \, \sum_i x_i = \sum_i y_i - c \, \sum_i {x_i}^2\,,
\end{equation}
\begin{equation}
a \, \sum_i x_i + b  \, \sum_i {x_i}^2 = \sum_i x_i \, y_i - 
c \, \sum_i {x_i}^3\,,
\end{equation}
so that 
\begin{equation}
b = b_0 + b_1 \, c 
\label{dif2}
\end{equation}
with
\begin{equation}
b_0 = \frac{k \, \sum_i x_i \, y_i - \sum_i x_i \, \sum_j y_j}
{k \, \sum_i {x_i}^2 - (\sum_i x_i)^2}\,,
\label{dif3}
\end{equation}
\begin{equation}
b_1 = \frac{-k \, \sum_i {x_i}^3 + \sum_i x_i \, \sum_j {x_j}^2}
{k \, \sum_i {x_i}^2 - (\sum_i x_i)^2}\,.
\end{equation}
Equation (\ref{dif3}) is the well-known result for the slope obtained
by linear regression. Finally, combining Eqs.\ (\ref{dif1}) and
(\ref{dif2}), we get the derivative approximation
\begin{equation}
\frac{\mbox{d} \varphi}{\mbox{d} x}(x) = b_0 + (b_1 + 2 \, x) \, c\,.
\label{dif4}
\end{equation}

Now, the decisive idea is to focus on that $x$ value, $x_{\rm a}$,
where we do not need the knowledge of $c$, which we only pretended to
have above. The prefactor of $c$ on the right-hand side of Eq.\ 
(\ref{dif4}) vanishes at 
\begin{equation}
x_{\rm a} = - \frac{b_1}{2} = 
\frac{k \, \sum_i {x_i}^3 - \sum_i x_i \, \sum_j {x_j}^2}
{2 \, (k \, \sum_i {x_i}^2 - (\sum_i x_i)^2)}\,.
\end{equation}

Thus, at $x_{\rm a}$, as in the case of the midpoint formula of
differentiation, the quadratic contribution to $\varphi(x)$ has no 
influence on the calculated value of $\mbox{d} \varphi / \mbox{d} x$.

Hence, for any quadratic polynomial $f(x)$, arbitrarily positioned
$x_i$, and negligible random errors of $y_i$, the linear regression
slope $b_0$ when related to the specific argument average $x_{\rm a}$
exactly equals the derivative value. (In case, the $x_i$ are
equidistant, $x_{\rm a}$ reduces to the mean value of smallest and
largest $x_i$ considered, which is in agreement with Ref.\
\onlinecite{Sav.Gol}. For $k = 2$, as well as for $k = 3$ with
equidistant $x_i$, our result is identical to the midpoint formula of
differentiation.)

In the second step, we consider an arbitrary analytic function $f(x)$,
but neglect the random errors of the $y_i$. Inserting the Taylor
expansion of $f(x)$ about $x_{\rm a}$ into Eq.\ (\ref{dif3}) yields
after a short calculation
\begin{eqnarray}
&& b_0 = \frac{\mbox{d} f}{\mbox{d} x} (x_{\rm a}) \, + \nonumber\\
&& \nonumber\\
&& \frac{1}{6} \, 
\frac{k \, \sum_i {\overline{x}_i}^4 - \sum_i \overline{x}_i \, 
\sum_j {\overline{x}_j}^3}
{k \, \sum_i {\overline{x}_i}^2 - (\sum_i \overline{x}_i)^2} 
\, \frac{\mbox{d}^3 f}{\mbox{d} x^3} (x_{\rm a}) \, +  \nonumber\\
&& \nonumber\\
&& ...\,,
\end{eqnarray}
where $\overline{x}_i = x_i - x_{\rm a}$. Hence, when rescaling all
$\overline{x}_i$ by a factor $\alpha$, one has to multiply the leading
term of the truncation error of  $\mbox{d} f / \mbox{d} x (x_{\rm a})$
by the factor $\alpha^2$. In this sense, when the number of data
points and all quotients of argument differences, $x_i - x_j$, are
kept fixed, the truncation error of the value of
$\mbox{d} f / \mbox{d} x(x_{\rm a})$ is proportional to the square of
the window width, $(\max \{x_i\} - \min \{x_i\})^2$. Thus, this
truncation error is usually far smaller than that of the value of
$\mbox{d} f / \mbox{d} x(x)$ for $x \ne x_{\rm a}$, which is
proportional to the window width itself. This result corresponds to
the properties of the truncation errors of two-point formulas
mentioned above.

Therefore, relating to $x_{\rm a}$ often enables the use of quite wide
windows including rather large numbers of data points. In this way,
the random errors of the derivative values, obtained by numerical
differentiation, are usually considerably reduced compared to the
random dispersion of values resulting from two-point formulas. --
Note, however, that these random errors are always correlated over the
window width. -- Furthermore, due to the averaging, the prefactor in
the truncation error of the derivative value obtained by our approach
is usually appreciably smaller than that for the two-point midpoint
formula considering the same argument interval. In consequence, a far
smaller total uncertainty of the value of $\mbox{d} f / \mbox{d} x$
can be reached by linear regression with relating the slope to
$x_{\rm a}$ than by applying common difference quotient approaches.

It is worth mentioning that the influence of the truncation error may
be diminished even further. For this aim, $f(x)$ has to be transformed
to a function $\phi(\xi)$ with weaker nonlinearity. After calculating
$\mbox{d} \phi / \mbox{d} \xi$ by means of the procedure described
above, one obtains the $\mbox{d} f / \mbox{d} x$ data using common
differentiation rules.

Finally, we would like to emphasize that the presented method is very
flexible. We have utilized it in various ways, sliding a window along
experimental relations $f(x)$ and obtaining pointwise
$\mbox{d} f / \mbox{d} x$ from all $(x_i,y_i)$ within the window. As
always with numerical differentiation, the art of applying this method
consists in the choice of an appropriate window width which ensures a
reasonable compromise between conflicting demands: On the one hand,
the random errors of the obtained derivative values of $f(x)$ should
be as small as possible, while, on the other hand, the truncation
errors should not be too large and the essential features of
$\mbox{d} f / \mbox{d} x (x)$ must not be smeared out too much.


\begin{thebibliography}{999}
\bibitem{cla}
  Classifying solids, which can be investigated only at finite 
  temperatures, according to a zero-temperature property may sound
  strange. At the current stage, however, it is the only possibility
  of an unambiguous and physically meaningful discrimination between
  metallic and non-metallic behaviour, see Subsection 2.1.
\bibitem{lin_approx}
  The effect of the inflection point becomes clear when, as a 
  simple example, linear approximations of $f(x) = \sin(x)$ and
  $g(x) = \cos(x)$ around $x = 0$ are compared: While 
  $\tilde{f}(x) = 0.99\, x$ satisfies $|f(x) - \tilde{f}(x)| < 0.001$
  as long as $|x| < 0.285$, $\tilde{g}(x) = 0.999$ fulfills
  $|g(x) - \tilde{g}(x)| < 0.001$ only within a far smaller interval,
  $|x| < 0.063$.
\end{thebibliography}
\end{document}